Michel Janssen and Jürgen Renn

# Einstein and the Perihelion Motion of Mercury

## Excerpts from *How Einstein Found His Field Equations. Sources and Interpretation* (forthcoming)

**Abstract** On November 23, 2021, the Einstein-Besso manuscript on the perihelion motion of Mercury will be auctioned at Christie's. Expected to fetch around $3M, it promises to be the most expensive scientific manuscript ever sold at auction. In this preprint, we present the parts of our forthcoming book, *How Einstein Found His Field Equations. Sources and Interpretation* (Springer, 2021) dealing with Einstein's attempts, in 1913 and in 1915, to account for the anomalous advance of Mercury's perihelion. In 1913, as documented in the Einstein-Besso manuscript, Einstein and his friend Michele Besso found that the Einstein-Grossmann or *Entwurf* (= outline or draft) theory, a preliminary version of general relativity, could only account for 18 of the 43 seconds-of-arc-per-century discrepancy between Newtonian theory and observation. In November 1915, however, putting the techniques developed in his collaboration with Besso to good use, Einstein showed that his new general theory of relativity could account for all missing 43″. After a brief introduction, we provide an annotated transcription of the key pages of the Einstein-Besso manuscript and an annotated new translation of the November 1915 perihelion paper. For permission to post these materials on the arXiv, we are grateful to our publisher, Springer, the Albert Einstein Archives at the Hebrew University of Jerusalem (and its director, Hanoch Gutfreund) and the Einstein Papers Project at Caltech (and its director, Diana Buchwald).

# Part I
# Essay: How Einstein Found His Field Equations

# Chapter 6
# Mercury's perihelion: From 18″ in the *Entwurf* theory to 43″ in general relativity

On November 18, 1915, Albert Einstein delivered a lecture on the perihelion motion of Mercury to the Prussian Academy of Sciences.[1] The paper based on this lecture was submitted that same day and published one week later (Einstein 1915c). Einstein was able to show that his new theory added 43″ to the more than 500″ per century that could be attributed to the perturbation of other planets, thereby closing the gap of $45'' \pm 5''$ between theory and observation.[2] As he told Arnold Sommerfeld on December 9, 1915 (CPAE8, Doc. 161), this gave him a whole new appreciation of "the pedantic precision of astronomy that I secretly used to make fun of in the past."[3] Years later, he told Adriaan Fokker that the result had given him heart palpitations (Pais 1982, p. 253). His excitement also comes through in a letter to his friend Heinrich Zangger of November 15, 1915: "*I have now derived the up to this point unexplained anomalies in the motion of the planets from the theory.* Imagine my good fortune! [*Stellen Sie sich mein Glück vor!*]" (CPAE8, Doc. 144a, in CPAE10). The day he submitted the paper, he shared the good news with his Göttingen competitor, David Hilbert. "Congratulations on conquering the perihelion motion," an astonished Hilbert wrote back the very next day,

> If I could calculate as fast as you can, the electron would be forced to surrender to my equations and the hydrogen atom would have to bring a note from home to be excused for not radiating (CPAE8, Doc. 149).

Einstein did not bother to tell his colleague that he had done very similar calculations before. He probably enjoyed giving Hilbert a taste of his own medicine. A few months later, he characterized Hilbert's style as "creating the impression of being superhuman by obfuscating one's methods."[4]

---

[1] This chapter is part of the introductory essay of our forthcoming book, *How Einstein Found His Field Equations. Sources and Interpretation* (Springer, 2021). The essay builds on our account of Einstein's path to the Einstein field equations in Janssen and Renn (2007, 2015), which in turn builds on earlier accounts of this episode, notably on classic papers by John Stachel (1989) and John Norton (1984).

[2] For the history of the problem of the anomalous motion of Mercury's perihelion, see Roseveare (1982), Earman and Janssen (1993), and Smith (2014, pp. 307–317).

[3] The German astronomy community did not endear itself to Einstein by making life difficult for his protégé Erwin Freundlich (Hentschel 1992, 1994; Janssen 1999). As he wrote to Zangger in a letter dated before December 4, 1915: "The astronomers behave like an anthill disturbed in its mindless activity by a hiker who has inadvertently stepped in it: they start biting at the hiker without making a dent in his shoes" (CPAE8, Doc. 159a, in CPAE10).

In the late 1980s, the Einstein Papers Project obtained a copy of a manuscript in possession of descendants of Einstein's close friend Michele Besso, consisting of about 50 pages of scratchpad calculations, showing that Einstein and Besso used the *Entwurf* field equations to calculate the perihelion motion of Mercury. Most of these notes can be dated to June 1913.[5] Einstein scholars were surprised to learn that Einstein had already calculated the perihelion motion of Mercury in 1913. That he had done so in collaboration with Besso surprised them even more. Besso had always been thought of as an important sounding board for Einstein, never as a serious scientific collaborator. Einstein eventually handed over the project to Besso. In a letter that can be dated to early 1914, he told his friend: "Here you finally have your manuscript package. It is really a shame if you do not bring the matter to completion" (CPAE5, Doc. 499). It is safe to assume that this refers to the Einstein-Besso manuscript and the perihelion problem. It is only because the manuscript ended up in Besso's hands rather than Einstein's, who almost certainly would have discarded it, that we can belatedly call Einstein's bluff trying to put one over on Hilbert.

Einstein's exhortation did not fall on deaf ears. Material made available to Einstein scholars by the Besso family in 1998 shows that Besso initially planned an ambitious paper on the perihelion problem.[6] It is not clear when exactly he abandoned the project but one surmises he had well before he and Einstein were scooped in December 1914. In a paper submitted that month, Johannes Droste (1915, p. 1010), like Fokker a student of Hendrik Antoon Lorentz, reported that the Leyden astronomer Willem de Sitter, using equations of motion derived by his colleague Lorentz, had found that the *Entwurf* field equations predict an additional advance of Mercury's perihelion of only 18″ per century (Röhle 2007, pp. 197–200).[7]

The calculations of 1913 undoubtedly helped Einstein in his calculation of the perihelion motion in November 1915. A comparison between the relevant pages of the Einstein-Besso manuscript and the perihelion paper shows that there are many similarities between the two sets of calculations (see Pt. II, Chs. 2 and 6). Yet, there are also important differences. The most important one perhaps is that, in 1915, Einstein determined the field of the sun in terms of the Christoffel symbols, whereas, in 1913, he determined it in terms of the metric tensor (Earman and Janssen 1993, secs. 5–7). This difference between the calculations of 1913 and 1915, we like to think, reflects his epiphany that the Christoffel symbols rather than the gradient of the metric field represents the gravitational field. These differences help to underscore that it remains an impressive feat that Einstein was able to produce the 1915 perihelion paper as fast as he did. Still, given the undeniable importance of his earlier calculations, one can legitimately ask why Einstein did not invite Besso as a co-author. As it happened, Besso did not even get an acknowledgment. We suspect that this is largely because of the race Einstein perceived himself to be in with Hilbert. He was in a hurry[8] and it probably never even occurred to him to ask Besso to write a joint paper on the topic.

---

[4] Einstein to Ehrenfest, May 24, 1916 (CPAE8, Doc. 220). In a letter a week later, on May 30, 1916, Einstein also complained about this to Hilbert himself, albeit in more polite terms: "Why do you make it so hard for poor mortals by withholding the technique behind your thinking?" (CPAE8, Doc. 223, quoted in Landsman 2021, p. 18).

[5] For more information on the Einstein-Besso manuscript, see Pt. II, Sec. 2.3

[6] For analysis of the pages Besso added to the Einstein-Besso manuscript in 1914, see Janssen (2007, pp. 790–806).

[7] On an undated page of Einstein's so-called Scratch Notebook (CPAE3, Appendix A, [p. 61]), a value of 17″ is given (CPAE4, p. 345). See our commentary on [p. 28] of the Einstein-Besso manuscript in Pt. II, Sec. 2.3, for discussion.

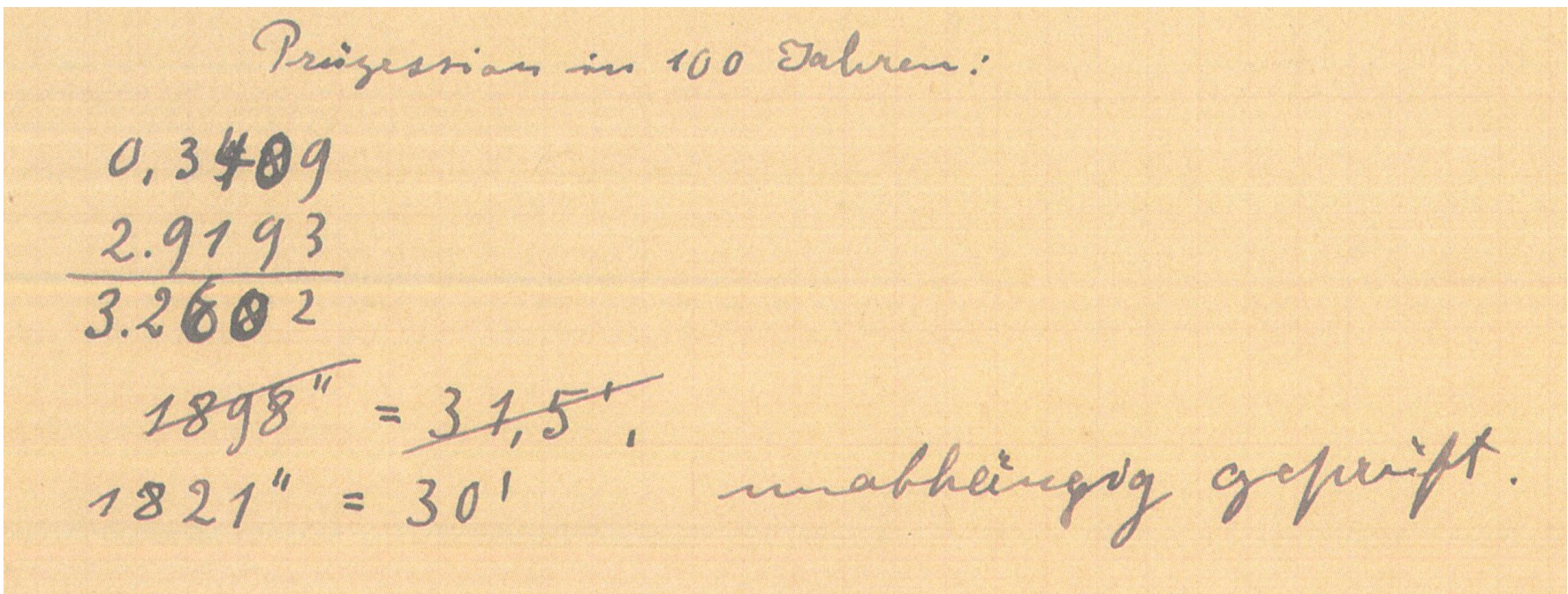

**Figure 6.1** On [p. 28] of the Einstein-Besso manuscript, Einstein recorded the value for the "precession in 100 years" (*Präzession in 100 Jahren*) of Mercury's perihelion produced by the field of the Sun that he and Besso found in June 1913 using the *Entwurf* theory. This theory predicts 5/12 the size of the effect predicted by general relativity, which comes to about 18″ per century. So the value that Einstein here claims was "independently checked" (*unabhängig geprüft*) is too large by a factor of 100. If this value were correct, the contribution of the field of the Sun to the perihelion motion of Mercury would be more than three times the effect of about 530″ per century coming from the perturbations of all other planets combined! Einstein almost certainly realized that his result was off by a factor of 100 but it looks as if Besso first found the source of the error (Figs. 6.2 and 6.3).

In fact, none of the three postcards of late 1915 in which Einstein enthusiastically reported his success with Mercury to Besso explicitly refer to their earlier collaboration.[9] The second and the third at least implicitly do. "You will be surprised," Einstein wrote in the second, "about the occurrence of $g_{11}$ ... $g_{33}$." In the third, he similarly wrote: "it is only because in first-order approximation $g_{11}$–$g_{33}$ do not occur in the equations of motion for a point particle that the Newtonian theory is so simple." Einstein reiterated this point

[8] An unusually high number of typos in the paper suggests that it was written in haste (see Pt. II, Sec. 6.3.3).

[9] Einstein to Besso, November 17, December 10, and December 21, 1915 (CPAE8, Docs. 147, 162 and 168)

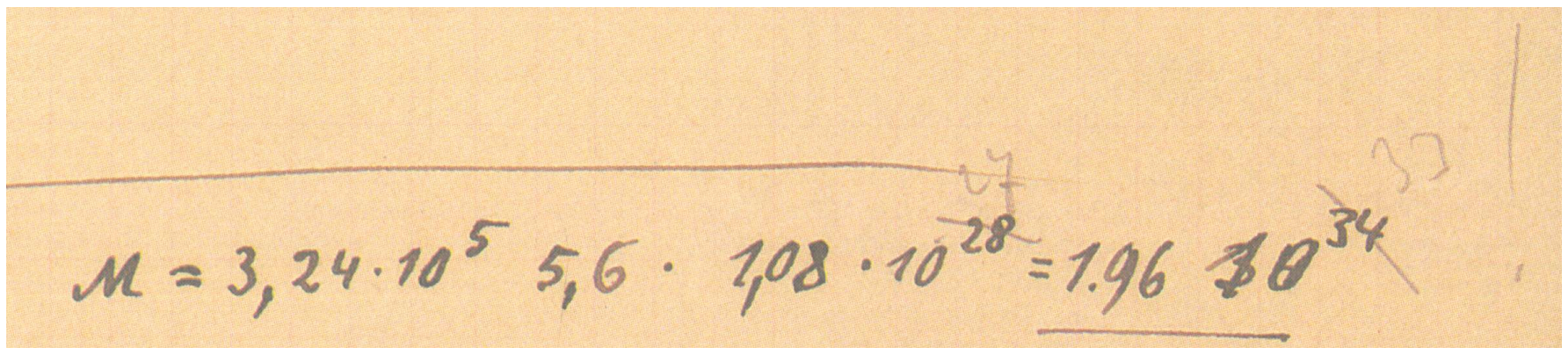

**Figure 6.2** On [p. 35] of the Einstein-Besso manuscript, Besso located the source of the error in the value Einstein reported on [p. 28] for the perihelion motion of Mercury (Fig. 6.1). The material in pen, the numerical calculation of $M$, the mass of the Sun, is in Einstein's hand but Besso added the corrections and exclamation point in pencil. The value for $M$ is found as the product of three factors, the ratio of the Sun's mass to the Earth's mass, the Earth's density and the Earth's volume. Besso noticed that the value for the Earth's volume was off by a factor of 10. Since $M$ occurs squared in the formula for the perihelion shift, this throws off the end result by a factor of 100.

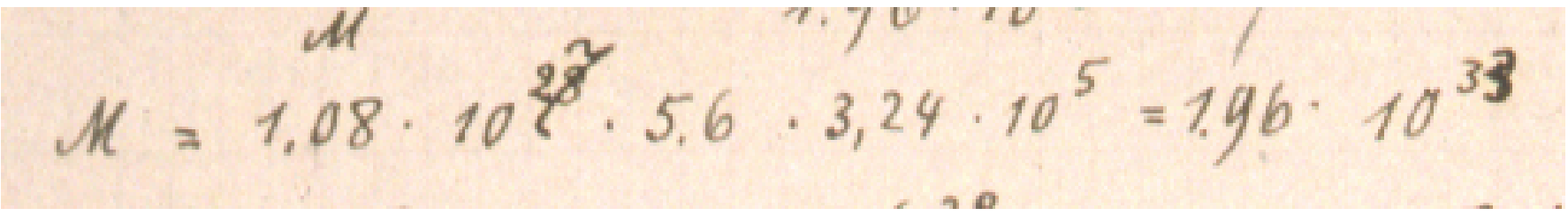

**Figure 6.3** On [p. 30] of the Einstein-Besso manuscript, we find the same calculation of $M$ that we found on [p. 35] (Fig. 6.2). It also has the same correction, this time in Einstein's hand. The exclamation point on [p. 35] suggests that Besso found the mistake first.

in a letter of January 3, 1916, in which he finally did explicitly refer to their earlier calculations:[10]

> That the effect is so much larger than in *our calculation* is because in the new theory the $g_{11}$–$g_{33}$ appear in first order and thus contribute to the perihelion motion (CPAE8, Doc. 178; our emphasis).

When Einstein and Besso collaborated in 1913, the metric they found as a first-order solution of the *Entwurf* field equations had only one variable component, viz. $g_{44}$, in accordance with Einstein's prejudice about the metric for weak static fields at the time. The new perihelion calculations freed him from this prejudice.

Besso made substantial contributions to the calculations in 1913. Einstein solved the *Entwurf* field equations for a point mass, using the same iterative approximation procedure he used in 1915. He was surprised when, only a month later, Karl Schwarzschild sent him the exact solution, which was published early the following year (Schwarzschild 1916).[11] Once Einstein had found the field, Besso derived the differential equation for the angle traversed by the radius of a planet moving in this field as a function of its distance to the Sun. Einstein integrated this equation to find the angle between the minimum and maximum value of this distance, i.e., between perihelion and aphelion. The deviation of this angle from $\pi$ gives the perihelion motion in radians per half a revolution. Converting this result to seconds of arc per century, he found a value of 1821 seconds or 30 *minutes* of arc! Next to this bizarre result, which is more than three times the total secular motion of Mercury's perihelion, he wrote: "independently checked" (Fig. 6.1). The correct value of 18″ is nowhere to be found in the manuscript. Both Besso and Einstein, however, located the error. The value they used for the mass of the Sun, which occurs squared in the final formula, was too large by a factor of 10 (Figs. 6.2 and 6.3).

---

[10] In a letter to Besso dated after December 6, 1916, Einstein once again mentions Mercury's perihelion without referring to their joint efforts. In this letter Einstein dismisses two attempts to account for Mercury's perihelion motion without general relativity, a new one by Emil Wiechert (1916) and an old one by Paul Gerber (1917), republished on the initiative of Ernst Gehrcke (CPAE8, Doc. 287a, in CPAE10). For Gehrcke's role in the anti-relativity movement, see Wazeck (2009).

[11] See Schwarzschild to Einstein, December 22, 1915 (CPAE8, Doc. 169) and Einstein to Schwarzschild, December 29, 1915 and January 9, 1916 (CPAE8, Docs. 176 and 181). "I

In the letter to Sommerfeld of November 28, 1915 (CPAE8, Doc. 153), Einstein listed the value of 18″ as one of his reasons for abandoning the *Entwurf* theory. It was generally assumed that he got that number from Droste's paper. The Einstein-Besso manuscript shows that he and Besso had already found it in 1913. Evidently, Einstein had not been too worried at the time about the discrepancy with the missing $45'' \pm 5''$.[12]

The Einstein-Besso manuscript thus allows us to answer a question that, before its discovery, would have been a purely hypothetical. What would Einstein have done had his theory predicted the wrong result for the perihelion motion of Mercury? If he had followed the prescriptions of Karl R. Popper (1959), he should have accepted that his theory had been falsified and gone back to the drawing board. Einstein did nothing of the sort. He kept quiet about the 18″ and continued to work on the *Entwurf* theory. This was perfectly rational. The theory would have been decisively refuted had it truly predicted 1800″; that it could only account for less than half the missing seconds was undoubtedly disappointing but not particularly troublesome. It does seem to diminish the importance of getting it exactly right, however, if being off by more than 50% is no big deal. The older Einstein might well have agreed with that assessment. In a letter to Max Born of May 12, 1952, the sage from Princeton had this to say on the three classic tests of general relativity:

---

would not have thought that the exact solution of the point problem would be so simple," he told Schwarzschild in the first of these two letters. Einstein used this exact solution (following a derivation given by Hermann Weyl 1918, pp. 202–205) when he covered the calculation of the perihelion advance of Mercury in lectures on general relativity in Berlin and Zurich in 1919 (see CPAE7, Docs. 19 and 20 and Janssen and Schulmann 1998).

[12] Einstein and Besso explored whether the *Entwurf* theory predicted additional effects on the perihelion motion that could account for another 25″ or so but came up empty handed. In particular, they considered the effect of the rotation of the Sun. Even though they overestimated it by at least two orders of magnitude, this effect was much too small. Worse yet, it was in the wrong direction (CPAE4, pp. 354–356).

> Even if there had been no deflection of light, no perihelion motion and no redshift, the gravitational equations would still be convincing because they avoid the inertial system (the phantom that affects everything but is not itself affected). It is actually rather curious that humans are mostly deaf to the strongest arguments, while they always tend to overestimate the accuracy of measurements (Einstein and Born 1969, Doc. 99, p. 192).

# Part II
# Sources

# Chapter 2

# The Einstein-Besso Manuscript on the Perihelion Motion of Mercury

## 2.2 Transcription

[p. 1] Einstein

Gleichungen der Grav. in erster Näherung (1)

[eq. 1] $$-\left(\frac{\partial^2\gamma_{\mu\nu}}{\partial x^2}+\cdot+\cdot-\frac{1}{c^2}\frac{\partial^2\gamma_{\mu\nu}}{\partial t}\right)=\kappa\rho_0\frac{dx_\mu}{\partial s}\frac{dx_\nu}{\partial s} \qquad ds^2=-dx^2-.-.+c_0^2dt^2$$

Statischer Fall

[eq. 2] $$-\Delta\gamma_{44}=\frac{\kappa\rho_0}{c_0^2}$$ $$\boxed{\kappa=K\frac{8\pi}{c_0^2}}$$ [eq. 3]

Tabelle des Schwerefeldes für die <u>erste Annäherung:</u>

[eq. 4] $$\gamma: \begin{matrix} -1 & 0 & 0 & 0\\ 0 & -1 & 0 & 0\\ 0 & 0 & -1 & 0\\ 0 & 0 & 0 & \frac{1}{c_0^2}(1+\frac{A}{r})\end{matrix} \qquad g: \begin{matrix} -1 & 0 & 0 & 0\\ 0 & -1 & 0 & 0\\ 0 & 0 & -1 & 0\\ 0 & 0 & 0 & c_0^2(1-\frac{A}{r})\end{matrix}$$ [eq. 5]

[eq. 6] $$\boxed{A=\frac{\kappa M}{4\pi}}=\frac{2KM}{c_0^2}$$ [eq. 7] $$g=-c_0^2(1-\frac{A}{r})$$

Zweite Annäherung.

$$-\left(\frac{\partial^2\gamma^x_{\mu\nu}}{\partial x^2}+\cdot+\cdot-\frac{1}{c_0^2}\frac{\partial^2\gamma^x_{\mu\nu}}{\partial t^2}\right) \quad \text{(struck through)}$$

1) $$\gamma^{\langle 0\rangle}_{\alpha\beta}\frac{\partial^2\gamma^{\langle x\rangle}_{\mu\nu}}{\partial x_\alpha\partial x_\beta}$$

$$\sum\gamma^0_{\alpha\alpha}\frac{\partial^2\gamma^x_{\mu\nu}}{\partial x_a^2}$$

2) $$\begin{matrix} 0 & 0 & 0 & 0\\ - & - & - & -\\ - & - & - & -\\ 0 & 0 & 0 & \frac{A^2}{2c_0^2r^4}\end{matrix}$$

3) $$-\sum\sum\gamma_{\alpha\beta}g_{\tau\rho}\frac{\partial\gamma_{\mu\tau}}{\partial x_\alpha}\frac{\partial\gamma_{\nu\rho}}{\partial x_\beta}$$

$$\begin{matrix} 0 & 0 & 0 & 0\\ - & - & - & -\\ - & - & - & -\\ 0 & 0 & 0 & +\frac{A^2}{c_0^2r^4}\end{matrix}$$

$$-\frac{1}{2}\sum\gamma_{\alpha\mu}\gamma_{\beta\nu}\frac{\partial g_{\tau\rho}}{\partial x_\alpha}\frac{\partial\gamma_{\tau\rho}}{\partial x_\beta}$$

$$\frac{A^2}{2}\ \Bigg|\ \begin{matrix} \left(\frac{\partial\frac{1}{r}}{\partial x}\right)^2 & \frac{\partial\frac{1}{r}}{\partial x}\frac{\partial\frac{1}{r}}{\partial y} & . & 0\\ . & . & . & 0\\ . & . & . & 0\\ 0 & 0 & 0 & 0\end{matrix}$$

$$+\frac{1}{4}\gamma_{\mu\nu}\gamma_{\alpha\beta}\frac{\partial g_{\langle\mu\nu\rangle\tau\rho}}{\partial x_\alpha}\frac{\partial\gamma_{\tau\rho}}{\partial x_\beta}$$

$$\frac{A^2}{4}\ \Bigg|\ \begin{matrix} -\frac{1}{r^4} & 0 & 0 & 0\\ 0 & -\frac{1}{r^4} & 0 & 0\\ 0 & 0 & -\frac{1}{r^4} & 0\\ 0 & 0 & 0 & \frac{1}{c_0^2r^4}\end{matrix}$$

[p. 7] Einstein

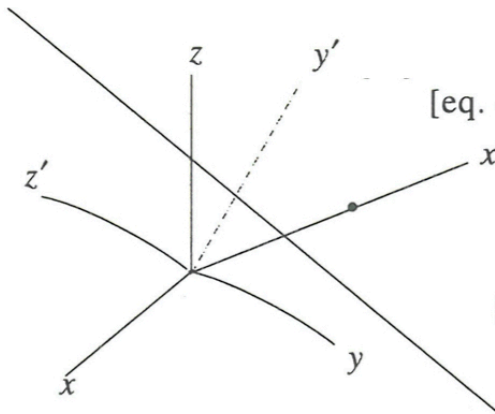

$$<T>g'_{11} = R$$

[eq. 43] $$<T>g'_{22} = <T>g'_{33} = T$$

$$<T>g'_{44} = \phi$$

$$\underline{dx_i} = \sum p_{ik} dx'_k = \sum p_{ik} \pi_{mk} dx_m$$

$$\underline{dx'_l} = \sum \pi_{m \cdot l} dx_m$$

$$= \sum_{mk} \pi_{ml} p_{mk} dx'_k$$

[eq. 44] $$g_{\mu\nu} = \sum \pi_{\mu\alpha} \pi_{\nu\beta} g'_{\alpha\beta}$$

[eq. 45] $$g'_{rs} = \sum p_{\mu r} p_{\nu s} g_{\mu\nu}$$

$$\pi_{<\mu>\alpha<\nu>r} \pi_{\beta s}$$

$$\sum_{rs} \pi_{\alpha r} \pi_{\beta s} g'_{rs} = \sum_{\mu\nu rs} p_{\mu r} \pi_{\alpha r} p_{\nu s} \pi_{\beta s} g_{\mu\nu}$$

$$g_{\alpha\beta}$$

$$g_{11} = \sum \pi_{1\alpha} \pi_{1\beta} g'_{\alpha\beta}$$

$$= \pi_{11}^2 R + (\pi_{12}^2 + \pi_{13}^2)\, T$$

$$= \pi_{11}^2 (R - T) + T \qquad \pi_{11} = \frac{x}{r}$$

[eq. 46] $$g_{11} = \frac{x^2}{r^2}(R - T) + T$$

| | | |
|---|---|---|
| $T$ | 0 | 0 |
| 0 | $T$ | 0 |
| 0 | 0 | $T$ |

$$g_{12} = \sum \pi_{1\alpha} \pi_{2\beta} g'_{\alpha\beta}$$

$$= \pi_{11}\pi_{21} R + (\pi_{12}\pi_{22} + \pi_{13}\pi_{23})\, T$$

$$-\pi_{11}\pi_{21}$$

$$= \pi_{11}\pi_{21}(R - T) = \frac{xy}{r^2}(R - T)$$ [eq. 47]

$$2r\sin\varphi\,(\underline{\sin\varphi} \cdot dr + r\cos\varphi\, d\varphi)$$

$$+2\,(\underline{r\cos\varphi} - 2e)\,(\underline{\cos\varphi} \cdot dr - \underline{r\sin\varphi d\varphi}) = -2\,(2a - r)\, dr$$

$$2r dr <+>^0 + 2e\,(\underline{\cos\varphi} dr - r\sin\varphi d\varphi) = +2\,(2a - r)\, dr$$

$$(e\cos\varphi - 2a)\, dr = r\sin\varphi d\varphi$$

$$d\varphi = \frac{e\cos\varphi - 2a}{r\sin\varphi} dr$$

## [p. 3] Einstein

Form der $\gamma_{\mu\nu}$ bei Radiussymmetrie.

[eq. 18] $R - T = <D><M>N$ (2)

[eq. 17]
$$\begin{matrix} \frac{x^2}{r^2}N+T & \frac{xy}{r^2}N & \frac{xz}{r^2}N & 0 \\ - & \frac{y^2}{r^2}N+T & - & 0 \\ - & - & - & 0 \\ 0 & 0 & 0 & \Phi \end{matrix}$$

$|g| = \Phi T^2 (N+T)$ [eq. 19]

Diff. Gl. für die $\gamma$.

$<11>\mu = 1 \quad \nu = 1$

[eq. 20] $$-\Delta(\gamma^x_{11}) = \frac{A^2}{2}\left[\left(\frac{\partial \frac{1}{r}}{\partial x}\right)^2 - \frac{1}{2}\frac{1}{r^4}\right]$$

$\mu = 1 \quad \nu = 2$

[eq. 21] $$-\Delta(\gamma^x_{12}) = \frac{A^2}{2}\frac{\partial \frac{1}{r}}{\partial x}\frac{\partial \frac{1}{r}}{\partial y}$$

$\mu = 4 \quad \nu = 4$

$$-\Delta(\gamma^x_{44}) + \frac{A^2}{2c_0^2 r^4} + \frac{A^2}{c_0^2 r^4} = \frac{A^2}{4c_0^2 r^4}$$

[eq. 22] $$\Delta(\gamma^x_{44}) = -\frac{<3>5}{4}\frac{A^2}{c_0^2 r^4}$$

$$\Delta T = \frac{\partial^2 T}{\partial r^2} + \frac{2}{r}\frac{\partial T}{\partial r}$$

$$\Delta\left(x^2\frac{N}{r^2}\right) = <6>2\frac{N}{r^2} + \frac{<x(>6x^2}{r}\frac{\partial}{\partial r}\left(\frac{N}{r^2}\right) + x^2\frac{\partial^2}{\partial r^2}\left(\frac{N}{r^2}\right)$$

Diff. Gl. für $N\ T\ \Phi$.

[eq. 23] $$\overbrace{\frac{<\partial>d^2 T}{\partial r^2} + \frac{2}{r}\frac{dT}{dr}}^{\Delta T} + 2\frac{N}{r^2} + \frac{x^2}{r^2}\left(-\frac{6}{r^2}N + \frac{2}{r}\frac{dN}{dr} + \frac{d^2N}{dr^2}\right) = -\frac{A^2}{2}\left(-\frac{1}{2r^4} + \frac{x^2}{r^6}\right)$$

[eq. 24] $$\frac{d^2T}{dr^2} + \frac{2}{r}\frac{dT}{dr} + \frac{2}{r^2}N = +\frac{A^2}{4r^4}$$

[eq. 25] $$-\frac{6}{r^2}N + \frac{2}{r}\frac{dN}{dr} + \frac{d^2N}{dr^2} = -\frac{A^2}{2r^4}$$

zweite Gleichung ergibt nichts Neues.

[eq. 26] $$\Delta\varphi = -\frac{<3>5}{4}\frac{A^2}{r^4}$$

Tabelle der $\gamma^{<x>}_{\mu\nu}$ in zweiter Näherung

$$\begin{matrix} -1 + \frac{A^2}{8}\frac{x^2}{r^{<2>4}} & \frac{A^2}{8}\frac{xy}{r^{<2>4}} & \cdot & 0 \\ \cdot & \cdot & \cdot & 0 \\ \cdot & \cdot & \cdot & 0 \\ 0 & 0 & 0 & \frac{1}{c_0^2}\left(1 + \frac{A}{r} + \frac{3}{8}\frac{A^2}{r^2}\right) \end{matrix}$$

[p. 4] Einstein

~~$\delta_x = -m$~~

2a)

$-\frac{1}{2}$

$-6+8$

$\varphi = \frac{\gamma}{r^2}$

$+\frac{3}{4}+1$

[eq. 27] $\Delta\left(T + x^2\frac{N}{r^2}\right)$

$$\frac{\partial T}{\partial x} = x<->\frac{1}{r}\frac{<\partial>dT}{dr}$$

$$\frac{d^2T}{dx^2} = \frac{1}{r}\frac{dT}{dr} + x^2\cdot\frac{1}{r}\frac{d}{dr}\left(\frac{1}{r}\frac{dT}{dr}\right)$$

[eq. 28] $\Delta T = \frac{3}{r}\frac{dT}{dr} + r\frac{\not 1}{\not r}\frac{d}{dr}\left(\frac{1}{r}\frac{dT}{dr}\right)$

---

$2\frac{x^2}{r}\frac{d}{dr}\left(\frac{N}{r^2}\right)$

$$\frac{\partial}{\partial x}\left(x^2\frac{N}{r^2}\right) = 2x\frac{N}{r^2} + x^3\cdot\frac{1}{r}\frac{d}{dr}\left(\frac{N}{r^2}\right)$$

$$\frac{\partial^2}{\partial x^2} = 2\frac{N}{r^2} + <3>5x^2\cdot\frac{1}{r}\frac{d}{dr}\left(\frac{N}{r^2}\right) + x^4\cdot\frac{1}{r}\frac{d}{dr}\left(\frac{1}{r}\frac{d}{dr}\left(\frac{N}{r^2}\right)\right)$$

$$\frac{\partial}{\partial y} = x^2y\cdot\frac{1}{r}\frac{d}{dr}\left(\frac{N}{r^2}\right)$$

$$\frac{\partial^2}{\partial y^2} = x^2\cdot\frac{1}{r}\frac{d}{dr}\left(\frac{N}{r^2}\right) + x^2y^2\cdot\frac{1}{r}\frac{d}{dr}\left(\frac{1}{r}\frac{d}{dr}<)>\left(\frac{N}{r^2}\right)\right)$$

[eq. 29] $\Delta\left(x^2\frac{N}{r^2}\right) = 2\frac{N}{r^2} + <5>7x^2\cdot\frac{1}{r}\frac{d}{dr}\left(\frac{N}{r^2}\right) + x^2\cdot\frac{\not 1}{\not r}<x>r\frac{d}{dr}\left(\frac{1}{r}\frac{d}{dr}\left(\frac{N}{r^2}\right)\right)$

[eq. 30] $\underbrace{\frac{3}{r}\frac{dT}{dr} + \frac{1}{r}\frac{d}{r}\left(\frac{1}{r}\frac{dT}{dr}\right) + 2\frac{N}{r^2}}_{+\frac{A^2}{4}\frac{1}{r^4}} + x^2\left(\underbrace{<5>7\frac{1}{r}\frac{d}{dr}\left(\frac{N}{r^2}\right) + r\frac{d}{dr}\left(\frac{1}{r}\frac{d}{dr}\frac{N}{r^2}\right)}_{-\frac{A^2}{2}\frac{1}{r^6}}\right)$

$$+\frac{A^2}{2}\left(\frac{1}{2}\frac{1}{r^4} - \frac{x^2}{r^6}\right)$$

$N = \frac{1}{8}\frac{A^2}{r^2}$

$<(1+\alpha)\,(1+\beta)\,(1+\gamma)>$

$\frac{N}{r^2} = \frac{1}{8}\frac{A^2}{r^4}$

$-\left(1 + \frac{1}{8}A^2\frac{x^2}{r^4} - \frac{1}{4}A^2\frac{1}{r^2}\right)$

$\frac{d}{dr}(\ ) = -\frac{1}{2}\frac{A^2}{r^5}$

$-\frac{3}{4}\frac{A^2}{r^2} + \frac{1}{8}\frac{A^2}{r^2} - \frac{5}{8}\frac{A^2}{r^2}$

$\frac{1}{r}(\quad) = -\frac{1}{2}\frac{A^2}{r^6}$ | $<5>7$

$bcd$

$abcd$

$\frac{d}{dr}(\quad) = 3\frac{A^2}{r^7}$

$-\frac{1}{1+\frac{x^2}{r^4}\frac{1}{8}A^2[-]-\varepsilon} = -(1- \quad +\varepsilon)$

$-1+ \quad -$

$r(\quad) = 3\frac{A^2}{r^6}$ | 1

# [p. 6] Einstein

Diff. Gl. für $N\ T$ & $\Phi$ (3)

[eq. 36] $$\text{<5>}7\cdot\frac{1}{r}\frac{d}{dr}\left(\frac{N}{r^2}\right)+r\frac{d}{dr}\left[\frac{1}{r}\left(\frac{d}{dr}\left(\frac{N}{r^2}\right)\right)\right] = -\frac{A^2}{2}\cdot\frac{1}{r^6}$$ $$N = +\frac{1}{8}\frac{A^2}{r^2}$$

[eq. 37] $$3\frac{1}{r}\frac{dT}{dr}+r\cancel{\frac{1}{r}}\frac{d}{dr}\left(\frac{1}{r}\frac{dT}{dr}\right)+2\frac{N}{r^2} = \frac{A^2}{4}\cdot\frac{1}{r^4}$$ $$T = \cancel{\frac{1}{4}\frac{A^2}{r^2}}\ 0$$ [eq. 39]

[eq. 38] $$3\cdot\frac{1}{r}\frac{\text{<}\partial\text{>}d\Phi}{dr}+r\cancel{\frac{1}{r}}\frac{d}{dr}\left(\frac{1}{r}\frac{d\Phi}{dr}\right) = +\frac{5}{\text{<2>}4}\frac{A^2}{c_0^2 r^4}$$ $$\Phi = +\frac{1}{c_0^2}\frac{5}{8}\frac{A^2}{r^2}$$

---

$$-\frac{1}{8}+\frac{3}{4}+\frac{5}{8}$$

$\gamma$

[eq. 40]
$$\begin{array}{cccc}
-1+\frac{1}{8}\frac{A^2}{r^2}\frac{x^2}{r^2}+\cancel{\frac{1}{4}\frac{A^2}{r^2}} & +\frac{1}{8}\frac{A^2}{r^2}\frac{xy}{r^2} & . & 0 \\
- & - & - & - \\
- & - & - & - \\
- & - & 0 & \frac{1}{c_0^2}\left(1+\frac{A}{r}+\frac{5}{8}\frac{A^2}{r^2}\right)
\end{array}$$

[eq. 41]
$$\gamma = \frac{1}{c_0^2}\left\{-1-\frac{A}{r}-\text{<->}\frac{1}{2}\frac{A^2}{r^2}\right\} = -\frac{1}{c_0^2}\left(1+\frac{A}{r}+\frac{1}{2}\frac{A^2}{r^2}\right)$$
$$g = -c_0^2\left\{1-\frac{A}{r}+\text{<->}\frac{1}{2}\frac{A^2}{r^2}\right\}$$

Tabelle der $g$.

[eq. 42]
$$\begin{array}{cccc}
-1-\frac{1}{8}\frac{A^2}{r^2}\frac{x^2}{r^2} & -\frac{1}{8}\frac{A^2}{r^2}\frac{xy}{r^2} & . & 0 \\
- & - & - & 0 \\
- & - & - & 0 \\
0 & 0 & 0 & c_0^2\left(1-\frac{A}{r}+\frac{3}{8}\frac{A^2}{r^2}\right)
\end{array}$$

[p. 8] Besso

<Für> <d>Die Bewegungsgleichungen des materiellen Punktes lauten: 4)

[eq. 48] 1 bis 3) $\frac{dJ_x}{dt} = \mathfrak{K}_x$ also (bei $m = 1$): [eq. 49] $-\frac{d}{dt}\frac{g_{11}\dot{x}_1 + \ldots + g_{14}}{\frac{ds}{dt}} = -\frac{1}{2}\sum_{\mu\nu}\frac{\partial g_{\mu\nu}}{\partial x_1}\cdot\frac{dx_\mu}{dt}\cdot\frac{dx_\nu}{dt}\cdot\frac{dt}{ds}$ (Seite 7. Gl. 7 & 8)

[eq. 50] 4) $E = +m\left(g_{41}\frac{dx_1}{ds} + \ldots + g_{44}\frac{dx_4}{ds}\right)$ also: [eq. 51] $E = -g_{44}\frac{dt}{ds}$ (Seite 7. Gl. 9)$_{\text{II}}$)

Dabei ist (<aus Gl. 5> S. <7>6 <)> auch S.7 Gl. 5), wenn man <relativ kleine Grössen,> die Grössen<ordnung von> <von der Ordnung> $\frac{A}{r}$ und <---> $\frac{q^2}{c_0^2} = \frac{\dot{x}^2 + \dot{y}^2 + \dot{z}^2}{c_0^2}$ als unendlich klein erster Ordnung betrachtet

(worin $\frac{A}{r} = \frac{M}{4\pi}\cdot\frac{8\pi}{c_0^2}\cdot\frac{K}{r} = \frac{2M}{c_0^2 r}K$, und wieder $M$ die Sonnenmasse, $c_0$ die Lichtgeschwindigkeit für $r$ – Abstand von der Sonne – unendlich), $K$ die gew. Gravitationskonstante ist – Kraft = $K\frac{Mm}{r^2}$ – bedeutet) und zudem berücksichtigt dass sich am Felde mit der Zeit nichts ändert, also $g_{\nu 4}$ für $\nu$ = 1 bis 3 <also> = 0 zu setzen sind, wenn man auch wieder <für> $x_1 = x$, $x_2 = y$, $x_3 = z$, $x_4 = t$ setzt

[eq. 52] $\frac{ds}{dt} = \sqrt{g_{11}\dot{x}^2 + <g>\ldots + 2g_{12}<g>\dot{x}_1\dot{x}_2 + \ldots 2g_{14}\dot{x} + \ldots + g_{44}} = \sqrt{g_{11}\dot{x}^2 + g_{22}\dot{y}^2 + g_{33}\dot{z}^2 + 2g_{12}\dot{x}\dot{y} + \ldots + g_{44}}$

Die $g_{\nu\nu \neq 4}$ sind aber, bis auf unendl. kl. zweiter Ordnung = – 1, die $g_{\nu\neq\mu} = 0$,

$g_{44} = c_0^2\left(1 - \frac{A}{r} + \frac{3}{8}\frac{A^2}{r^2}\right)$. Daher reducier<en>t sich <die Gleichngen> der Ausdruck für $\frac{ds}{dt}$ auf

1) <bis 3) auf> [eq. 53] $\frac{ds}{dt} = \sqrt{c_0^2 - q^2 - c_0^2\frac{A}{r} + \frac{3}{8}c_0^2\frac{A^2}{r^2}} = c_0^2\left(1 - \frac{1}{2}\frac{q^2}{c_0^2} - \frac{1}{2}\frac{A}{r}\right)$ ; die Gleichngen werden

1) bis 3) $\frac{d}{dt}\cdot\frac{\dot{x}}{\sqrt{c_0^2 - q^2 - \frac{A}{r}}} = c_0^2\frac{d}{dt}\left[\dot{x}\left(1 + \frac{1}{2}\frac{q^2}{c_0^2} + \frac{1}{2}\frac{A}{r}\right)\right] =$

[eq. 54] $\frac{d}{dt}\frac{\dot{x}}{\frac{ds}{dt}} = -\frac{1}{2}\cdot\sum_{<44>}\cdot\frac{\partial g_{44}}{\partial x}\cdot\frac{dt}{dt}\cdot\frac{dt}{dt}\cdot\frac{1}{\frac{ds}{dt}} = -\frac{1}{2}\cdot\frac{<\partial>dg_{44}}{<\partial>dr}\frac{\partial r}{\partial x}\Big/\frac{ds}{dt} = -\frac{1}{2}\frac{dg_{44}}{dr}\frac{<r>x}{r}\Big/\frac{ds}{dt}$

4) $E = +g_{44}\Big/\frac{ds}{dt}$ · [eq. 55]

Aus <das> $y$ · Gl. 1) – $x$ · Gl. 2) ergibt sich der Flächensatz

[eq. 56] $y\frac{d}{dt}\left(\dot{x}\Big/\frac{ds}{dt}\right) - x\frac{d}{dt}\left(\dot{y}\Big/\frac{ds}{dt}\right) = \frac{d}{dt}\frac{y\dot{x} - x\dot{y}}{\frac{ds}{dt}} = 0$

Nimmt man die <--> Bahnebene als $xy$ Ebene, so ergibt sich daraus

[eq. 57]
I) $2\dot{f} = y\dot{x} - x\dot{y} = B\frac{ds}{dt}$ (Flächensatzkonstante = $Bc_0$) und $\dot{f}$ = Flächengeschwindigkeit

II) $g_{44} = E\frac{ds}{dt}$ ($E \sim c_0$) $E = c_0$

Da nun $\frac{ds}{dt}$ <so> eine Wurzel ist, so ist es bequemer mit den quadrierten Gleichngen (in Polarcoordinaten) zu operieren:

[p. 9] Besso

$$2\dot f = \dot\phi r^2 = BW \quad \text{[eq. 58]}$$

$$E = \frac{g_{44}}{W} \quad \text{[eq. 59]}$$

[eq. 60] $$E\dot\phi r^2 = Bg_{44} = Bc_0^2\left(1-\frac{A}{r}+\frac{3}{8}\frac{A^2}{r^2}\right)$$

[eq. 61] $$\dot\phi r^2 = \cancel{Bc}\ F\langle'\rangle\left(1-\frac{A}{r}\right), \qquad F = \frac{Bc_0^2}{E}$$

[eq. 62] $$\langle E\rangle c_0^4\left(1-2\frac{A}{r}+\frac{7}{4}\frac{A^2}{r^2}\right) = E^2c_0^2\left(1-\frac{q^2}{c_0^2}-\frac{A}{r}+\frac{3}{8}\frac{A^2}{r^2}\right)$$

[eq. 63] 5)

$$1-\frac{E^2}{c_0^2} = \varepsilon$$

$$\frac{E^2}{c_0^2} = 1-\varepsilon$$

$$\frac{c_0^2}{E^2} = 1+\varepsilon$$

$$\frac{c_0^2}{E^2}-\langle E\rangle 1 = \varepsilon$$

$$\frac{1}{1-\alpha} = 1+\alpha+\alpha^2$$

$$\cancel{\frac{E^2-c_0^2}{c_0^2} = \varepsilon \qquad \frac{E^2-c_0^2}{E^2} = \frac{c_0^2}{E^2}\varepsilon}$$

sondern so: $$\cancel{1-\frac{A}{r}+\frac{q^2}{c_0^2}+\frac{11}{8}\frac{A^2}{r^2}+\left(\frac{q^2}{c_0^2}+\frac{A}{r}\right)^2 = \frac{E^2}{c_0^2}-1} \qquad \frac{E^2}{c_0^2}-1 = \varepsilon$$

$$\cancel{\dot\phi r^2 = f} \quad \frac{1}{E^2}c_0^2\left(1-2\frac{A}{r}+\frac{7}{4}\frac{A^2}{r^2}\right) = \cancel{E^2}\left(1-\frac{q^2}{c_0^2}-\frac{A}{r}+\frac{3}{8}\frac{A^2}{r^2}\right)$$

$$\cancel{\frac{c_0^2}{E^2} = 1-\frac{q^2}{c_0^2}+\frac{A}{r}+\frac{3}{8}\frac{A^2}{r^2}-\frac{14}{8}\frac{A^2}{r^2}}$$

nicht so: $$\cancel{\frac{E^2}{c^2}\left(1-\frac{q^2}{c_0^2}-\frac{A}{r}+\frac{3}{8}\frac{A^2}{r^2}\right) = 1-2\frac{A}{r}+\frac{7}{4}\frac{A^2}{r^2}}$$

[eq. 64] $$\frac{q^2}{c_0^2} = 1-\frac{c_0^2}{E^2}+\frac{A}{r}\left(2\frac{c_0^2}{E^2}-1\right)+\frac{A^2}{r^2}\left(\frac{3}{8}-\frac{7}{4}\frac{c_0^2}{E^2}\right) \qquad 2\left(\frac{c_0^2}{E^2}-1\right)+1$$

$$= -\varepsilon+\frac{A}{r}(2\varepsilon+1)+\frac{A^2}{r^2}\left(-\frac{\langle 3\rangle 11}{8}\right)$$

[eq. 65] $$\frac{dr^2+r^2d\phi^2}{dt^2} = c_0^2\left[-\varepsilon+\frac{A}{r}(1+2\varepsilon)-\frac{A^2}{r^2}\cdot\frac{11}{8}\right]$$

[eq. 66] $$r^{\langle 2\rangle 4}\frac{d\phi^2}{dt^2} = F^2\left(1-2\frac{A}{r}\right) \ \cancel{\#}$$

$$\frac{1}{c_0^2}(dr^2+r^2d\phi^2)F^2\left(1-2\frac{A}{r}\right) = -r^4d\phi^2\cdot\cancel{c_0^2}\left(\varepsilon-(1+2\varepsilon)\frac{A}{r}+\frac{11}{8}\frac{A^2}{r^2}\right)$$

$$-r^2d\phi^2\frac{1}{c_0^2}F^2\left(1-2\frac{A}{r}\right)$$

$$\frac{1}{c_0^2}dr^2F^2\left(1-2\frac{A}{r}\right) = d\phi^2\left\{-\varepsilon\cancel{c_0^2}\cdot r^4+\cancel{c_0^2}(1+2\varepsilon)A\cdot r^3-\left[\frac{11}{8}A^2\cancel{r^2}+\frac{F^2}{c_0^2}\left(1\cancel{-2\frac{A}{r}}\right)\right]r^2+2\frac{F^2}{c_0^2}Ar\right\}$$

— — — — — — — — — — — — — — — — —

$$d\phi^{\langle 2\rangle} = \frac{\frac{F}{c_0}\left(1-\frac{A}{r}\right)dr}{\cancel{F\left(1-\frac{A}{r}\right)}\sqrt{-\varepsilon r^4+(1+2\varepsilon)\cdot A\cdot r^3-\left(\frac{11}{8}A^2+\frac{F^2}{c_0^2}\right)r^2+2\frac{F^2}{c_0^2}A^2\cdot r}} = \quad \text{[eq. 67]}$$

$$\int d\phi = \frac{F}{c_0\sqrt{\varepsilon}}\int\frac{1-\frac{A}{r}}{\sqrt{-r^4+\left(\frac{1}{\varepsilon}+2\right)A\cdot r^3-\frac{1}{\varepsilon}\left(\frac{F^2}{c_0^2}+\frac{11}{8}A^2\right)r^2+\frac{2}{\varepsilon}\frac{F^2}{c_0^2}A\cdot r}}dr \quad \text{[eq. 68]}$$

[p. 10] Einstein

$$\int \frac{dr}{\sqrt{(r-r')\,(r-r'')\,(r-r_1)\,(r_2-r_{<2>})}} = \frac{2\pi}{\sqrt{r_1 r_2}}\left[1+\frac{1}{2}\left(\frac{1}{r_1}+\frac{1}{r_2}\right)\right] \quad \text{[eq. 69]}$$

$$= \frac{2\pi}{\sqrt{r_1 r_2}}\left[1+\frac{1}{<2>4}\left(\frac{1}{r_1}+\frac{1}{r_2}\right)\underbrace{(r'+r'')}_{-\frac{d}{c}}\right]$$

$$= \int \frac{dr}{r\sqrt{(r-r_1)\,(r_2-r_{<2>})}\sqrt{1-\frac{r'}{r}}\sqrt{1-\frac{r''}{r}}} \quad \text{[eq. 70]}$$

$$= \int \frac{\left(1+\frac{r'+r''}{2}\frac{1}{r}\right)dr}{r\sqrt{(r-r_1)\,(r-r_2)}} \quad \text{[eq. 71]}$$

---

$$\int \frac{dr}{r\sqrt{(r_{<1>}-r_1)\,(r-r_2}} = \frac{2\pi}{\sqrt{r_1 r_2}} \quad \text{[eq. 72]}$$

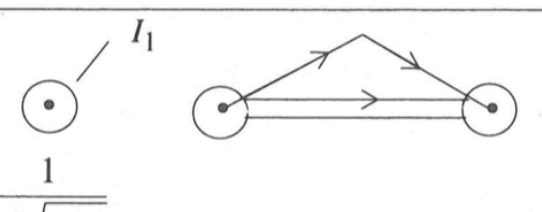

$$\frac{1}{-i\sqrt{r_1 r_2}}$$

$$\underbrace{\begin{matrix}-2\pi \\ -\pi\end{matrix}}_{-\frac{3}{2}\pi}$$

$$= -\int_{+} \frac{1}{-i\sqrt{r_1 r_2}}\frac{dr}{r} = \frac{1}{i\sqrt{r_1 r_2}}\underbrace{\int i\,d\varphi}_{2\pi i}$$

---

[eq. 73] [eq. 74]

$$\int \frac{dr}{r^2\sqrt{(r-r_1)\,(r_2-r)}} \quad \Bigg| \quad \frac{1}{-i\sqrt{r_1 r_2}\sqrt{\left(1-\frac{r}{r_1}\right)\left(1-\frac{r}{r_2}\right)}} = \frac{1}{-i\sqrt{r_1 r_2}}\left(1+\frac{1}{2}\left(\frac{1}{r_1}+\frac{1}{r_2}\right)r\right)$$

$$-\int_{+0\text{-Kreis}} \frac{1}{-i\sqrt{r_1 r_2}}\frac{1}{2}\left(\frac{1}{r_1}+\frac{1}{r_2}\right)\underbrace{\int \frac{dr}{r}}_{2\pi i}$$

---

$$\int \frac{dr}{\sqrt{-ar^4+br^3+cr^2+dr+e}} \quad \text{[eq. 75]}$$

Unendlich kleine Wurzeln aus der Gleichung

$$\text{[eq. 76]} \quad cr^2+dr+e = 0 = c\,(r-r')\,(r-r'')$$

$$d = -c\,(r'+r'')$$

$$\text{[eq. 77]} \quad r'+r'' = -\frac{d}{c}$$

---

$$ar^4<->-br^3-cr^2-dr-e = a\,(r-r_1)\,(r-r_2)\,(r-r')\,(r-r'') \quad \text{[eq. 78]}$$

$$= ar^4$$

$$-(r_1+r_2+r'+r'')\,r^3$$

$$+(r_1r_2+r_1r'+r_2r'+.\ +.\ +r'r'')\,r^2$$

Für endliche Grössen

$$\text{[eq. 79]} \quad r_1+r_2 = \frac{b}{a}$$

$$\text{[eq. 80]} \quad r_1r_2 = -\frac{c}{a}$$

## [p. 11] Einstein

6)

$$\int \frac{dr}{\sqrt{(r-r')\,(\;)\,(\;)\,(\;)}} = \frac{2\pi}{\sqrt{-\frac{c}{<a>1}}}\left[1+\frac{1}{4}\frac{\frac{b}{<a>1}}{-\frac{c}{<a>1}}\cdot - \frac{d}{c}\right] \quad \text{[eq. 81]}$$

$$= \frac{2\pi}{\sqrt{-<\frac{c}{a}>c}}\left[1+\frac{1}{4}\frac{bd}{c^2}\right]$$

---

$$\int \frac{dr\,(1+\alpha\frac{1}{r}+\beta\frac{1}{r^2})}{\sqrt{(r-r_1)\,(r_2-r)\,(r-r')\,(r-r'')}} = I \quad \text{[eq. 82]}$$

$$= \int \frac{dr}{r\sqrt{(r-r_1)\,(r_2-r)}}\;\Big|\;(1-\frac{r'}{r})^{-\frac{1}{2}}\,(1-\frac{r''}{r})^{-\frac{1}{2}}\,(1+\alpha\frac{1}{r}+<)>\beta\frac{1}{r^2})$$

Faktor auf ∞ Kl. erster Ordnung berechnet

$$\left[1+\frac{1}{2}\,(r'+r'')\cdot\frac{1}{r}\right](1+\alpha\frac{1}{r})$$

[eq. 83] $$1+\underbrace{(\alpha+\frac{r'+r''}{2})}_{\alpha'}\frac{1}{r}$$

---

[eq. 84]

$$I = \int \frac{dr\,(1+\alpha'\frac{1}{r})}{r\sqrt{(r-r_1)\,(r_2-r)\,\cancel{(r-r')\,(r-r'')}}} = \cancel{\int \frac{dr\,(1+\alpha'\frac{1}{r})}{r\sqrt{(r-r_1)\,(r_2-r)}}}$$

$$I = \int \frac{1+\alpha\frac{1}{r}}{\sqrt{-r^4+ar^3+br^2+cr+d}}\,dr \quad \text{[eq. 85]}$$

$$a = r_1+r_2+r'+r''$$
$$-b = r_1r_2+(r_1+r_2)\,(r'+r'')$$
$$c = r_1r_2(r'+r'')$$
$$-d = 0$$

$$= \frac{2\pi}{\sqrt{r_1r_2}}\left[1+\frac{1}{2}\,(\frac{1}{r_1}+\frac{1}{r_2})\,(\alpha+\frac{r'+r''}{2})\right]$$

---

[eq. 86] $$c_0B\int \frac{(dr\,(1-A\frac{1}{r})}{\sqrt{(E^2-c_0^2)\,r^4+A\,(2c_0^2-E^2)\,r^3+\left[A^2\,(\frac{7}{4}c_0^2-\frac{3}{8}E^2)-B^2c_0^2\right]r^2+2B^2c_0^2Ar - \frac{7}{4}B^2c_0^2A^2}}$$

$$= \frac{c_0B}{\sqrt{c_0^2-E^2}}\cdot\frac{2\pi}{\sqrt{r_1r_2}}\left[1+\frac{1}{2}\,(\frac{1}{r_1}+\frac{1}{r_2})\,(-A+\;-\frac{1}{2}\underbrace{\frac{d}{c}}_{-2A})\right]$$

[eq. 87] $$\sqrt{r_1r_2} = \sqrt{-c} = \sqrt{\frac{c_0^2B^2}{c_0^2-E^2}}$$

Faktor $\frac{1}{2}$ vor dem ganzen Integral vergessen.

## [p. 14] Besso

<6>7)

[eq. 98] $$r' + r'' = -\frac{\frac{2}{\varepsilon}\frac{F^2}{c_0^2}A}{-\frac{1}{\varepsilon}\frac{F^2}{c_0^2}} = \cancel{+\frac{2}{\varepsilon}}A + 2A$$

[eq. 99] $$r_1 + r_2 = +\left(\frac{1}{\varepsilon}+2\right)A - 2A = +\frac{1}{\varepsilon}A$$

$$\xcancel{r_1 r_2 = \frac{1}{\varepsilon}\left(\frac{F^2}{c_0^2}+\frac{11}{8}A^2\right) - \left\{-\left(\frac{1}{\varepsilon}-2\right)A + 2A\right\}(-2A)}$$

$$\xcancel{= \frac{1}{\varepsilon}\frac{F^2}{c_0^2} + \left(\frac{11}{8}\frac{1}{\varepsilon}+4\right)A^2 - \left(\frac{1}{\varepsilon}-2\right)2A^2}$$

$$\xcancel{= \frac{1}{\varepsilon}\frac{F^2}{c_0^2} + \left(-\frac{5}{8}\frac{1}{\varepsilon}+8\right)A^2}$$

$$\xcancel{= \frac{1}{\varepsilon}\left(\frac{F^2}{c_0^2} - \frac{5}{8}A^2\right)}$$

[eq. 100] $$r_1 r_2 = \frac{1}{\varepsilon}\left(\frac{F^2}{c_0^2}\text{<->} + \frac{\text{<5>}11}{8}A^2\right)\text{<+>} - 2A\cdot\frac{1}{\varepsilon}A$$

$$= \frac{1}{\varepsilon}\left(\frac{F^2}{c_0^2}\text{<+>} - \frac{5}{8}A^2\right)$$

[eq. 101] $$\int d\varphi = \frac{F}{c_0\cancel{\sqrt{\varepsilon}}}\frac{\pi\cancel{\sqrt{\varepsilon}}}{\sqrt{\frac{F^2}{c_0^2}-\frac{5}{8}A^2}}\left[1 + \xcancel{\frac{1}{2}\left(\frac{1}{r_1}+\frac{1}{r_2}\right)(-A+A)}\right]$$

$$= \cancel{\frac{F}{c_0}}\frac{\pi}{\cancel{\frac{F}{c_0}}\sqrt{1-\frac{5}{8}\frac{A^2c_0^2}{F^2}}} = \pi\left(1+\frac{5}{16}\frac{A^2c_0^2}{F^2}\right)\text{<=>} \sim \pi\left(1+\frac{5}{8}\frac{A}{R^{\text{<>}}}\right) = \pi\left(1+\frac{5}{8}\frac{8\pi^2R^{[-]}}{c_0^2T^2}\right) \Bigg\| A = \frac{\kappa M}{4\pi} = \frac{2KM}{c_0^2}$$

$$= \pi\left(1+\frac{5}{8}\frac{A}{a(1-e^2)}\right)$$

[eq. 102] $$F = \frac{Bc_0^2}{E} = Bc_0 = \cancel{\frac{1}{2}}2\dot{f} = 2\frac{\pi\cdot a^2\cdot\sqrt{1-e^2}}{\cancel{2}T}$$

$ae = c$

$b = a\sqrt{1-e^2}$ $\quad \cancel{b^2 = a^2 - a^2e^2}$

$$A^2c_0^2 = 4K^2\cdot\frac{M^2}{c_0^2}$$

$$\frac{A^2c_0^2}{F^2} = 16\pi^2\xcancel{\left(\frac{r}{c_0T}\right)}^2 \quad ?$$

$$A = K\frac{\overset{2}{\cancel{8\pi}}}{c_0^2}\cdot\frac{M^2}{\cancel{4\pi}} = 2K\frac{M}{c_0^2}$$

$$\xcancel{\frac{A^2c_0^2}{F^2} = \frac{\cancel{4}K^2M^2T^2}{\pi^2c_0^2a^4(1-e^2)} = \frac{4K^2M^2}{c_0^2\cdot a^2(1-e)^2v_{max}^2}}$$

$$\xcancel{F = a(1-e)\,v_{max} \qquad = \frac{4}{a^2(1-e)^2}\frac{K^2M^2}{c_0^2}\cdot\frac{1}{v_m^2/c_0^2}}$$

[p. 26] Einstein

<Zwischenrechgen>

5a)
I

[eq. 174] $r' + r'' = +2A$

[eq. 175] $r_1 + r_2 = \frac{\frac{1}{\varepsilon}+2}{} \left(\frac{1}{\varepsilon}+2\right)A - 2A = \frac{1}{\varepsilon}A$

$r_1 r_2 = \left\{\frac{1}{\varepsilon}\left(\frac{F^2}{c_0^2} + \frac{11}{8}A^2\right) - \frac{1}{\varepsilon}A\right.$

[eq. 176] $r_1 r_2 = \frac{1}{\varepsilon}\left(\frac{F^2}{c_0^2} + \frac{11}{8}A^2\right) - \frac{1}{\varepsilon}A \cdot 2A = \frac{1}{\varepsilon}\left(\frac{F^2}{c_0^2} - \frac{5}{8}A^2\right)$

[eq. 177] $\int d\varphi = \frac{F}{c_0\sqrt{\varepsilon}} \frac{\pi c_0 \sqrt{\varepsilon}}{F} \Big/ \sqrt{1 - \frac{5}{8}\frac{A^2 c_0^2}{F^2}} = \pi\left(1 + \frac{5}{16}\frac{A^2 c_0^2}{F^2}\right)$

[eq. 178] $F = \frac{\pi a^2 \sqrt{1-e^2} \cdot 2}{T}$

[eq. 179] $A = \frac{2KM}{c_0^2}$

$K = \frac{1}{3862^2}$

$\lg K = -2 \cdot 3.58681$

$= -7, 17362$

$\lg M = 5, 51108$ | 5.6 = Dichte d. E

28, 78265
28.29373
34

$\lg V = 28.03456$
$\lg 5.6 = 74819$
$\lg M_e = 28, 78265$

$\lg T = \lg\ 87.97 \cdot 24 \cdot 60 \cdot 60$

= 1.94433 3
1.38021 1, 77815
3.55630
6.88084

$\lg c_0 = 10, 47712$ Mittlerer Abst. = grosse Halbachse gesetzt.

$\lg \pi = 0, 49715$

$\lg a = .58782 - 1$
13.17164
12.75946

$\lg a^2 = 25.51892$

$\lg\ (1 - e^2) = \lg =$ 1.2056 / .08120 98124 − 1
0.7944 − 1 90004 − 1
~~000 − 1~~

$\lg \sqrt{1-e^2} = 9906 - 1$

[eq. 180]

$\frac{A c_0}{F} = K \cdot \frac{2M}{c_0^2} \cdot \frac{c_0 T}{\pi a^2 \sqrt{1-e^2} \cdot 2}$

34, 2937 7.1736
6.8808 10.4771
41.1745 0.4971
43 6573 25.5189
0.5172 − 3 9906 − 1
43.6573

lg Präzession pro halbem Umlauf.
zu mult. mit

$2 \cdot \frac{T_{\text{erde}}}{T_{\text{Merkur}}} \cdot 100$

2.30103
2.5626
4.8636
1.9443
2.9193

0.0151 − 8
2.9193
0.9344 − 6 Winkelabschnitt.
5.3145
0.2489

[p. 28] Einstein

I.

$0.5172 - 3 = \lg \frac{Ac_0}{F}$

$0,0344 - 5 = \lg \left(\frac{Ac_0}{F}\right)^2$

0.6990

0.7334 − 5

1.2041

$0.5293 - 6 = \lg \frac{5}{16} (\ )^2 = 3.4 \cdot 10^{-6}$

~~5.8116~~

~~0.35~~

~~2.2~~

2.2553
1.7781
1.7782

5.8116

$2 \cdot \frac{365,2}{87,97} \cdot 100$

2.3010
2.5625

4.8635
1.9443

2.9192

Präzession pro halbem Umlauf in Bogensekunden

0.5293 − 6
5.8116

0, 3409

Präzession in 100 Jahren:

0, 3409
2.9193

3.2602

~~1898″ = 31, 5′~~
1821″ = 30′ unabhängig geprüft.

[p. 41, top half] Einstein

$$\begin{matrix} -1 & 0 & 0 & -2\alpha y \\ 0 & -1 & 0 & -2\alpha x \\ 0 & 0 & -1 & 0 \\ -2\alpha y & -2\alpha x & 0 & \text{Const.} - \alpha^2 \langle r^2 \rangle \rho^2 \end{matrix}$$ [eq. 267]

[eq. 268] $\gamma_{14} = -2\alpha y$, $\gamma_{24} = 2\alpha x$

$D_{\mu\nu}(g) + k\ t_{\mu\nu} = 0$ [eq. 269]

[eq. 270] $D_{44}(g) = -\Delta g_{44} - 8\alpha^2$

[eq. 271] $\kappa\ t_{\mu\nu} = \sum -\frac{1}{2} \frac{\partial g_{\tau\rho}}{\partial x_\mu} \frac{\partial \gamma_{\tau\rho}}{\partial x_\nu} + \frac{1}{4} g_{\mu\nu} \gamma_{\alpha\beta} \frac{\partial g_{\tau\rho}}{\partial x_\alpha} \frac{\partial \gamma_{\tau\rho}}{\partial x_\beta}$

$c^2 \quad -1$

$\frac{1}{c^2} \cdot 16\alpha^2$

$\tau = 1 \quad \rho = 4 \quad \alpha = 2$
$\tau = 2 \quad \rho = 4 \quad \alpha = 1$
$\tau = 4 \quad \rho = 1 \quad \alpha = 2$
$\tau = 4 \quad \rho = 2 \quad \alpha = 1$

[eq. 272] $-\Delta g_{44} - 8\alpha^2 + 4\alpha^2 = 0$

[eq. 273] $\Delta g_{44} = -4\alpha^2$

[eq. 274] $g_{44} = \text{konst} - \alpha^2 \langle r^2 \rangle \rho^2$ stimmt.

[p. 42, left part] Einstein

[eq. 281] $$-\Delta\gamma_{44}<+>-\frac{8\alpha^2}{c^4} - \underbrace{\frac{1}{4}\cdot\frac{1}{c^2}(-1)\cdot 2\cdot 8\alpha^2\cdot\frac{1}{c^2}}_{+4\frac{\alpha^2}{c^4}} = 0$$ [190]

[eq. 282] $$g_{14} = -2\alpha y \qquad g_{24} = 2\alpha x$$

[eq. 283] $$\Delta\gamma_{44} = -\frac{4\alpha^2}{c^4}$$

$$\frac{\partial\gamma_{44}}{\partial x} = \frac{\partial\gamma_{44}}{\partial\rho}\,\frac{x}{\rho}$$

$$\frac{\partial^2\gamma_{44}}{\partial x^2} = \frac{\partial^{<2>}}{\partial x_\rho}\left(\frac{1}{\rho}\,\frac{\partial\gamma_{44}}{\partial\rho}\right)\frac{x^2}{\rho} + \frac{1}{\rho}\,\frac{\partial\gamma_{44}}{\partial\rho}$$

[eq. 284] $$\gamma^x_{44} = \beta\rho^2$$

[eq. 285] $$\Delta\gamma_{44} = <2>\rho\frac{\partial}{\partial\rho}\left(\frac{1}{\rho}\,\frac{\partial\gamma_{44}}{\partial\rho}\right) + \frac{2}{\rho}\,\frac{\partial\gamma_{44}}{\partial x_\rho}$$

[eq. 286] $$\Delta\gamma_{44} = 4\beta = -\frac{4\alpha^2}{c^{<2>4}}$$

[eq. 287] $$\beta = -\frac{\alpha^2}{c^4}$$

[eq. 288] $$\gamma_{44} = \frac{1}{c_0^2} \not{+} - 3\frac{\alpha^2}{c_0^4}\rho^2$$

[eq. 289] $$g_{44} = c_0^2\left(1 + <3>\frac{\alpha^2}{c_0^2}\rho^2\right)$$

[eq. 290] $$-m\frac{d}{dt}\left(-\frac{dx}{ds} + g_{14}\frac{dt}{ds}\right) = -\frac{1}{2}m\frac{\partial g_{44}}{\partial x}\cdot\frac{dt}{ds}$$

[eq. 291] $$m\ddot{x} <=> + \cdot = 3\alpha^2 x$$

## 2.3 Commentary

The Einstein-Besso manuscript is a sheaf of loose sheets with research notes documenting (for the most part) attempts by Einstein and his friend Michele Besso to account for the anomalous advance of the perihelion of Mercury on the basis of the *Entwurf* theory (cf. Pt. I, Ch. 6). The manuscript can thus be dated to the reign of that theory, which was roughly from May 1913 to October 1915. In fact, as we shall see shortly, many pages can be dated more precisely.

The bulk of the manuscript—53 pages, including the 14 pages presented here—was published in CPAE4, in facsimile as Appendix B and in transcription as Doc. 14. Three sets of numbers were added to the transcription, one numbering pages ("[p. 1]," etc.: Einstein and Besso's own numbering is erratic at best), one numbering editorial notes ("[1]," etc.), and one numbering equations ("[eq. 1]," etc.). For this volume, we used the transcription from CPAE4 without the editorial-note numbers but with the equation numbers. We will refer to those in our commentary, which is based on the editorial note, "The Einstein-Besso manuscript on the motion of the perihelion of Mercury," the annotation of the transcription in CPAE4 (pp. 344–359, pp. 360–473), and Janssen (1996, 1999).

The published part of the manuscript comes from two sources. Two of its 53 pages, [pp. 16–17], mostly in Besso's hand, are recto and verso of a letter from Charles-Eugène Guye to Einstein, May 31, 1913 (CPAE5, Doc. 443). The remaining 51 pages are on 37 sheets found in the Besso *Nachlass*. Michele's son Vero presented these to Pierre Speziali, editor and translator (into French) of the correspondence between Einstein and Besso (1972). Speziali made them available to the editors of the Einstein Papers Project in the late 1980s. A copy of these 51 pages is in the supplementary archive to the Einstein Archive, assembled and used by the editors of the Einstein Papers Project. The designation of this copy of the manuscript is EA 79 896. Of its 51 pages, 24 are in Einstein's hand (with an occasional entry by Besso), 24 in Besso's (most of them without any contribution from Einstein), while 3 pages, i.e., [p. 33] and [pp. 41–42], contain substantial contributions from both.

These 51 pages were sold at auction at Christie's in 1996 and again in 2002, fetching $360K and $500K, respectively (Coover 1996, 2002). One of us wrote an essay on the manuscript for the auction catalogs (Janssen 1996). In 2002, the manuscript was acquired by Aristophil in Paris. This company published, in a limited edition, a boxed set with a reprint of the auction-catalog essay (illustrated by Laurent Taudin) and a facsimile of the manuscript (Einstein and Besso 2003). Closely matching the original in physical appearance, this facsimile comes as a folder with 37 loose sheets. It contains four pages not deemed significant enough for inclusion in CPAE4: a printed letter, dated "Ende April 1913," soliciting subscriptions to a *Festschrift* for the Swiss historian Gerald Meyer von Knonau (1843–1931), which has [p. 2] on the verso; the verso of a large folded sheet with [pp. 10–11] on the recto; and the verso of [p. 20], which has a figure and parts of an equation related to rotation.

In the editorial note in CPAE4 (p. 358), the 53 pages of the Einstein-Besso manuscript are divided into two parts: Part One written in close collaboration during a well-documented visit of Besso to Einstein in Zurich in June 1913 (42 pages) [we know that on June 18 Besso went back to Gorizia, where he was living in at the time; CPAE4, p. 357, note 57]; Part Two produced by Besso alone after Einstein sent him some papers, presumed to be Part One, in early 1914 (11 pages). In his (undated) cover letter (CPAE5, Doc. 499), Einstein wrote: "Here you finally have your manuscript bundle. It is really a shame if you do not bring the matter to completion."

Since the publication of the Einstein-Besso manuscript in CPAE4 in 1995, more pages of Part Two have come to light. These pages are all in Besso's hand and do not contain a single contribution from Einstein. They were part of another sheaf of loose sheets from the Besso *Nachlass* that Laurent Besso, Vero's son, made available to Robert Schulmann, then director of the Einstein Papers Project, in 1998. They include 14 additional pages of the Einstein-Besso manuscript, which can all be dated to 1913–14, and 8 pages with notes on general relativity, which can all be dated to 1916 (Janssen 2007, sec. 2).

The new material strongly suggests that Einstein and Besso met again in August 1913 to resume work on their joint project (see Janssen 2007, secs. 2.1.2 and 2.2, on the content and the dating of the new pages, respectively). Part One of the Einstein-Besso manuscript should accordingly be divided into two parts, the first part written during Besso's visit in June 1913, the second part during his visit in August 1913. Some of the Einstein pages could have been written any time between June 1913 and early 1914, when Einstein presumably sent the manuscript to Besso.

The most important item in the new part of the Einstein-Besso manuscript is the so-called "Besso Memo". Unlike any other pages of the manuscript, Besso dated the first of its four pages (which are written on a folded sheet): August 28, 1913. Two pages of the memo are reproduced in facsimile in Janssen (2007, p. 786, p. 789). In this memo, Besso appears to have recorded some of his discussions with Einstein at the time about the problem of rotation and about what would become the hole argument (see Janssen 2007, secs. 3 and 4).

With the exception of [pp. 41–42], the 14 pages presented in this chapter can reliably be dated to June 1913 and thus belong to Part One. The Besso material on [pp. 41–42] (which is not important for our purposes here) belongs to Part Two. Most likely, the Einstein material, which again bears on the problem of rotation (Janssen 1999), was written in June 1913 but it cannot be ruled out that Einstein only added it later that year (Janssen 2007, p. 809).

The other 12 pages selected for inclusion here, 9 by Einstein and 3 by Besso, deal with the perihelion motion of a single planet in the static field of the sun. Both the *Entwurf* theory and general relativity in its final form predict that the field of the sun contributes to a planet's perihelion motion. In Newton's theory, any perihelion motion is due to perturbations of other planets (see the tables in Misner, Thorne, and Wheeler (1973, p. 1113) or Smith (2014, p. 310 and 314) for the actual numbers). In fact, Newton had shown that a planet's perihelion would be stationary as long as there is only a central inverse-square force acting on it (Harper 2011, pp. 120–121).

On [pp. 1, 3–4, 6–7], Einstein solved the *Entwurf* field equations for the static field of the sun using an iterative approximation procedure. On [pp. 8–9], Besso took over and derived a differential equation for the angle between perihelion and aphelion of a planet moving in this field. The deviation of this angle from $\pi$ gives the perihelion advance in radians per half a revolution. On [pp. 10–11], Einstein performed some contour integrations to find this angle. On [p. 14] Besso expressed the formula Einstein found in terms of observable parameters. On [pp. 26 and 28], using one of the formulae Besso derived, Einstein calculated the perihelion advance of Mercury in radians per half a revolution and converted the result to seconds of arc per century. It can no longer be established from which source(s) Einstein got the various numbers he needed. It is clear, however, from several pages of the manuscript (e.g., [p. 31], not included here, and the bottom half of [p. 41]) that Besso consulted Simon Newcomb's (1895) *The Elements of the Four Inner Planets.*

Even at a superficial glance, one notices the striking difference between pages by Besso and pages by Einstein. The former have much more explanatory text to go with the calculations and many more deletions. The pages we selected are representative of the manuscript as a whole in this respect. Besso comes across as tentative and unsure of himself, Einstein as supremely confident and sure-footed. Besso nonetheless took responsibility for key parts of their joint project and corrected some errors in Einstein's calculations (see, in particular, [p. 4, eq. 28], [p. 6, eqs. 36–37]; [p. 35, eq. 211] (see Fig. 6.2 in Pt. I, Ch. 6); and [p. 14] where he corrected Einstein's overly crude approximations on [pp. 10–11]). He also raised at least one important question. On [p. 16] (not included here), in the course of summarizing Einstein's iterative procedure for solving the *Entwurf* field equations, he asked: "Is the static gravitational field [that Einstein had found for the sun] a particular solution or is it the general solution expressed in particular coordinates?" This same question comes up in the perihelion paper two years later (Einstein 1915c, p. 832; see Sec. 6.3). As one of us (JR) first suggested, Besso's original question may have been what inspired Einstein's hole argument (Janssen 2007, p. 820).

## [P. 1] Einstein

For his iterative calculation of the metric field of the sun, Einstein used the *Entwurf* field equations in their 'contravariant' form $\Delta_{\mu\nu}(\gamma) = \kappa(\Theta_{\mu\nu} + \vartheta_{\mu\nu})$ (Einstein and Grossmann 1913, p. 17, eq. (18)). $\Theta_{\mu\nu}$ is Einstein's notation for the contravariant components $T^{\mu\nu}$ of the energy-momentum tensor of matter. $\vartheta_{\mu\nu}$ is his notation for the 'contravariant' components of a quantity representing gravitational energy-momentum. We use scare quotes to indicate that this is not a generally covariant tensor. Neither is $\Delta_{\mu\nu}(\gamma)$.

Einstein expanded the metric field as a power series in the small parameter $A/r$, of the order of magnitude of the Newtonian potential. In our notation,

$$g^{\mu\nu} = \overset{(0)}{g}{}^{\mu\nu} + \overset{(1)}{g}{}^{\mu\nu} + \overset{(2)}{g}{}^{\mu\nu} + \ldots$$

For the first term, Einstein simply took the standard diagonal Minkowski metric, $\eta^{\mu\nu} = \mathrm{diag}(-1,-1,-1,1/c^2)$, in an arbitrary Lorentz frame.

To find the next term, he considered the *Entwurf* field equations for the weak static field produced by slow moving dust,

$$-\Box \overset{(1)}{g}{}^{\mu\nu} = \kappa\rho \frac{dx^\mu}{ds}\frac{dx^\nu}{ds}, \qquad \text{[eq. 1]}$$

where $\Box = \Delta - (\partial/\partial t)^2$ and the right-hand side is the product of Einstein's gravitational constant $\kappa$ and the energy-momentum tensor for slow moving dust. Einstein used the notation $\gamma_{\mu\nu}$ for the contravariant metric $g^{\mu\nu}$ (see Sec. 1.3, note 1). The only non-trivial component of this equation is the 44-component:

$$-\Delta \overset{(1)}{g}{}^{44} = \frac{\kappa\rho}{c^2}, \qquad \text{[eq. 2]}$$

where we used that $dx^4/ds = (dx^4/dt)(dt/ds) \approx 1/c$ (note that $x^\mu \equiv (x,y,z,t)$). Einstein explicitly wrote $c_0$ for the velocity of light in vacuo. This may be a remnant of his 1912 theory for static gravitational fields in which the velocity of light played the role of the gravitational potential and thus, in general, was not a constant but a function of $(x,y,z)$.

The solution of this equation for the special case of a spherical body of mass $M$, representing the sun, at some coordinate distance $r \equiv \sqrt{x^2+y^2+z^2}$ from the center, is given by

$$\overset{(1)}{g}{}^{44} = \frac{1}{c^2}\left(1 + \frac{A}{r}\right),$$

where

$$A \equiv \frac{\kappa M}{4\pi}. \qquad \text{[eq. 6]}$$

Inverting the matrix for $\overset{(1)}{g}{}^{\mu\nu}$ we find:

$$\overset{(1)}{g}{}_{44} = c^2\left(1 - \frac{A}{r}\right).$$

These are the 44-components of the matrices for $\overset{(1)}{g}_{\mu\nu}$ and $\overset{(1)}{g}{}^{\mu\nu}$ in [eqs. 4–5] under the header "table of the gravitational field for the first approximation."

To find the relation between $\kappa$ and Newton's gravitational constant $K$ (see [eqs. 3 and 6]), Einstein considered the equation of motion for a unit point mass in a static field (see [p. 13], not included here). The equation of motion can be found in the Zurich notebook (p. 5R/p. 11), in the *Entwurf* paper (Einstein and Grossmann 1913, p. 7, Eqs. 7 and 8), and on [p. 8, eqs. 48–49, 54], where Besso cites the *Entwurf* paper. For the static metric $\overset{(1)}{g}{}^{\mu\nu}$, it reduces to

$$\frac{d}{dt}\left(\frac{\dot{x}}{ds/dt}\right) = -\frac{1}{2}\frac{\partial \overset{(1)}{g}_{44}}{\partial x}\frac{dx^4}{ds}\frac{dx^4}{dt},$$

where the dot indicates a time derivative. Multiplying both sides by $ds/dt$, which can be taken to be constant, and using that $dx^4/dt = 1$, we arrive at

$$\ddot{x} = -\frac{1}{2}\frac{\partial \overset{(1)}{g}_{44}}{\partial x} = -\frac{\partial}{\partial x}\left(-\frac{\kappa M c^2}{8\pi r}\right).$$

Comparing this to Newton's second law for a unit point mass moving in the gravitational field of the sun,

$$\ddot{x} = -\frac{\partial \varphi}{\partial x} = -\frac{\partial}{\partial x}\left(-\frac{KM}{r}\right),$$

we conclude that:

$$\kappa = \frac{8\pi K}{c^2}. \qquad \text{[eq. 3]}$$

To find the field in second approximation, we need to include the terms quadratic in first-order derivatives of the metric in the *Entwurf* field equations that could be neglected in the first-order approximation. Inserting the zeroth- and first-order terms $\overset{(0)}{g}{}^{\mu\nu}$ and $\overset{(1)}{g}{}^{\mu\nu}$ into these quadratic terms, we arrive at the following equation for $\overset{(2)}{g}{}^{\mu\nu}$ (CPAE4, p. 349, Eq. (2)):

$$-\Delta \overset{(2)}{g}{}^{\mu\nu} + \frac{\overset{(0)}{g}{}^{\alpha\beta}}{\sqrt{-\overset{(0)}{g}}}\sqrt{-\overset{(1)}{g}}_{,\alpha}\,\overset{(1)}{g}{}^{\mu\nu}_{,\beta} - \overset{(0)}{g}{}^{\alpha\beta}\,\overset{(0)}{g}_{\tau\rho}\,\overset{(1)}{g}{}^{\mu\tau}_{,\alpha}\,\overset{(1)}{g}_{\nu\rho,\beta}$$
$$= -\frac{1}{2}\overset{(0)}{g}{}^{\alpha\mu}\,\overset{(0)}{g}{}^{\beta\nu}\,\overset{(1)}{g}_{\tau\rho,\alpha}\,\overset{(1)}{g}{}^{\tau\rho}_{,\beta} + \frac{1}{4}\overset{(0)}{g}{}^{\mu\nu}\,\overset{(0)}{g}{}^{\alpha\beta}\,\overset{(1)}{g}_{\tau\rho,\alpha}\,\overset{(1)}{g}{}^{\tau\rho}_{,\beta} \qquad (2.1)$$

The five terms in the equation above can be found on the bottom half of [p. 1], under the heading "Second approximation," the three terms on the left-hand side to the left of the vertical line, the two terms on the right-hand side to the right. Note that Einstein used subscripts '0' and '×' to mark the zeroth- and second-order terms $\overset{(0)}{g}{}^{\mu\nu}$ and $\overset{(2)}{g}{}^{\mu\nu}$, respectively.

Einstein substituted the expressions for $\overset{(1)}{g}{}^{\mu\nu}$ in [eqs. 4–5] into the five terms in Eq. (2.1) and recorded the results. We only present the calculations of the second term on the left-hand side. Only the 44-component of that term is non-vanishing. Using that

$$\sqrt{-\overset{(0)}{g}} = c, \quad \overset{(0)}{g}{}^{ij} = -\delta^{ij}, \quad \delta_{ij}x^i x^j = r^2, \quad \frac{\partial r}{\partial x^i} = \frac{x^i}{r},$$

$$\sqrt{-\overset{(1)}{g}}_{,i} = c\frac{\partial}{\partial x^i}\sqrt{1-\frac{A}{r}} = \frac{c}{2\sqrt{1-\frac{A}{r}}}\frac{A}{r^2}\frac{x^i}{r} \approx \frac{cA}{2r^2}\frac{x^i}{r},$$

$$\text{and}\quad \overset{(1)}{g}{}^{44}_{,i} = \frac{1}{c^2}\frac{\partial}{\partial x^i}\left(1+\frac{A}{r}\right) = -\frac{A}{c^2 r^2}\frac{x^i}{r},$$

we find that, to second order in $A/r$,

$$\frac{1}{\sqrt{-\overset{(0)}{g}}}\,\overset{(0)}{g}{}^{ij}\,\sqrt{-\overset{(1)}{g}}_{,i}\,\overset{(1)}{g}{}^{44}_{,j} = -\frac{1}{c}\delta_{ij}\left(\frac{cAx^i}{2r^3}\right)\left(-\frac{Ax^j}{c^2r^3}\right) = \frac{A^2}{2c^2r^4}.$$

The expressions on the bottom half of [p. 1] for the other terms in Eq. (2.1) are recovered in similar fashion.

## [P. 7] Einstein

On [p. 7], the verso of [p. 6], Einstein derived the general form of a spherically symmetric metric in Cartesian coordinates (on [p. 5], not included here, Besso more laboriously went through the same derivation; cf. Droste (1915, pp. 999–1000) and Earman and Janssen (1993, p. 144)). It is unclear why Einstein crossed out this entire calculation. Maybe it is because he realized he needed the general form of the contravariant $\overset{(2)}{g}{}^{\mu\nu}$ but derived the general form of a covariant metric on this page. However, the covariant and the contravariant metrics have the same form in this case (Droste 1915, p. 1000). The material in the lower-right corner is in Besso's hand.

As the figure in the top-left corner illustrates, Einstein first determined the metric at an arbitrary point $P$ in a Cartesian coordinate system $(x', y', z')$ in which $P$ lies on the $x'$-axis and then transformed to an arbitrary Cartesian coordinate system $(x, y, z)$. In this primed coordinate system, the metric at $P$, with coordinates $(x', y', z') = (r, 0, 0)$, has a very simple form. Since it is static, $g_{i4} = g_{4i} = 0$. Since it is spherically symmetric, $g'_{ij} = 0$ whenever $i \neq j$ and $g'_{22} = g'_{33}$. Einstein set

$$g'_{11} = R, \quad g'_{22} = g'_{33} = T, \quad g'_{44} = \Phi, \qquad \text{[eq. 43]}$$

where $R$, $T$, and $\Phi$ are yet to be determined functions of $r$. He then wrote down the transformation law for the metric from the primed to the unprimed coordinates:

$$g_{\mu\nu} = \pi_{\mu\alpha}\,\pi_{\nu\beta}\,g'_{\alpha\beta}, \qquad \text{[eq. 44]}$$

where $\pi_{\mu\alpha}$ is the notation he used during this period for the transformation matrix $\partial x'^{\alpha}/\partial x^{\mu}$ (cf. Sec. 1.3, note 20). Since the transformation only affects the spatial coordinates, $g_{44} = \pi_{44}\,\pi_{44}\,g'_{44} = g'_{44}$. Einstein only explicitly computed $g_{11}$ and $g_{12}$. Using [eq. 43] and $\pi_{i1} = \partial x'^1/\partial x^i = \partial r/\partial x^i = x^i/r$, he found that

$$g_{11} = \pi_{11}^2\,g'_{11} + \pi_{12}^2\,g'_{22} + \pi_{13}^2\,g'_{33} = \frac{x^2}{r^2}R + \frac{y^2}{r^2}T + \frac{z^2}{r^2}T = \frac{x^2}{r^2}(R-T) + T. \quad \text{[eq. 46]}$$

For $g_{12}$, [eqs. 43 and 44] give:

$$g_{12} = \pi_{11}\,\pi_{21}\,R + \pi_{12}\,\pi_{22}\,T + \pi_{13}\,\pi_{23}\,T.$$

Since $\pi_{ij}$ is the matrix of a rotation, its inverse is just the transposed matrix, $\pi^{-1}_{ij} = \pi^{\mathrm{T}}_{ij} = \pi_{ji}$. Hence, $\sum_k \pi_{ik}\pi_{jk} = \sum_k \pi_{ik}\pi^{-1}_{kj} = \delta_{ij}$. Einstein could thus use that $\sum_k \pi_{1k}\,\pi_{2k} = 0$ to substitute $-\pi_{11}\,\pi_{21}$ for $\pi_{12}\,\pi_{22} + \pi_{13}\,\pi_{23}$ and rewrite $g_{12}$ as

$$g_{12} = \pi_{11}\,\pi_{21}\,(R-T) = \frac{xy}{r^2}(R-T). \quad \text{[eq. 47]}$$

The other components of $g_{\mu\nu}$ can be found in the same way. The end result is given in [eq. 17] at the top of [p. 3].

## [P. 3] Einstein

At the top of [p. 3], under the header "Form of $\gamma_{\mu\nu}$ [i.e., $g^{\mu\nu}$] in the case of spherical symmetry," Einstein wrote down the matrix for a spherically symmetric (contravariant) metric $g^{\mu\nu}$ in Cartesian coordinates. A derivation of (the covariant form of) this metric can be found on [p. 7]. At this point, Einstein introduced the function $N \equiv R - T$ ([eq. 18]). With the help of this function, the spherically symmetric metric can be written as:

$$g^{ij} = \frac{x^i x^j}{r^2} N(r) + \delta^{ij} T(r), \quad g^{44} = \Phi(r), \quad g^{i4} = g^{4i} = 0. \qquad \text{[eq. 17]}$$

Einstein noted that the determinant $|g^{\mu\nu}|$—which in his notation is $\gamma$ and not $g \equiv |g_{\mu\nu}|$ (cf. [p. 6, eq. 41])—is given by

$$|g^{\mu\nu}| = \Phi T^2 (N + T). \qquad \text{[eq. 19]}$$

This follows directly from the diagonal form in which the metric is introduced on [p. 7] in the primed coordinate system (see [p. 7, eq. 43]): $|g^{\mu\nu}| = |g'^{\mu\nu}| = RT^2\Phi$. Substituting $R = N + T$, we recover [eq. 19].

On [p. 1], Einstein already determined that $\overset{(0)}{g}{}^{\mu\nu} = \eta^{\mu\nu}$ and that

$$\overset{(1)}{g}{}^{\mu\nu} = \text{diag}\left(0, 0, 0, \frac{1}{c^2}\frac{A}{r}\right).$$

To first order in $A/r$, the three functions determining the metric are therefore given by

$$\overset{(0)}{N} = \overset{(1)}{N} = 0, \quad \overset{(0)}{T} = -1, \quad \overset{(1)}{T} = 0, \quad \overset{(0)}{\Phi} = \frac{1}{c^2}, \quad \overset{(1)}{\Phi} = \frac{1}{c^2}\frac{A}{r} \qquad (2.2)$$

To find the second-order contributions to these functions, Einstein substituted [eq. 17] for $\overset{(2)}{g}{}^{\mu\nu}$ in the first term on the left-hand side of Eq. (2.1). But first he substituted the zeroth- and first-order solutions $\overset{(0)}{g}{}^{\mu\nu}$ and $\overset{(1)}{g}{}^{\mu\nu}$ in the remaining terms of Eq. (2.1). Under the header "Differential equations for $\gamma$" (i.e., $\overset{(2)}{g}{}^{\mu\nu}$), he wrote down the result for a few components.

$$-\Delta \overset{(2)}{g}{}^{11} = \frac{A^2}{2}\left(\left(\frac{\partial}{\partial x}\left(\frac{1}{r}\right)\right)^2 - \frac{1}{2r^4}\right) = -\frac{A^2}{4r^4} + x^2\frac{A^2}{2r^6}. \qquad \text{[eq. 20]}$$

$$-\Delta \overset{(2)}{g}{}^{12} = \frac{A^2}{2}\frac{\partial}{\partial x}\left(\frac{1}{r}\right)\frac{\partial}{\partial y}\left(\frac{1}{r}\right) = xy\frac{A^2}{2r^6} \qquad \text{[eq. 21]}$$

(the material to the right of [eq. 21] is in Besso's hand). For the 44-component, finally, he found (modulo an erroneous minus sign):

$$\Delta \overset{(2)}{g}{}^{44} = \frac{5}{4}\frac{A^2}{c^2 r^4}. \qquad \text{[eq. 22] (corrected)}$$

On [p. 4], Einstein derived an expression for the action of $\Delta$ on the 11-component of the metric in [eq. 17]. On the bottom half of [p. 3], the resulting equations for $\overset{(2)}{N}$ and $\overset{(2)}{T}$ are stated without derivation (see [eqs. 23–25]). [Eq. 26] for $\overset{(2)}{\Phi}$ follows directly from [eq. 22] and inherits the sign error from it. Moreover, the factor $1/c^2$ got lost.

Einstein crossed out the calculation on the bottom half of [p. 3] and started over at the top of [p. 6]. We will briefly return to the crossed-out material on [p. 3] after we have covered the derivation on [p. 4].

## [P. 4] Einstein

On [p. 4] Einstein derived the equations for the contributions of second order in $A/r$ to the functions $N(r)$, $T(r)$, and $\Phi(r)$ introduced in [p. 3, eq. 17]. Rewriting the left-hand side of [p. 3, eq. 20] in terms of these functions, we find:

$$\Delta \overset{(2)}{g}{}^{11} = \Delta\left(\frac{x^2}{r^2}\overset{(2)}{N} + \overset{(2)}{T}\right). \qquad \text{[eq. 27]}$$

Einstein now derived equations for $\Delta\overset{(2)}{T}$ and $\Delta(x^2\overset{(2)}{N}/r^2)$ (see [eqs. 28–29]). In our reconstruction of the calculations on [p. 4], we will suppress these superscripts, as they would further clutter the already tedious though straightforward algebra, but all $N$'s and $T$'s below should be read as $\overset{(2)}{N}$'s and $\overset{(2)}{T}$'s.

Einstein began by computing $\partial^2 T/\partial x^2$:

$$\frac{\partial^2 T}{\partial x^2} = \frac{\partial}{\partial x}\left(\frac{dT}{dr}\frac{x}{r}\right) = \frac{d^2T}{dr^2}\frac{x^2}{r^2} + \frac{1}{r}\frac{dT}{dr} - \frac{dT}{dr}\frac{x}{r^2}\frac{x}{r}.$$

The first and the third term on the right-hand side combine to form

$$\frac{x^2}{r}\frac{d}{dr}\left(\frac{1}{r}\frac{dT}{dr}\right).$$

Using similar expressions for $\partial^2 T/\partial y^2$ and $\partial^2 T/\partial z^2$, we find that

$$\Delta T = \frac{3}{r}\frac{dT}{dr} + r\frac{d}{dr}\left(\frac{1}{r}\frac{dT}{dr}\right) \qquad \text{[eq. 28]}$$

(the second term on the right-hand side originally had $1/r$ instead of $r$: the correction is in Besso's hand).

Next, Einstein turned his attention to $\frac{\partial^2}{\partial x^2}\left(x^2\frac{N}{r^2}\right)$. The first derivative is given by

$$\frac{\partial}{\partial x}\left(x^2\frac{N}{r^2}\right) = 2x\frac{N}{r^2} + \frac{x^2}{r^2}\frac{dN}{dr}\frac{x}{r} - \frac{2x^2}{r^3}\frac{x}{r}N.$$

The last two terms combine to form

$$\frac{x^3}{r}\frac{d}{dr}\left(\frac{N}{r^2}\right).$$

The second derivative is thus given by

$$\frac{\partial^2}{\partial x^2}\left(x^2\frac{N}{r^2}\right) = \frac{\partial}{\partial x}\left(\frac{2xN}{r^2} + \frac{x^3}{r}\frac{d}{dr}\left(\frac{N}{r^2}\right)\right).$$

Working out the first term on the right-hand side, we get

$$\frac{\partial}{\partial x}\left(\frac{2xN}{r^2}\right) = \frac{2N}{r^2} + \frac{2x}{r^2}\frac{dN}{dr}\frac{x}{r} - \frac{4xN}{r^3}\frac{x}{r} = \frac{2N}{r^2} + \frac{2x^2}{r}\frac{d}{dr}\left(\frac{N}{r^2}\right)$$

(where in the last step we combined two terms same way we did above); working out the second term, we get

$$\frac{\partial}{\partial x}\left(\frac{x^3}{r}\frac{d}{dr}\left(\frac{N}{r^2}\right)\right) = \frac{3x^2}{r}\frac{d}{dr}\left(\frac{N}{r^2}\right) - \frac{x^3}{r^2}\frac{x}{r}\frac{d}{dr}\left(\frac{N}{r^2}\right) + \frac{x^3}{r}\frac{\partial}{\partial x}\left(\frac{d}{dr}\left(\frac{N}{r^2}\right)\right).$$

We examine the third term on the right-hand side separately:

$$\begin{aligned}\frac{\partial}{\partial x}\left(\frac{d}{dr}\left(\frac{N}{r^2}\right)\right) &= \frac{\partial}{\partial x}\left(\frac{1}{r^2}\frac{dN}{dr} - \frac{2N}{r^3}\right)\\ &= \frac{1}{r^2}\frac{d^2N}{dr^2}\frac{x}{r} - \frac{2}{r^3}\frac{x}{r}\frac{dN}{dr} - \frac{2}{r^3}\frac{dN}{dr}\frac{x}{r} + \frac{6}{r^4}\frac{x}{r}N.\end{aligned}$$

Collecting all terms proportional to $x^2$ in $\frac{\partial^2}{\partial x^2}\left(x^2\frac{N}{r^2}\right)$, we find $\frac{5x^2}{r}\frac{d}{dr}\left(\frac{N}{r^2}\right)$; collecting all terms proportional to $x^4$, we find

$$-\frac{x^4}{r^3}\left(\frac{1}{r^2}\frac{dN}{dr} - \frac{2N}{r^3}\right) + \frac{x^4}{r}\left(\frac{1}{r^3}\frac{d^2N}{dr^2} - \frac{4}{r^4}\frac{dN}{dr} + \frac{6N}{r^5}\right),$$

which reduces to

$$\frac{x^4}{r}\left(\frac{1}{r^3}\frac{d^2N}{dr^2} - \frac{5}{r^4}\frac{dN}{dr} + \frac{8N}{r^5}\right).$$

One readily verifies that the factor in parentheses can be written more compactly:

$$\begin{aligned}\frac{d}{dr}\left(\frac{1}{r}\frac{d}{dr}\left(\frac{N}{r^2}\right)\right) &= \frac{d}{dr}\left(\frac{1}{r^3}\frac{dN}{dr} - \frac{2N}{r^4}\right)\\ &= \frac{1}{r^3}\frac{d^2N}{dr^2} - \frac{3}{r^4}\frac{dN}{dr} - \frac{2}{r^4}\frac{dN}{dr} + \frac{8N}{r^5}.\end{aligned}$$

Einstein thus arrived at the following intermediate result:

$$\frac{\partial^2}{\partial x^2}\left(x^2\frac{N}{r^2}\right) = \frac{2N}{r^2} + \frac{5x^2}{r}\frac{d}{dr}\left(\frac{N}{r^2}\right) + \frac{x^4}{r}\frac{d}{dr}\left(\frac{1}{r}\frac{d}{dr}\left(\frac{N}{r^2}\right)\right).$$

He similarly found that

$$x^2\frac{\partial^2}{\partial y^2}\left(\frac{N}{r^2}\right) = \frac{x^2}{r}\frac{d}{dr}\left(\frac{N}{r^2}\right) + \frac{x^2y^2}{r}\frac{d}{dr}\left(\frac{1}{r}\frac{d}{dr}\left(\frac{N}{r^2}\right)\right).$$

If $y$ is replaced by $z$ this turns into the expression for $\partial^2/\partial z^2$ acting on $x^2N/r^2$.

Adding the expressions for $\partial^2/\partial x^2$, $\partial^2/\partial y^2$, and $\partial^2/\partial z^2$ acting on $x^2N/r^2$, we obtain:

$$\Delta\left(x^2\frac{N}{r^2}\right) = \frac{2N}{r^2} + \frac{7x^2}{r}\frac{d}{dr}\left(\frac{N}{r^2}\right) + x^2r\frac{d}{dr}\left(\frac{1}{r}\frac{d}{dr}\left(\frac{N}{r^2}\right)\right). \qquad \text{[eq. 29]}$$

Adding [eq. 28] to [eq. 29] and substituting the resulting expression for $\Delta\left(\frac{x^2}{r^2}N + T\right)$ (cf. [eq. 27]) into [p. 3, eq. 20], Einstein arrived at

$$\underbrace{\frac{3}{r}\frac{dT}{dr}+r\frac{d}{dr}\left(\frac{1}{r}\frac{dT}{dr}\right)+\frac{2N}{r^2}}_{=\frac{A^2}{4r^4}}+x^2\underbrace{\left(\frac{7}{r}\frac{d}{dr}\left(\frac{N}{r^2}\right)+r\frac{d}{dr}\left(\frac{1}{r}\frac{d}{dr}\left(\frac{N}{r^2}\right)\right)\right)}_{=-\frac{A^2}{2r^6}}. \qquad \text{[eq. 30] (a), (b)}$$

As indicated by the upbrackets, the equation for $\overset{(2)}{g}{}^{11}$ splits into two equations, one for the terms proportional to $x^2$ and one for the remaining terms (the material in the lower-left corner, under [eq. 30] (a), is in Besso's hand).

[Eq. 30] is fully equivalent to [p. 3, eq. 23]. Like [eq. 30], [eq. 23] splits into two parts:

$$\frac{d^2T}{dr^2}+\frac{2}{r}\frac{dT}{dr}+\frac{2}{r^2}N=\frac{A^2}{4r^4}, \qquad \text{[p. 3, eq. 24]}$$

$$-\frac{6}{r^2}N+\frac{2}{r}\frac{dN}{dr}+\frac{d^2N}{dr^2}=-\frac{A^2}{2r^4}. \qquad \text{[p. 3, eq. 25]}$$

One readily verifies the equivalence of [eq. 30] (a)–(b) and [p. 3, eqs. 24–25].

As Einstein noted on [p. 3] ("second equation gives nothing new"), [eq. 21] for $\overset{(2)}{g}{}^{12}$ leads to equations for $\overset{(2)}{T}$ and $\overset{(2)}{N}$ that are equivalent to [eqs. 24–25]/[eqs. 30] (a)–(b).

The equation for $\overset{(2)}{\Phi}$, the second-order contribution to the third function determining the metric, is not given on [p. 4] but can be found (with a spurious minus sign and a factor $1/c^2$ missing) in the crossed out part at the bottom of [p. 3]:

$$\Delta\overset{(2)}{\Phi}=\frac{5}{4}\frac{A^2}{c^2r^4}. \qquad \text{[p. 3, eq. 26]}$$

## [P. 6] Einstein

On [p. 6], Einstein gave the end result of his derivation of the spherically symmetric metric field of the sun in Cartesian coordinates to second order in the parameter $A/r$ proportional to the Newtonian gravitational potential.

At the top of [p. 6], we once again find the equations derived on [pp. 3–4] for the second-order contributions to the functions $N(r)$, $T(r)$, and $\Phi(r)$. These equations, [eqs. 36 and 37], are identical to [p. 4, eqs. 30] (b) and (a), respectively, and equivalent to [p. 3, eqs. 25 and 24], respectively. [Eq. 38] is (the corrected version of) [p. 3, eq. 26] for $\overset{(2)}{\Phi}$ with the Laplacian $\Delta$ expressed in spherical coordinates (cf. [p. 4, eq. 28] for $\Delta\overset{(2)}{T}$):

$$\frac{3}{r}\frac{d\overset{(2)}{\Phi}}{dr}+r\frac{d}{dr}\left(\frac{1}{r}\frac{d\overset{(2)}{\Phi}}{dr}\right)=\frac{5}{4}\frac{A^2}{c^2r^4}. \qquad \text{[eq. 38]}$$

As in [p. 4, eq. 28], Besso corrected a factor $1/r$ to $r$ in the expressions for $\Delta\overset{(2)}{T}$ and $\Delta\overset{(2)}{\Phi}$ in [eqs. 37–38].

To the right of the vertical line drawn next to [eqs. 36–38] for $\overset{(2)}{N}$, $\overset{(2)}{T}$, and $\overset{(2)}{\Phi}$, Einstein listed the solutions of these equations:

$$\overset{(2)}{N}=\frac{1}{8}\frac{A^2}{r^2}, \quad \overset{(2)}{T}=0, \quad \overset{(2)}{\Phi}=\frac{1}{c^2}\frac{5}{8}\frac{A^2}{r^2}. \qquad \text{[eq. 39]}$$

As expected, these solutions result in contributions of second order in $A/r$ to the metric. One readily verifies that these are indeed solutions by inserting them into [eq. 38] and [eqs. 36–37].

Adding these second-order contributions to $N$, $T$, and $\Phi$ to the zeroth- and first-order contributions found earlier (see Eq. 2.2), we find

$$N = \frac{1}{8}\frac{A^2}{r^2}, \quad T = -1, \quad \Phi = \frac{1}{c^2}\left(1 + \frac{A}{r} + \frac{5}{8}\frac{A^2}{r^2}\right).$$

Substituting these expressions into [p. 3, eq. 17], Einstein arrived at the following result for the contravariant components of the metric field of the sun to second order in $A/r$:

$$g^{ij} = -\delta^{ij} + \frac{1}{8}\frac{A^2}{r^2}\frac{x^i x^j}{r^2}, \quad g^{44} = \frac{1}{c^2}\left(1 + \frac{A}{r} + \frac{5}{8}\frac{A^2}{r^2}\right), \quad g^{i4} = g^{4i} = 0. \qquad \text{[eq. 40]}$$

Inverting this matrix, Einstein found the covariant components of this metric field to second order in $A/r$:

$$g_{ij} = -\delta^{ij} - \frac{1}{8}\frac{A^2}{r^2}\frac{x^i x^j}{r^2}, \quad g_{44} = c^2\left(1 - \frac{A}{r} + \frac{3}{8}\frac{A^2}{r^2}\right), \quad g_{i4} = g_{4i} = 0. \qquad \text{[eq. 42]}$$

As becomes clear on [pp. 8–9], only $g_{44}$ is needed to second order in $A/r$ to calculate the motion of a planet's perihelion in the field of the sun. On [p. 2], not included here, Einstein went through a calculation similar to the one on [pp. 1, 3–4, 6–7] but starting from the 'covariant' form of the *Entwurf* field equations, $-D_{\mu\nu} = \kappa(t_{\mu\nu} + T_{\mu\nu})$ (Einstein and Grossmann 1913, p. 17, eq. (21)) and deriving only the expression for $g_{44}$ to order $A^2/r^2$. [P. 2] is the verso of a printed letter dated "Ende April 1913." Einstein, in all likelihood, did the calculations on [pp. 1, 3–4, 6–7] before those on [p. 2]. It cannot be determined, of course, exactly when Einstein cannibalized this piece of paper for his calculations. But it is plausible that this would have been around the time of Besso's visit in June 1913.

## [P. 8] Besso

On [pp. 8–9], Besso used the equations of motion of a test particle in a metric field to derive a differential equation for the angle between perihelion and aphelion of the orbit of a single planet in the field of the sun given by [p. 6, eq. 42].

He started from the Lagrangian for a test particle of rest mass $m$ in an arbitrary metric field. Following Einstein and Grossmann (1913, p. 7, eq. 5), Besso used $H$ instead of $L$ for the Lagrangian:

$$H = -m\frac{ds}{dt} = -m\sqrt{\frac{ds^2}{dt^2}} = -m\sqrt{g_{\mu\nu}\dot{x}^\mu \dot{x}^\nu}.$$

The Euler-Lagrange equations are (p. 7, eq. 6):

$$\frac{d}{dt}\left(\frac{\partial H}{\partial \dot{x}^\mu}\right) - \frac{\partial H}{\partial x^\mu} = 0.$$

If the generalized momentum and the gravitational four-force are defined as $J^\mu \equiv \partial H/\partial \dot{x}^\mu$ and $\mathfrak{K}^\mu \equiv \partial H/\partial x^\mu$, respectively (p. 7, eqs. 7–8), the $x$-component of the Euler-Lagrange equations can be written as $dJ_x/dt = \mathfrak{K}_x$ ([eq. 48]), where

$$J_x = \frac{\partial H}{\partial \dot{x}^1} = -\frac{m}{2\sqrt{ds^2/dt^2}} 2g_{1\nu}\dot{x}^\nu = -\frac{mg_{1\nu}dx^\nu}{ds},$$

and

$$\mathfrak{K}_x = -\frac{\partial H}{\partial x^1} = -\frac{m}{2\sqrt{ds^2/dt^2}} \frac{\partial g_{\mu\nu}}{\partial x^1} \dot{x}^\mu \dot{x}^\nu = -\frac{1}{2} m \frac{\partial g_{\mu\nu}}{\partial x^1} \frac{dx^\mu}{ds} \frac{dx^\nu}{dt}.$$

Under the header, "The equations of motion for the material point are," Besso could thus write down the $x$-component of the equations of motion as

$$\frac{d}{dt}\left(\frac{g_{1\nu}\dot{x}^\nu}{ds/dt}\right) = \frac{1}{2}\frac{\partial g_{\mu\nu}}{\partial x^1}\frac{dx^\mu}{dt}\frac{dx^\nu}{dt}\frac{dt}{ds}. \qquad \text{[eq. 49]}$$

The energy $E$ of a test particle is given by the Hamiltonian, i.e., the Legendre transform $\mathbf{J}\cdot\dot{\mathbf{x}} - H$ of the Lagrangian $H$ (Einstein and Grossmann 1913, p. 7, eq. 9):

$$E = -\frac{mg_{i\nu}dx^\nu}{ds}\dot{x}^i + m\frac{ds}{dt} = m\frac{-g_{i\nu}dx^i dx^\nu + ds^2}{dsdt} = m\frac{g_{4\nu}dx^\nu}{ds}, \qquad \text{[eq. 50]}$$

where in the last step we used that $ds^2 = g_{\mu\nu}dx^\mu dx^\nu$ and that

$$g_{\mu\nu}dx^\mu dx^\nu - g_{i\nu}dx^i dx^\nu = g_{4\nu}dtdx^\nu.$$

It follows that, for a stationary metric ($g_{i4} = g_{4i} = 0$),

$$E = mg_{44}\frac{dt}{ds} \qquad \text{[eq. 51] (corrected)}$$

(the manuscript has a minus sign on the right-hand side and omits the factor $m$).

Besso was interested in a slow-moving planet in the spherically symmetric weak static field of the sun. The virial theorem told Besso that potential and kinetic energy, and hence the quantities $A/r$ and $q^2/c^2$ (with $q \equiv \dot{x}^2 + \dot{y}^2 + \dot{z}^2$), are of the same order of magnitude. This allowed him to write down [eqs. 52–53] for $ds/dt$ to second order in $A/r$. The convoluted sentence, replete with deletions, introducing these equations reads: "Here [i.e., in [eq. 51]] is, if one considers the quantities $A/r$ and $q^2/c^2$ as infinitesimally small quantities of the first order ... and in addition takes into account that the field is constant in time, so that $g_{\nu 4}$ for $\nu = 1 \text{ to } 3$ should be set $= 0$, and also once again sets $x_1 = x$, $x_2 = y$, $x_3 = z$, $x_4 = t$:"

$$\begin{aligned}\frac{ds}{dt} &= \sqrt{\left(\overset{(0)}{g}_{ij} + \overset{(1)}{g}_{ij}\right)\dot{x}^i\dot{x}^j + \overset{(0)}{g}_{44} + \overset{(1)}{g}_{44} + \overset{(2)}{g}_{44}} \\ &= c\sqrt{1 - \frac{q^2}{c^2} - \frac{A}{r} + \frac{3}{8}\frac{A^2}{r^2}}, \qquad \text{[eqs. 52, 53]}\end{aligned}$$

where we used [p. 6, eq. 42] to set $\overset{(1)}{g}_{ij} = 0$ and

$$g_{44} = c^2\left(1 - \frac{A}{r} + \frac{3}{8}\frac{A^2}{r^2}\right). \qquad \text{[p. 6, eq. 42]}$$

In general relativity in its final November 1915 form, it is no longer true that $\overset{(1)}{g}_{ij} = 0$ (see Einstein 1915c, p. 834; cf. Sec. 6.3.3, note 8, and Pt. I, Ch. 6).

Again neglecting terms smaller than second order in $A/r$, Besso could write [eq. 49] and the corresponding $y$- and $z$-components of the equations of motion for this special case as

$$\frac{d}{dt}\left(\frac{\dot{x}^i}{ds/dt}\right) = -\frac{1}{2}\frac{\partial g_{44}}{\partial x^i}\frac{dt}{ds} = -\frac{1}{2}\frac{dg_{44}}{dr}\frac{x^i}{r}\frac{dt}{ds}, \qquad \text{[eq. 54]}$$

and, silently correcting the sign error in [eq. 51], the energy $E$ per unit rest mass as

$$E = \frac{g_{44}}{ds/dt}. \qquad \text{[eq. 55]}$$

Besso now derived the analogue in the *Entwurf* theory of Kepler's area law. In Newtonian theory, the areal velocity, the area swept out in unit time by the vector $\mathbf{x}$ giving the position of the test particle in some arbitrary Cartesian coordinate system, is given by $\frac{1}{2}(\mathbf{x}\times\dot{\mathbf{x}})$. The angular momentum, $\mathbf{L} \equiv m(\mathbf{x}\times\dot{\mathbf{x}})$, is thus equal to $2m$ times the areal velocity. For a central force, the angular momentum and hence the areal velocity are conserved:

$$\frac{d\mathbf{L}}{dt} = m\frac{d}{dt}(\mathbf{x}\times\dot{\mathbf{x}}) = \dot{\mathbf{x}}\times\dot{\mathbf{x}} + \mathbf{x}\times\ddot{\mathbf{x}} = 0,$$

where the second term vanishes on account of the equations of motion, $\ddot{\mathbf{x}} = -(d\varphi/dr)(\mathbf{x}/r)$. It follows that the plane of the orbit of a particle moving under the influence of a central force remains fixed and that the areal velocity is a constant (i.e., Kepler's area law).

Besso picked Cartesian coordinates in which the planetary orbit lies in the $xy$-plane. The angular momentum and the areal velocity vector are thus in the $z$-direction. Besso introduced the notation $\dot{f}$ for the magnitude of the areal velocity (*Flächengeschwindigkeit*). He set $2\dot{f}$ equal to $y\dot{x} - x\dot{y}$, which is actually *minus* the $z$-component of $\mathbf{x}\times\dot{\mathbf{x}}$, but this sign error does not materially affect the rest of his argument (at the bottom of [p. 13], not included here, Einstein went through the same argument using the correct sign). Besso noted that

$$\frac{d}{dt}\left(\frac{y\dot{x} - x\dot{y}}{ds/dt}\right) = y\frac{d}{dt}\left(\frac{\dot{x}}{ds/dt}\right) - x\frac{d}{dt}\left(\frac{\dot{y}}{ds/dt}\right) = 0, \qquad \text{[eq. 56]}$$

where in the last step we used [eq. 54]. Besso concluded that $(y\dot{x} - x\dot{y})(dt/ds)$ should be some constant, $B$. He called the product of $B$ and $c$, the leading term in $ds/dt$ (if $g_{\mu\nu} = \eta_{\mu\nu}$, $ds = cdt\sqrt{1-q^2/c^2}$), the "area law constant" (*Flächensatzkonstante*). Correcting the sign error in Besso's definition of the areal velocity, we can thus write:

$$2\dot{f} = x\dot{y} - y\dot{x} = B\frac{ds}{dt}. \qquad \text{[eq. 57] (corrected)}$$

At the bottom of the page, Besso noted: "Since $ds/dt$ is a square root, it is more convenient to work with the squared equations in polar coordinates." This is what he proceeded to do at the top of [p. 9].

Note that the two equations for the perihelion motion, [eqs. 55 and 57], only involve the $g_{44}$ component to second order in $A/r$ (cf. [eq. 53] for $ds/dt$). This may be why Einstein rederived the metric field for the sun on [p. 2], calculating only $g_{44}$ to order $A^2/r^2$ (cf. the observation at the end of our commentary for [p. 6]).

## [P. 9] Besso

On [p. 9], Besso finished the derivation he began on [p. 8] of a differential equation for the angle between perihelion and aphelion. First, he rewrote [p. 8, eqs. 57 and 55]—expressing essentially conservation of angular momentum and energy, respectively—as

$$\dot{\varphi}r^2 = BW \qquad \text{[eq. 58]}$$

(where the left-hand side is $x\dot{y} - y\dot{x}$ in polar coordinates and $W \equiv ds/dt$), and

$$E = \frac{g_{44}}{W}. \qquad \text{[eq. 59]}$$

Using [eq. 59] to eliminate $W$ from [eq. 58] and using [p. 6, eq. 42] for $g_{44}$, Besso found

$$E\dot{\varphi}r^2 = Bg_{44} = Bc^2\left(1 - \frac{A}{r} + \frac{3}{8}\frac{A^2}{r^2}\right). \qquad \text{[eq. 60]}$$

Besso neglected the last term. Although this is not clear at this point, we will see later that he was justified in doing so (see our comments following Eq. (2.10) below). Introducing $F \equiv Bc^2/E$, he wrote [eq. 60] as

$$\dot{\varphi}r^2 = F\left(1 - \frac{A}{r}\right). \qquad \text{[eq. 61]}$$

On the next line, Besso inserted [p. 6, eq. 42] for $g_{44}$ and [p. 8, eq. 53] for $W = ds/dt$ into the equation $g_{44}^2 = E^2W^2$ (i.e., [eq. 59] squared). Neglecting terms smaller than of order $A^2/r^2$, he found:

$$c^4\left(1 - \frac{2A}{r} + \frac{A^2}{r^2} + \frac{6}{8}\frac{A^2}{r^2}\right) = E^2c^2\left(1 - \frac{q^2}{c^2} - \frac{A}{r} + \frac{3}{8}\frac{A^2}{r^2}\right). \qquad \text{[eq. 62]}$$

Dividing both sides by $E^2c^2$, putting $q^2/c^2$ on the left- and everything else on the right-hand side, he rewrote this as:

$$\frac{q^2}{c^2} = 1 - \frac{c^2}{E^2} + \frac{A}{r}\left(\frac{2c^2}{E^2} - 1\right) + \frac{A^2}{r^2}\left(\frac{3}{8} - \frac{7}{4}\frac{c^2}{E^2}\right). \qquad \text{[eq. 64]}$$

[Eq. 59] shows that $c^2/E^2$ is of the form $1+\varepsilon$, where $\varepsilon$ is of order $A/r$ (see the calculation in the upper-right corner of the page). Besso could thus substitute $1+\varepsilon$ for $c^2/E^2$ in the term with $A/r$ on the right-hand side of [eq. 64] and 1 in the term with $A^2/r^2$. Expressing $q^2$ on the left-hand side in polar coordinates, he arrived at:

$$\frac{dr^2 + r^2d\varphi^2}{c^2dt^2} = -\varepsilon + \frac{A}{r}(1+2\varepsilon) - \frac{11}{8}\frac{A^2}{r^2}. \qquad \text{[eq. 65]}$$

Besso now used [eq. 61] squared (to order $A/r$),

$$r^4\frac{d\varphi^2}{dt^2} = F^2\left(1 - \frac{2A}{r}\right), \qquad \text{[eq. 66]}$$

to eliminate $dt^2$ from [eq. 65]:

$$\frac{1}{c^2}\left(dr^2 + r^2d\varphi^2\right)F^2\left(1 - \frac{2A}{r}\right) = r^4d\varphi^2\left(-\varepsilon + \frac{A}{r}(1+2\varepsilon) - \frac{11}{8}\frac{A^2}{r^2}\right).$$

Collecting all terms with $d\varphi^2$ on the left- and all terms with $dr^2$ on the right-hand side, we can rewrite this equation as

$$\frac{r^2F^2}{c^2}\left(1 - \frac{2A}{r}\right)d\varphi^2 - r^4\left(-\varepsilon + \frac{A}{r}(1+2\varepsilon) - \frac{11}{8}\frac{A^2}{r^2}\right)d\varphi^2 = -\frac{F^2}{c^2}\left(1 - \frac{2A}{r}\right)dr^2.$$

Rearranging terms on the left-hand side, grouping them by powers of $r$, we find:

$$-\varepsilon\left(-r^4 + A\left(\frac{1}{\varepsilon} + 2\right)r^3 - \frac{1}{\varepsilon}\left(\frac{F^2}{c^2} + \frac{11}{8}A^2\right)r^2 + \frac{2AF^2}{\varepsilon c^2}r\right)d\varphi^2.$$

Dividing both sides of the equation above by the expression multiplying $d\varphi^2$ and taking the square root of the resulting equation, we find the differential equation for the angle $\varphi$ as a function of $r$ that Besso was after:

$$d\varphi = \frac{F}{c\sqrt{\varepsilon}} \frac{1-\dfrac{A}{r}}{\sqrt{-r^4 + A\left(\dfrac{1}{\varepsilon}+2\right)r^3 - \dfrac{1}{\varepsilon}\left(\dfrac{F^2}{c^2}+\dfrac{11}{8}A^2\right)r^2 + \dfrac{2AF^2}{\varepsilon c^2}r}}\,dr. \quad \text{[eq. 67]}$$

Integrating the right-hand side from the minimum value $r_1$ at perihelion to the maximum value $r_2$ at aphelion, we find the angle between perihelion and aphelion (cf. [eq. 68]).

## [P. 10] Einstein

On [pp. 10–11], Einstein performed two contour integrations to evaluate $\int_{r_1}^{r_2} d\varphi$ (see [p. 9, eq. 68] or, equivalently, [p. 11, eq. 86]) to find the angle between perihelion and aphelion. He made two mistakes. First, as he noticed at the bottom of [p. 11], he dropped a "factor $\frac{1}{2}$ in front of the whole integral." He noticed this because he found an angle between perihelion and aphelion close to $2\pi$, whereas this angle clearly should be close to $\pi$. Second, the approximation he used in the final steps of his derivation of the formula for this angle was so crude that he did not find any deviation from $\pi$ and hence no perihelion motion. On [p. 14] Besso repeated these steps more carefully and did find a small deviation from $\pi$.

The fourth-order polynomial under the square-root sign in [p. 9, eq. 68] can be factorized as (cf. [eq. 69])

$$(r-r_1)(r_2-r)(r-r')(r-r''). \tag{2.3}$$

The roots $r_1$ and $r_2$ are the values of $r$ at perihelion and aphelion, respectively. They are both much larger than the roots $r'$ and $r''$. It is therefore convenient to expand the integrand of [p. 9, eq. 68] in $r/r'$ and $r/r''$, with the help of (cf. [eq. 70])

$$\frac{1}{\sqrt{(r-r')(r-r'')}} = \frac{1}{r\sqrt{\left(1-\dfrac{r'}{r}\right)\left(1-\dfrac{r''}{r}\right)}} = \frac{1}{r}\left(1+\frac{r'+r''}{2r}\right).$$

The angle between perihelion and aphelion can then be written as the sum of two integrals:

$$\int_{r_1}^{r_2} d\varphi = CI_f + C\left(\tfrac{1}{2}(r'+r'')-A\right)I_g \tag{2.4}$$

where

$$I_f \equiv \int_{r_1}^{r_2} dr\, f(r), \quad \text{with} \quad f(r) \equiv \frac{1}{r\sqrt{(r-r_1)(r_2-r)}},$$

$$I_g \equiv \int_{r_1}^{r_2} dr\, g(r), \quad \text{with} \quad g(r) \equiv \frac{1}{r^2\sqrt{(r-r_1)(r_2-r)}}.$$

The factor $C$ is our abbreviation for $F/c_0\sqrt{\varepsilon}$ in [p. 9, eq. 68]. In our discussion of [pp. 10–11, 13] we will follow Einstein and Besso in writing the velocity of light in vacuo as $c_0$ to avoid confusion with the coefficient $c$ of $r^2$ in the polynomial under the square-root sign in [p. 9, eq. 68].

Below the first horizontal line on [p. 10] (see [eq. 72]), Einstein evaluated $I_f$; below the second horizontal line (see [eq. 73]), he evaluated $I_g$. He used that $2I_f$ and $2I_g$ are equal to integrals in the complex plane over the contour $A$ with negative (i.e., clockwise) orientation (the integrals just above and just below the real axis are over different sheets of the functions $f(z)$ and $g(z)$ and therefore have opposite signs). As illustrated in Fig. 2.1, an enhanced version of Einstein's own figure to the right of [eq. 72], these integrals can be replaced by integrals over the contour $C$, a small circle around $z = 0$ in the complex plane, with a positive (i.e., counterclockwise) orientation.

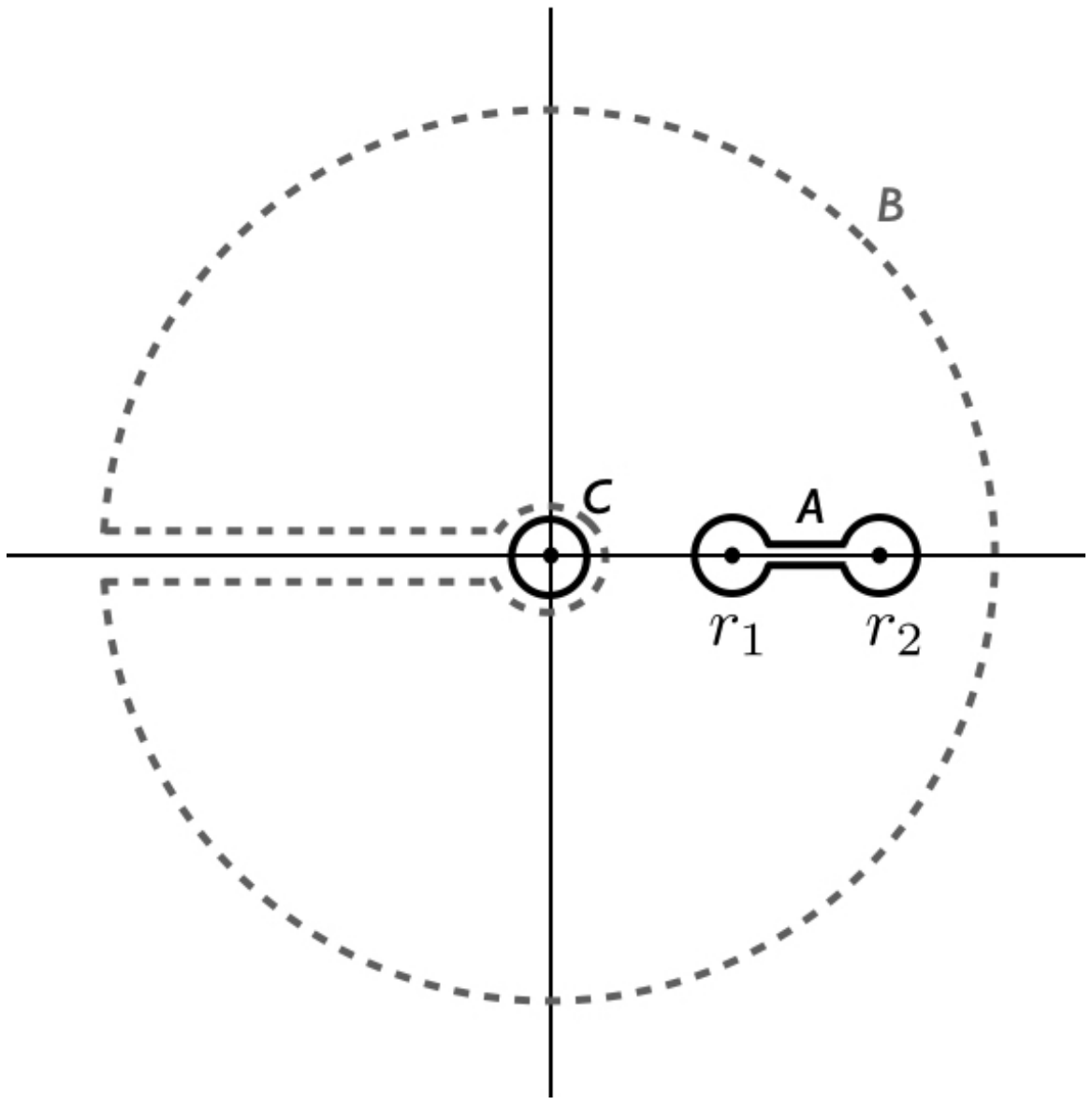

**Figure 2.1** The contour $A$ in the complex plane can continuously be deformed into the contour $B$ without passing through the poles 0, $r_1$, and $r_2$ on the real axis of the functions $f(z)$ and $g(z)$ being integrated on [p. 10]. If the radius of $B$ goes to infinity, the only contributions to these integrals come from the small circle $C$ around the pole $z = 0$ (the integral from $-\infty$ to 0 just below the real axis cancels the integral from 0 to $-\infty$ just above the real axis). Einstein's own drawing on [p. 10] shows $A$ and $C$. With the help of $B$ one readily sees that the integrals over $A$ (clockwise) and $C$ (counterclockwise) give the same result.

Although Einstein did not appeal to it in the manuscript, Cauchy's residue theorem tells us that

$$2I_f = 2\pi i \operatorname*{Res}_{z=0} f(z), \qquad 2I_g = 2\pi i \operatorname*{Res}_{z=0} g(z),$$

where the residues are

$$\operatorname*{Res}_{z=0} f(z) = \Big(z f(z)\Big)\Big|_{z=0} = \frac{1}{i\sqrt{r_1 r_2}},$$

$$\operatorname*{Res}_{z=0} g(z) = \frac{d}{dz}\Big(z^2 g(z)\Big)\Big|_{z=0} = \frac{1}{2i\sqrt{r_1 r_2}}\left(\frac{1}{r_1} + \frac{1}{r_2}\right).$$

We thus find that

$$2I_f = \frac{2\pi}{\sqrt{r_1 r_2}}, \qquad 2I_g = \frac{\pi}{\sqrt{r_1 r_2}}\left(\frac{1}{r_1} + \frac{1}{r_2}\right). \tag{2.5}$$

Einstein omitted the factors of 2 on the left-hand side, which is the source of the error he noted at the bottom of [p. 11].

Substituting the results for $I_f$ and $I_g$ into Eq. (2.4), we find:

$$\int_{r_1}^{r_2} d\varphi = \frac{C\pi}{\sqrt{r_1 r_2}} \left(1 + \frac{1}{2}\left(\frac{1}{r_1} + \frac{1}{r_2}\right)\left(\frac{r' + r''}{2} - A\right)\right). \tag{2.6}$$

The next step is to express the roots of the polynomial in Eq. (2.3) in terms of its coefficients (cf. [eq. 75]):

$$-a\,r^4 + b\,r^3 + c\,r^2 + d\,r + e. \tag{2.7}$$

The roots $r'$ and $r''$ are so small that their third and fourth powers are negligible. These roots should thus satisfy (or, as Einstein wrote, "infinitesimally small roots from the equation"):

$$c\left(r^2 + \frac{d}{c}r + \frac{e}{c}\right) = c(r - r')(r - r'') = c(r^2 - (r' + r'')\,r + r'r'').$$

It follows that

$$r' + r'' = -\frac{d}{c}. \qquad \text{[eq. 77]}$$

Writing the polynomial in Eq. (2.7) (with $a = -1$) as (cf. [eq. 78])

$$-r^4 + (r_1 + r_2 + r' + r'')\,r^3 - (r_1 r_2 + (r_1 + r_2)(r' + r''))\,r^2 + r_1 r_2 (r' + r'')\,r - r_1 r_2 r' r'' \tag{2.8}$$

and neglecting any and all terms containing $r'$ or $r''$, Einstein found

$$-r^4 + (r_1 + r_2)\,r^3 - r_1 r_2\,r^2,$$

from which he concluded that:

$$r_1 + r_2 = b, \quad r_1 r_2 = -c. \qquad \text{[eq. 79, 80]}$$

It follows that

$$\frac{1}{r_1} + \frac{1}{r_2} = \frac{r_1 + r_2}{r_1 r_2} = -\frac{b}{c}. \tag{2.9}$$

On [p. 14] Besso revisited this identification of roots and coefficients neglecting only terms of second order in $r'$ and $r''$.

## [P. 11] Einstein

Inserting [p. 10, eq. 77] and Eq. (2.9) into Eq. (2.6), we arrive at

$$\int_{r_1}^{r_2} d\varphi = \frac{C\pi}{\sqrt{-c}} \left(1 + \frac{1}{2}\frac{b}{c}\left(\frac{d}{2c} + A\right)\right). \qquad \text{[eq. 81] (corrected)}$$

Besides the spurious factor 2, Einstein omitted the factor $C$ (our abbreviation for $F/c_0\sqrt{\varepsilon}$; cf. [p. 9, eq. 67]) and the term with $A$ on the right-hand side.

The part of [p. 11] between the double horizontal line under [eq. 81] and the horizontal line above [eq. 86] mostly repeats parts of the argument on [p. 10]. If we set $C = 1$ and $A = -\alpha$ in Eq. (2.6), the right-hand side turns into the final expression for the integral $I$ in [eq. 84]. [Eq. 85] and the four lines below it are in Besso's hand (cf. Eq. (2.8)). Besso may also have crossed out the factor 2 in [eq. 84].

[Eq. 86] is Einstein's version of the right-hand side of [p. 9, eq. 68] derived by Besso for the angle between perihelion and aphelion. On [pp. 25, 27, and 33], not

included here, Besso checked whether these two equations are equivalent. No pages with Einstein's derivation of [eq. 86] survive. Einstein probably discarded them before mailing the manuscript to Besso in early 1914. Using Eq. (2.7), with $a = -1$, for the polynomial under the square-root sign, we can write [eq. 86] as:

$$\int_{r_1}^{r_2} d\varphi = \frac{c_0 B}{\sqrt{c_0^2 - E^2}} \int_{r_1}^{r_2} \frac{1 - A/r}{\sqrt{-r^4 + br^3 + cr^2 + dr + e}} dr, \tag{2.10}$$

where the coefficients $b$, $c$, $d$, and $e$ that we read off of Einstein's [eq. 86] are

$$b = A\left(\frac{2c_0^2 - E^2}{c_0^2 - E^2}\right), \quad c = \frac{A^2\left(\frac{7}{4}c_0^2 - \frac{3}{8}E^2\right) - B^2 c_0^2}{c_0^2 - E^2},$$

$$d = \frac{2B^2 c_0^2 A}{c_0^2 - E^2}, \qquad e = -\frac{7}{4}\frac{B^2 c_0^2 A^2}{c_0^2 - E^2}.$$

The zeroth-order term, $e$, is absent from Besso's result, [p. 9, eq. 68]. This is because Besso neglected the $A^2/r^2$-term in [p. 9, eq. 60]. Since $e = -r_1 r_2 r' r''$ (see Eq. (2.8)), Besso was justified in setting $e = 0$.

The same cannot be said for the approximations in [p. 10, eqs. 79–80] on which Einstein relied in [eq. 81]. Setting $c = -\frac{B^2 c_0^2}{c_0^2 - E^2}$ (he was justified in neglecting the term with $A^2$), he found $\frac{d}{c} = -2A$ and $\sqrt{-c} = \frac{Bc_0}{\sqrt{c_0^2 - E^2}}$ ([eq. 87]). When these expressions along with $C = \frac{c_0 B}{\sqrt{c_0^2 - E^2}}$ are inserted into [eq. 81], the result is that the angle between aphelion and perihelion is equal to $\pi$, which would mean that the *Entwurf* theory predicts no perihelion motion at all in the case of a single planet orbiting the sun.

Because of the factor 2 he dropped on [p. 10] (see Eq. (2.5)), Einstein found $2\pi$ instead of $\pi$. He immediately recognized he was off by a factor 2. As he wrote directly below [eq. 87]: "Factor $\frac{1}{2}$ forgotten in front of the whole integral." He left it to Besso to find the deviation from $\pi$.

## [P. 14] Besso

On [p. 14], Besso went through a more careful version of Einstein's derivation at the bottom of [p. 10] of the relation between roots and coefficients of the fourth-order polynomial under the square-root sign in the integral giving the angle between perihelion and aphelion. Besso used [p. 9, eq. 67] for this integral, which he himself derived, rather than [p. 11, eq. 86], which Einstein had found. Writing the polynomial in the form of Eq. (2.7), we thus arrive at the following identification of its coefficients:

$$a = -1, \quad b = \left(\frac{1}{\varepsilon} + 2\right)A, \quad c = -\frac{1}{\varepsilon}\left(\frac{F^2}{c_0^2} + \frac{11}{8}A^2\right), \quad d = \frac{2F^2 A}{\varepsilon c_0^2}, \quad e = 0. \tag{2.11}$$

Using [p. 10, eq. 77] and neglecting the term with $A^2$ in $c$, we find:

$$r' + r'' = -\frac{d}{c} = 2A. \qquad \text{[eq. 98]}$$

Comparing Eq. (2.7) and Eq. (2.8), we read off the following relations between roots and coefficients of the polynomial:

$$r_1 + r_2 + r' + r'' = b, \tag{2.12}$$

$$r_1 r_2 + (r_1 + r_2)(r' + r'') = -c. \tag{2.13}$$

In [p. 10, eqs. 79–80], Einstein neglected the terms with $r'$ and $r''$ in these relations.

Using Eq. (2.11) for $b$ and [eq. 98] for $r' + r''$ in Eq. (2.12), we find:

$$r_1 + r_2 = \left(\frac{1}{\varepsilon} + 2\right) A - 2A = \frac{A}{\varepsilon}. \qquad \text{[eq. 99]}$$

Similarly, using Eq. (2.11) for $c$ (now without neglecting the term with $A^2$), [eq. 98] for $r' + r''$, and [eq. 99] for $r_1 + r_2$ in Eq. (2.13), we find:

$$r_1 r_2 = \frac{1}{\varepsilon}\left(\frac{F^2}{c_0^2} + \frac{11}{8}A^2\right) - \frac{2A^2}{\varepsilon} = \frac{1}{\varepsilon}\left(\frac{F^2}{c_0^2} - \frac{5}{8}A^2\right). \qquad \text{[eq. 100]}$$

Substituting [eq. 98] into Eq. (2.6), we see that $\frac{1}{2}(r' + r'') - A = 0$. Eq. (2.6) thus reduces to $\int_{r_1}^{r_2} d\varphi = \frac{C\pi}{\sqrt{r_1 r_2}}$. Inserting $C = \frac{F}{c_0\sqrt{\varepsilon}}$ (cf. [p. 9, eq. 67]) and [eq. 100] for $r_1 r_2$, we find:

$$\int_{r_1}^{r_2} d\varphi = \frac{\frac{F}{c_0\sqrt{\varepsilon}}\pi}{\sqrt{\frac{1}{\varepsilon}\left(\frac{F^2}{c_0^2} - \frac{5}{8}A^2\right)}} = \frac{\pi}{\sqrt{1 - \frac{5}{8}\frac{A^2 c_0^2}{F^2}}} \approx \pi\left(1 + \frac{5}{16}\frac{A^2 c_0^2}{F^2}\right). \qquad \text{[eq. 101]}$$

For a single planet orbiting the sun, the *Entwurf* theory thus predicts a perihelion advance of

$$\frac{5}{16}\pi\left(\frac{Ac_0}{F}\right)^2 \tag{2.14}$$

radians per half a revolution.

On the remainder of [p. 14], Besso expressed $F$ and $A$ in terms of parameters for which one could easily find numerical values.

In [p. 9, eq. 61], $F$ was defined as $Bc_0^2/E$. Using the definition of $\varepsilon$ in [p. 9, eq. 63] to set $c_0/E = 1$ (as is certainly allowed when evaluating $F$ in expression (2.14) for the perihelion advance), $F = Bc_0$, which is the leading term in $2\dot{f}$ in [p. 8, eq. 57]. $F$ is thus twice the areal velocity which we can take to be a constant when evaluating expression (2.14):

$$F = 2\dot{f} = 2\,\frac{\pi ab}{T} = \frac{2\pi a^2\sqrt{1 - e^2}}{T}, \qquad \text{[eq. 102]}$$

where $a$, $b$, $T$, and $e$ are the planetary orbit's semi-major axis, semi-minor axis, period, and eccentricity, respectively. Besso used the triangle he drew next to [eq. 102] to help him reconstruct the formula $\pi ab = \pi a^2\sqrt{1 - e^2}$ for the area of an ellipse.

In [p. 1, eqs. 3 and 6], $A$ was defined as $2KM/c_0^2$. Setting the centripetal force $m\omega^2 r$, with $\omega = 2\pi/T$, equal to the gravitational force $KMm/r^2$ and setting $r$ equal to the major axis $a$ we find $4\pi^2 a^3/T^2 = KM$, which is just Kepler's third law (the ratio $a^3/T^2$ is the same for all planets in the solar system). It follows that

$$A = \frac{8\pi^2 a^3}{c_0^2 T^2}. \tag{2.15}$$

Using this relation along with [eq. 102], Besso could rewrite expression (2.14) for the perihelion advance in radians per half a revolution as

$$\frac{5}{16}\pi\frac{A^2c_0^2}{F^2} = \frac{5}{16}\pi A\left(\frac{8\pi^2a^3}{T^2}\right)\left(\frac{T^2}{4\pi^2a^4(1-e^2)}\right) = \frac{5}{8}\pi\frac{A}{a(1-e^2)}. \tag{2.16}$$

In the 1915 perihelion paper, Einstein (1915c, p. 838, Eq. (13)) found that his new theory predicted a perihelion advance of

$$3\pi\frac{\alpha}{a(1-e^2)}$$

radians per revolution, where $\alpha$ is just a different notation for $A$. The *Entwurf* theory thus predicts a perihelion advance $\frac{5/8}{3/2} = \frac{5}{12}$ the size of that predicted by general relativity, which in the case of Mercury amounts to 18 seconds of arc per century.

In the 1915 perihelion paper, Einstein switches from $r$ to $x = \frac{1}{r}$ and integrates an equation for $\frac{d\varphi}{dx}$ between the minimum and maximum values $\alpha_1$ and $\alpha_2$ of $x$ instead of integrating an equation for $\frac{d\varphi}{dr}$ of the form of [eq. 67] on [p. 9] between the minimum and maximum values $r_1$ and $r_2$ of $r$. This greatly simplifies the calculation. Einstein now only has to evaluate an integral of the form

$$\int_{\alpha_1}^{\alpha_2}\frac{dx}{\sqrt{-(x-\alpha_1)(x-\alpha_2)}}$$

(see Sec. 6.3.3, note 21). This integral does not have a pole at $x = 0$. The contour $A$ in the complex plane in Fig. 2.1 can thus be continuously deformed to a large circle with radius $R \gg \alpha_2$ and the integral above can be replaced by twice the integral of $\frac{1}{iz}$ over this circle. Writing $z = Re^{i\varphi}$ and $dz = iRe^{i\varphi}\,d\varphi$, we have

$$\int_{\alpha_1}^{\alpha_2}\frac{dx}{\sqrt{-(x-\alpha_1)(x-\alpha_2)}} = \frac{1}{2}\int_0^{2\pi}\frac{iRe^{i\varphi}\,d\varphi}{iRe^{i\varphi}} = \frac{1}{2}\int_0^{2\pi}d\varphi = \pi. \tag{2.17}$$

## [P. 26] Einstein

On [pp. 26 and 28], Einstein computed the numerical value of expression (2.14) derived by Besso on [p. 14] for the perihelion advance in radians per half a revolution for Mercury and converted the result to seconds of arc per century.

The top half of [p. 26] repeats equations derived by Besso on [p. 14]. [Eqs. 174–175] are the same as [p. 14, eqs. 98–100] for the relation between roots and coefficients of the fourth-order polynomial under the square-root sign in [p. 9, eq. 68] for the angle between perihelion and aphelion. [Eq. 177] for this angle is the same as [p. 14, eq. 101]. [Eq. 178] is the same as [p. 14, eq. 102]. [Eq. 179], originally introduced in [p. 1, eq. 6], was also used by Besso on [p. 14] (see Eq. (2.15)).

With the help of [eqs. 178 and 179], Einstein rewrote part of Besso's expression (2.14) as

$$\frac{Ac}{F} = c\frac{2KM}{c^2}\frac{T}{2\pi a^2\sqrt{1-e^2}} = \frac{KMT}{c\pi a^2\sqrt{1-e^2}} \qquad \text{[eq. 180]}$$

(where we dropped the subscript '0' in the velocity of light $c_0$).

The two columns immediately below [eq. 180] contain the logarithms of the quantities on the right-hand side of this equation. The first two numbers in the left column give the logarithms of $M$ and $T$, respectively; the first five numbers in the right column give the logarithms of $K^{-1}$, $c$, $\pi$, $a^2$, and $\sqrt{1-e^2}$, respectively. Subtracting the sum of these five numbers from the sum of $\log M$ and $\log T$, one arrives at $\log(Ac/F) = .5172 - 3$, which is the starting point of the calculation on [p. 28]. The calculations in the three lines immediately above [eq. 180] and the calculations to the left of the vertical line on [p. 26] show how the values for the various factors in $Ac/F$ were found that were used in the calculation to the right of the vertical line. With the exception of the value for $M$, these values are reasonably close to the modern values.

The value of $K$ is given as $1/3862^2$, the value of $T$ as 87.97 days. From the calculation here and from $a = .387 \cdot 1.48 \cdot 10^{13} = 5.72 \cdot 10^{12}$ cm ([eq. 266] on [p. 40], not included here), it can be inferred that $a$ was computed via the relation $a = (\bar{r}_m/\bar{r}_e)\bar{r}_e$, where $\bar{r}_m$ and $\bar{r}_e$ are the mean distances from the sun of Mercury and the earth, respectively. As Einstein jotted down next to the result for $\log c$: "Average distance set equal to semi-major axis"). Before calculating the logarithm of $1-e^2$, he wrote it as $(1-e)(1+e)$. He used the value .2056 for $e$.

As we already noted in Pt. I, Ch. 6, the value Einstein inserted for $M$ is off by a factor 10. As can be inferred from the calculation of $\log M$ on [p. 26], an almost identical calculation on [p. 34] (by Einstein, not included here), and the equation

$$M = 3.24 \cdot 10^5 \cdot 5.6 \cdot 1.08 \cdot 10^{2\langle 8 \rangle 7} = 1.96 \cdot 10^{3\langle 4 \rangle 3},$$

which occurs several times in the manuscript—[p. 30, eq. 185, Einstein] (see Pt. I, Fig. 6.3), [p. 35, eq. 211, Einstein with corrections by Besso] (see Pt. I, Fig. 6.2), and, without the corrections (and not included here), [p. 40, Einstein, eq. 264]—Einstein found the value for $M$ as the product of three factors, the ratio of the sun's mass to the earth's mass, the earth's density and the earth's volume. It is the value for the earth's volume that is a factor 10 too large.

Due to the error in $M$, the value Einstein found for $Ac/F$ is a factor 10 too large. Since the expression (2.14) involves the square of this quantity, this leads to a value for the motion of Mercury's perihelion that is a factor 100 too large.

In the crossed-out passage in the lower-right corner, Einstein began to convert the perihelion advance from radians per half a revolution to seconds of arc per century. He made a fresh start with this conversion on [p. 28].

## [P. 28] Einstein

Starting from the value for $\log(Ac/F)$ he had computed on [p. 26], which is off by a factor 10, Einstein found that

$$\frac{5}{16}\left(\frac{Ac}{F}\right)^2 = 3.4 \cdot 10^{-6},$$

which is off by a factor 100. Comparison with expression (2.14) shows that this should give the perihelion motion of Mercury in fractions of $\pi$ per half a revolution. Einstein recorded this same result on [p. 29] and at the bottom of [p. 40] (pages not included here). On [p. 30] (also not included here), Einstein once again computed the perihelion advance in fractions of $\pi$ per half a revolution. He made several mistakes in this calculation and arrived at a value of $1.65 \cdot 10^{-6}$. Right underneath this result, however, he wrote the correct result, $3.4 \cdot 10^{-8}$. This suggests that he had come to realize at that point that the result on [p. 28] was off by a factor of 100. [P. 28] gives no indication that Einstein realized his mistake. The mistake was first found, it seems, by Besso (see [p. 35, eq. 211] shown in Fig. 6.2 in Pt. I, Ch. 6).

To convert his result to seconds of arc per century, Einstein computed the logarithms of $180 \cdot 60 \cdot 60$, the number of seconds of arc in $\pi$ radians, and $2(365.2/87.97)100$, the number of half revolutions of Mercury in a century. He thus found the "precession per half a revolution in seconds of arc" and, finally, the "precession in 100 years" in seconds of arc: "$1821'' = 30'$." Next to this bizarre result, he wrote: "independently checked" (see the discussion in Pt. I, Ch. 6). The correct result of $18''$ is not stated anywhere in the manuscript, although, as we have seen, there are clear indications that Einstein and Besso discovered the source of the erroneous factor 100 in the result stated on [p. 28].

On an undated page of his so-called Scratch Notebook (CPAE3, Appendix A, [p. 61]), Einstein gave, without derivation, the following expression for the "advance per revolution":

$$10\pi^3 \left(\frac{a}{cT'}\right)^2.$$

One arrives at this expression by substituting Eq. (2.15) for $A$ and [p. 14, eq. 102] for $F$ into expression (2.14) for the perihelion advance in radians per half a revolution and multiplying the result by 2:

$$2 \cdot \frac{5}{16}\pi \left(\frac{Ac}{F}\right)^2 = \frac{5}{8}\pi \left(\frac{8\pi^2 a^3}{cT^2}\,\frac{T}{2\pi a^2\sqrt{1-e^2}}\right)^2 = 10\pi^3 \left(\frac{a}{cT'}\right)^2,$$

where $T' \equiv T\sqrt{1-e^2}$. Two years later, Einstein (1915c, p. 839, Eq. (14)) gave the final formula for the perihelion advance in the new theory in this same form. The only difference is that the new theory replaces the factor 10 by 24. On [p. 61] of the Scratch Notebook, Einstein used the formula of the *Entwurf* theory to compute the perihelion advance of Mercury and converted the result to seconds of arc per century. This time he arrived at essentially the correct (though disappointing) result: $17''$.

## [P. 41, top half] Einstein

On [pp. 41–42], Einstein used the same iterative approximation procedure used on [pp. 1–6] to find the metric field of the sun to check whether the rotation metric, the Minkowski metric in a slowly rotating Cartesian coordinate system, is a vacuum solution of the *Entwurf* equations (Janssen 1999).

On the final page of the Scratch Notebook mentioned above (CPAE3, Appendix A, [p. 66]), Einstein wrote three components of the rotation metric

$$g_{44} = 1 - \omega^2 r^2, \quad g_{14} = \omega y, \quad g_{24} = -\omega x,$$

where $\omega$ is the angular frequency of the rotation and $r \equiv \sqrt{x^2+y^2}$ (the term 1 in $g_{44}$ should be $c^2$). Right next to these components he wrote: "Is the first equation a consequence of the last two on the basis of the theory?" (pp. 137–138).

On [pp. 41–42] of the Einstein-Besso manuscript, Einstein tried to answer this question for the *Entwurf* theory by substituting the components of the rotation metric of first order in $\omega$ in the 44-component of the *Entwurf* field equations and solving for the $\omega^2$-term in $g_{44}$. The solution he found was

$$-\alpha^2\rho^2, \qquad \text{cf. [eq. 274]}$$

which is just the $\omega^2$-term in the 44-component of the rotation metric. (We will follow the notation of [pp. 41–42] in using $\rho$ rather than $r$ for $\sqrt{x^2+y^2}$ but the notation of the Scratch Notebook in using $\omega$ rather than $\alpha$ for the angular frequency.) Next to [eq. 274], he thus wrote: "Is correct" [*stimmt*].

Einstein, however, made several mistakes in this calculation. As can be inferred from [eq. 267] (in Besso's hand), [eq. 268] and [p. 42, eq. 282], he incorrectly read off the components of the metric from the line element, resulting in spurious factors of 2. He also lost a minus sign at some point. When these mistakes are corrected, the $\omega^2$-term in $g_{44}$ changes from $-\omega^2\rho^2$ to $-\frac{3}{4}\omega^2\rho^2$.

On [p. 42], Einstein similarly calculated the $\omega^2$-term in the 44-component of the contravariant metric. He discovered that the expressions he found for the 44-components of the covariant and the contravariant metric are incompatible with one another. However, he still did not correct the expression for $g_{44}$ he had found on [p. 41]. Instead, it appears, he convinced himself that he must have made an error in the calculation of $g^{44}$ on [p. 42]. All this strongly suggests that, when he did these calculations (probably in June 1913), he was firmly convinced that the rotation metric is a solution of the *Entwurf* field equations. Despite warnings recorded in the Besso memo of August 1913 (Janssen 2007, p. 785), it was only in September 1915 that Einstein finally accepted that it is not (see the letter to Freundlich presented in Ch. 4).

We reconstruct the derivations on [pp. 41–42], modernizing and enhancing Einstein's notation the same way we did in our commentary on [p. 1] (cf. Eq. (2.1)). Our reconstruction follows Janssen (1999, pp. 139–148).

On [p. 41], Einstein used the vacuum *Entwurf* field equations in their 'covariant' form:

$$D_{\mu\nu}(g) + \kappa t_{\mu\nu} = 0 \qquad \text{[eq. 269]} \tag{2.18}$$

(square quotes because neither the left-hand side as a whole nor the two terms it consists of are generally covariant tensors). In modernized notation, the first term is defined as

$$D_{\mu\nu}(g) \equiv \frac{1}{\sqrt{-g}}\left(g^{\alpha\beta}\sqrt{-g}\, g_{\mu\nu,\beta}\right)_{,\alpha} - g^{\alpha\beta} g^{\tau\rho} g_{\mu\tau,\alpha}\, g_{\nu\rho,\beta}, \tag{2.19}$$

while the second, representing ($\kappa$ times) the energy-momentum density of the gravitational field, is defined as

$$\kappa t_{\mu\nu} \equiv -\frac{1}{2} g_{\tau\rho,\mu}\, g^{\tau\rho}{}_{,\nu} + \frac{1}{4} g_{\mu\nu}\, g^{\alpha\beta}\, g_{\tau\rho,\alpha}\, g^{\tau\rho}{}_{,\beta} \qquad \text{[eq. 271]} \tag{2.20}$$

(Einstein and Grossmann 1913, pp. 16–17, Eqs. (21), (16), and (14), respectively).

It is easily seen that, to first order in $\omega$, the rotation metric is a solution of [eq. 269]. The spatial derivatives of the $\omega$-terms in the rotation metric are all $\omega$ or $\omega/c^2$. In first-order approximation, the one term in the *Entwurf* field equations with second-order derivatives of the metric thus vanishes and terms with products of first-order derivatives either vanish or are of order $\omega^2$ and can be neglected.

Einstein considered the 44-component of the field equations to second order,

$$\overset{(2)}{D}_{44}(g) + \kappa \overset{(2)}{t}_{44} = 0, \tag{2.21}$$

and inserted the $\omega$-terms of the rotation metric,

$$g^{14} = g^{41} = -\frac{\omega}{c^2}y, \quad g^{24} = g^{42} = \frac{\omega}{c^2}x, \qquad \text{[eq. 268] (corrected)}$$

$$g_{14} = g_{41} = -\omega y, \quad g_{24} = g_{42} = \omega x, \qquad \text{[p. 42, eq. 282] (corrected)}$$

into this equation. Einstein's expressions for these components are all off by a factor 2.

With the help of Eq. (2.19) and using that $\overset{(1)}{g}_{44} = 0$ and $g_{\mu\nu,4} = 0$ for the rotation metric, we find that

$$\overset{(2)}{D}_{44}(g) = \overset{(0)}{g}{}^{ij}\,\overset{(2)}{g}_{44,ij} - \overset{(0)}{g}{}^{\alpha\beta}\,\overset{(0)}{g}{}^{\tau\rho}\,\overset{(1)}{g}_{4\tau,\alpha}\,\overset{(1)}{g}_{4\rho,\beta}.$$

The first term on the right-hand side is just $-\Delta\overset{(2)}{g}_{44}$; the second gives two identical contributions, one for the index combination $(\alpha = \beta = 1, \tau = \rho = 2)$ and one for $(\alpha = \beta = 2, \tau = \rho = 1)$:

$$\left(\overset{(1)}{g}_{42,1}\right)^2 = \left(\overset{(1)}{g}_{41,2}\right)^2 = \omega^2.$$

Hence:

$$\overset{(2)}{D}_{44}(g) = -\Delta\overset{(2)}{g}_{44} - 2\,\omega^2. \qquad \text{[eq. 270] (corrected)}$$

Because of the erroneous factors 2 in [eq. 268] and [p. 42, eq. 282], [eq. 270] has $-8\omega^2$ instead of $-2\omega^2$.

With the help of Eq. (2.20), we find that

$$\kappa\,\overset{(2)}{t}_{44} = -\frac{1}{2}\,\overset{(1)}{g}_{\tau\rho,4}\,\overset{(1)}{g}{}^{\tau\rho}{}_{,4} + \frac{1}{4}\,\overset{(0)}{g}_{44}\,\overset{(0)}{g}{}^{\alpha\beta}\,\overset{(1)}{g}_{\tau\rho,\alpha}\,\overset{(1)}{g}{}^{\tau\rho}{}_{,\beta}.$$

The first term on the right-hand side vanishes; the second gives four identical contributions for the four index combinations written next to [eqs. 270–271]:

$$\tau = 1,\ \rho = 4,\ \alpha = \beta = 2;$$
$$\tau = 2,\ \rho = 4,\ \alpha = \beta = 1;$$
$$\tau = 4,\ \rho = 1,\ \alpha = \beta = 2;$$
$$\tau = 4,\ \rho = 2,\ \alpha = \beta = 1.$$

For the first of these index combinations, for instance, we find

$$\frac{1}{4}\,\overset{(0)}{g}_{44}\,\overset{(0)}{g}{}^{22}\,\overset{(1)}{g}_{14,2}\,\overset{(1)}{g}{}^{14}{}_{,2} = -\frac{c^2}{4}\,\frac{\omega^2}{c^2}.$$

Hence

$$\kappa\,\overset{(2)}{t}_{44} = -\omega^2.$$

The evaluation of the various factors in the second term of [eq. 271], written underneath this term, shows that Einstein, because of the erroneous factors 2 in [eq. 268] and [p. 42, eq. 282], set $\kappa\,\overset{(2)}{t}_{44}$ equal to

$$\frac{1}{4}\cdot c^2 \cdot -1 \cdot \frac{1}{c^2}\cdot 16\alpha^2 = -4\alpha^2$$

rather than $-\alpha^2$ (or, in our notation, $-\omega^2$).

Combining the correct expressions for $\overset{(2)}{D}_{44}(g)$ and $\kappa\,\overset{(2)}{t}_{44}$ for the rotation metric, we find

$$-\Delta\overset{(2)}{g}_{44} - 2\,\omega^2 - \omega^2 = 0 \qquad \text{[eq. 272] (corrected)}$$

Because of the spurious factors of 2 and a sign error, the last term on left-hand side of [eq. 272] is $+4\omega^2$ rather than $-\omega^2$. The sign error was probably the result of wishful thinking on Einstein's part. It ensures, as he probably foresaw at this point, that $\overset{(2)}{g}_{44}$ is just the $\omega^2$-term in the 44-component of the rotation metric.

The equation for $\overset{(2)}{g}_{44}$ thus reduces to

$$\Delta \overset{(2)}{g}_{44} = -3\,\omega^2. \qquad \text{[eq. 273] (corrected)}$$

To solve this equation, we need a formula for the action of the Laplacian $\Delta$ on a function of $\rho \equiv \sqrt{x^2+y^2}$. On [p. 42], in the two lines leading up to [eq. 285], we see Einstein derive this formula in the same way that he derived one for the action of $\Delta$ on a function of $r \equiv \sqrt{x^2+y^2+z^2}$ in [p. 4, eq. 28]. For an arbitrary function $f(\rho)$ we have:

$$\frac{\partial^2 f}{\partial x^2} = \frac{x^2}{\rho}\frac{d}{d\rho}\left(\frac{1}{\rho}\frac{df}{d\rho}\right) + \frac{1}{\rho}\frac{df}{d\rho}.$$

Adding a similar expression for $\partial^2 f/\partial y^2$, we find that

$$\Delta f = \rho\frac{d}{d\rho}\left(\frac{1}{\rho}\frac{df}{d\rho}\right) + \frac{2}{\rho}\frac{df}{d\rho}. \qquad \text{[p. 42, eq. 285] with } f = \gamma_{44}$$

With the help of this formula, one immediately sees that

$$\overset{(2)}{g}_{44} = -\frac{3}{4}\omega^2\rho^2 \tag{2.22}$$

is the solution of (the corrected version of) [eq. 273]. This means that the rotation metric is *not* a vacuum solution of the *Entwurf* field equations. Since Einstein's version of [eq. 273] has $-4\,\omega^2$ rather than $-3\,\omega^2$, he concluded that $\overset{(2)}{g}_{44} = -\omega^2\rho^2$ and that the rotation metric *is* a vacuum solution.

## [P. 42, top half, left part] Einstein

On [p. 42], Einstein used the vacuum *Entwurf* field equations in their 'contravariant' form (cf. [p. 41, eq. 269]):

$$\Delta_{\mu\nu}(\gamma) - \kappa\,\vartheta_{\mu\nu} = 0,$$

where

$$\Delta_{\mu\nu}(\gamma) \equiv \frac{1}{\sqrt{-g}}\left(g^{\alpha\beta}\sqrt{-g}\,g^{\mu\nu}_{\ \ ,\beta}\right)_{,\alpha} - g^{\alpha\beta}\,g_{\tau\rho}\,g^{\mu\tau}_{\ \ ,\alpha}\,g^{\nu\rho}_{\ \ ,\beta},$$

and

$$\kappa\,\vartheta_{\mu\nu} \equiv -\frac{1}{2}g^{\alpha\mu}g^{\beta\nu}g_{\tau\rho,\alpha}\,g^{\tau\rho}_{\ \ ,\beta} + \frac{1}{4}g^{\mu\nu}\,g^{\alpha\beta}\,g_{\tau\rho,\alpha}\,g^{\tau\rho}_{\ \ ,\beta}$$

(Einstein and Grossmann 1913, pp. 15–17, Eqs. (18), (15), and (13), respectively).

Einstein considered the 44-component of the field equations to second order,

$$\overset{(2)}{\Delta}_{44}(\gamma) - \kappa\,\overset{(2)}{\vartheta}_{44} = 0, \tag{2.23}$$

inserted the components of the rotation metric of first order in $\omega$ into this equation, and solved for the second-order contribution to $g^{44}$. He used [eq. 282] and [p. 41, eq. 268] for the first-order components, which are off by a factor of 2. Using the correct expressions for these $\omega$-terms, we find:

$$\overset{(2)}{\Delta}_{44}(\gamma) = -\Delta\overset{(2)}{g}{}^{44} - \overset{(0)}{g}{}^{\alpha\beta}\,\overset{(0)}{g}_{\tau\rho}\,\overset{(1)}{g}{}^{4\tau}_{\ \ ,\alpha}\,\overset{(1)}{g}{}^{4\rho}_{\ \ ,\beta} = -\Delta\overset{(2)}{g}{}^{44} - 2\frac{\omega^2}{c^4},$$

$$\kappa\,\overset{(2)}{\vartheta}_{44} = \frac{1}{4}\overset{(0)}{g}{}^{44}\,\overset{(0)}{g}{}^{\alpha\beta}\,\overset{(1)}{g}_{\tau\rho,\alpha}\,\overset{(1)}{g}{}^{\tau\rho}_{\ \ ,\beta} = -\frac{\omega^2}{c^4}.$$

The equation for $\overset{(2)}{\vartheta}{}^{44}$ thus becomes:

$$-\Delta \overset{(2)}{g}{}^{44} - 2\frac{\omega^2}{c^4} + \frac{\omega^2}{c^4} = 0, \qquad \text{[eq. 281] (corrected)}$$

or

$$\Delta \overset{(2)}{g}{}^{44} = -\frac{\omega^2}{c^4}. \qquad \text{[eq. 283] (corrected)}$$

In the manuscript, the coefficients of the $\omega^2$-terms in [eq. 281] and [eq. 283] are a factor 4 too large because of the erroneous factors of 2 in [eq. 282] and [p. 41, eq. 268].

With the help of [eq. 285], we see that

$$\overset{(2)}{g}{}^{44} = -\frac{1}{4}\frac{\omega^2}{c^4}\rho^2 \tag{2.24}$$

is the solution of (the corrected version of) [eq. 283]. The rotation metric has $\overset{(2)}{g}{}^{44} = 0$. So this calculation shows, once again, that the rotation metric is not a vacuum solution of the *Entwurf* field equations.

Einstein found $\overset{(2)}{g}{}^{44} = -\omega^2\rho^2/c^4$. In [eq. 284], he wrote $\gamma^{\times}_{44}$, his notation for $\overset{(2)}{g}{}^{44}$, as $\beta\rho^2$. From [eqs. 283, 285–286] he concluded that $\beta = -\alpha^2/c^4$, where $\alpha$ is his notation for $\omega$ ([eq. 287]). For $g^{44}$, Einstein would then have arrived at:

$$g^{44} = \frac{1}{c^2} - \frac{\omega^2}{c^4} \qquad \text{[eq. 288] without extra factor 3}$$

We conjecture that the factor 3 was only added to the last term of [eq. 188] after Einstein had found [eq. 289] for $g_{44}$, which must have been puzzling and disappointing to him. The reasoning behind this conjecture is as follows (Janssen 1999, pp. 147–148).

It looks as if Einstein used his result for $\overset{(2)}{g}{}^{44}$ and the condition $g_{\mu\alpha}g^{\alpha\nu} = \delta^{\nu}_{\mu}$ to derive an expression for $\overset{(2)}{g}_{44}$. Collecting terms of second order in $\omega$ in $g_{4\alpha}g^{\alpha 4} = 1$, we find:

$$\overset{(1)}{g}_{41}\overset{(1)}{g}{}^{14} + \overset{(1)}{g}_{42}\overset{(1)}{g}{}^{24} + \overset{(0)}{g}_{44}\overset{(2)}{g}{}^{44} + \overset{(2)}{g}_{44}\overset{(0)}{g}{}^{44} = \frac{\omega^2\rho^2}{c^2} + c^2\overset{(2)}{g}{}^{44} + \frac{1}{c^2}\overset{(2)}{g}_{44} = 0. \tag{2.25}$$

Inserting Eq. (2.24) for $\overset{(2)}{g}{}^{44}$, we simply recover Eq. (2.22) for $\overset{(2)}{g}_{44}$.

For Einstein, because of the extra factors of 2, Eq. (2.25) would have taken the form:

$$c^2\overset{(2)}{g}{}^{44} + \frac{\overset{(2)}{g}_{44}}{c^2} = -\frac{4\omega^2\rho^2}{c^2}.$$

Inserting $-\omega^2\rho^2/c^4$ for $\overset{(2)}{g}{}^{44}$ into this equation, one finds

$$\overset{(2)}{g}_{44} = -3\omega^2\rho^2.$$

This explains, we submit, why Einstein wrote

$$g_{44} = c^2\left(1 + \langle 3\rangle \frac{\omega^2}{c^2}\rho^2\right) \qquad \text{[eq. 289]} \tag{2.26}$$

on the line following [eq. 288] for $g^{44}$ (the plus sign in [eq. 289] should clearly have been a minus sign). This expression for $g_{44}$ differs from the one in [p. 41, eq. 274], next to which Einstein had written “is correct.” In fact, the factor 3 in [eq. 289] is correct: it is what Einstein would have found had he not made the sign error in [p. 41, eq. 272].

We conjecture that Einstein crossed out the factor 3 in [eq. 289] for $g_{44}$ and added it in [eq. 288] for $g^{44}$, thereby reassuring himself that the rotation metric really is a vacuum solution of the *Entwurf* field equations. It apparently did not bother him that he got the wrong result for $g^{44}$.

In [Eq. 290] (cf. [p. 8, eq. 49]), Einstein started to evaluate the equation of motion for a test particle in the metric field he found on [pp. 41–42] but this calculation breaks off after one line. Besso took up this same calculation on the bottom half of [p. 41]. He took it a few steps further than Einstein but it remains unclear what the purpose of the calculation was.

# Chapter 6
# The November 1915 Papers

## 6.3 Explanation of the perihelion motion of Mercury on the basis of the general theory of relativity

### 6.3.2 Translation

### *Explanation of the Perihelion Motion of Mercury from the General Theory of Relativity*

In a paper recently published in these proceedings, I introduced gravitational field equations that are covariant under arbitrary transformations with determinant 1. In a supplement I showed that these equations correspond to generally covariant ones if the scalar of the energy tensor of "matter" vanishes, and I showed that the introduction of this hypothesis, which robs time and space of the last vestige of objective reality, faces no fundamental objections.[1,[1]]

In this paper I find an important corroboration of this most radical theory of relativity, for it turns out that it explains qualitatively and quantitatively the secular precession of the orbit of Mercury (in the direction of the orbital motion), which was discovered by LEVERRIER and which amounts to 45″ per century.[2,[2]]

It also turns out that the theory entails a bending of light rays by gravitational fields that is twice the size of that found in my earlier investigations.

#### § 1. The Gravitational Field

From my last two communications it follows that the gravitational field in a vacuum has to satisfy, given a properly chosen reference frame, the equations

$$\sum_{\alpha} \frac{\partial \Gamma^{\alpha}_{\mu\nu}}{\partial x_{\alpha}} + \sum_{\alpha\beta} \Gamma^{\alpha}_{\mu\beta} \Gamma^{\beta}_{\nu\alpha} = 0, \tag{1}$$

---

[1] In a forthcoming communication it will be shown that one can dispense with this hypothesis. The only thing that matters is that it is possible to choose the reference frame in such a way that the determinant $|g_{\mu\nu}|$ takes on the value $-1$. The following investigation is independent of this.

[2] E. FREUNDLICH recently wrote a noteworthy article on the impossibility of satisfactorily explaining the anomalies in the motion of Mercury on the basis of the NEWTONian theory (*Astronomische Nachrichten 201*, 49 [1915]).

where $\Gamma_{\mu\nu}^{\alpha}$ is defined by the equations

$$
\begin{aligned}
\Gamma_{\mu\nu}^{\alpha} &= -\begin{Bmatrix}\mu\nu\\ \alpha\end{Bmatrix} = -\sum_{\beta} g^{\alpha\beta}\begin{bmatrix}\mu\nu\\ \beta\end{bmatrix}\\
&= -\frac{1}{2}\sum_{\beta} g^{\alpha\beta}\left(\frac{\partial g_{\mu\beta}}{\partial x_{\nu}} + \frac{\partial g_{\nu\beta}}{\partial x_{\mu}} - \frac{\partial g_{\mu\nu}}{\partial x_{\beta}}\right).
\end{aligned} \tag{2}
$$

If, moreover, we adopt the hypothesis argued for in my last communication, that the scalar of the energy tensor of "matter" always vanishes, we get an additional equation for the determinant:

$$|g_{\mu\nu}| = -1. \tag{3}$$

Suppose a point mass (the sun) is located at the origin of the coordinate system. The gravitational field this point mass produces can be calculated from these equations through iterative approximation.[3]

One should keep in mind, however, that the $g_{\mu\nu}$ for a given solar mass are not fully determined mathematically by equations (1) and (3). This is because these equations are covariant under arbitrary transformations with determinant 1. Yet it might be justified to assume that all these solutions can be reduced to one another by such transformations and that (for given boundary conditions) they differ only formally, not physically. Following this conviction, I am satisfied for the time being to derive *one* solution, without entering into a discussion of whether it is the only possible one.[4]

We proceed as follows. Let the $g_{\mu\nu}$ in "zeroth approximation" be given by the following scheme corresponding to the original theory of relativity:

$$
\left.\begin{matrix}
-1 & 0 & 0 & 0\\
0 & -1 & 0 & 0\\
0 & 0 & -1 & 0\\
0 & 0 & 0 & +1
\end{matrix}\right\}, \tag{4}
$$

or, abbreviated,

$$
\left.\begin{aligned}
g_{\rho\sigma} &= \delta_{\rho\sigma}\\
g_{\rho 4} = g_{4\rho} &= 0\\
g_{44} &= 1
\end{aligned}\right\}. \tag{4a}
$$

Here $\rho$ and $\sigma$ refer to the indices 1, 2, 3; $\delta_{\rho\sigma}$ is equal to 1 or to 0 depending on whether $\rho = \sigma$ or $\rho \neq \sigma$.

In the following, we assume that the $g_{\mu\nu}$ only differ from the values in equation (4a) by quantities that are small compared to unity. We will treat these deviations as small quantities of "first order" and functions of the $n$th degree in these deviations as "quantities of $n$th order". Equations (1) and (3) enable us, starting with equation (4a), to calculate through successive approximations the gravitational field up to quantities of $n$th order. The equation (4a) form the "zeroth approximation".[5]

The solution given below has the following properties, which fix the coordinate system:

1. All components are independent of $x_4$.
2. The solution is spatially symmetric around the origin of the coordinate system, in the sense that we get back to the same solution if we subject it to a linear orthogonal (spatial) transformation.
3. The equations $g_{\rho 4} = g_{4\rho} = 0$ hold exactly (for $\rho = 1, 2, 3$).
4. At infinity, $g_{\mu\nu}$ has the values in equation (4a).

### First approximation

It is easy to verify that, up to quantities of first order, equations (1) and (3) as well as the four conditions just listed are satisfied if we posit[6]

$$\left.\begin{aligned} g_{\rho\sigma} &= -\delta_{\rho\sigma} + \alpha\left(\frac{\partial^2 r}{\partial x_\rho \partial x_\sigma} - \frac{\delta_{\rho\sigma}}{r}\right) = -\delta_{\rho\sigma} - \alpha\frac{x_\rho x_\sigma}{r^3} \\ g_{44} &= 1 - \frac{\alpha}{r} \end{aligned}\right\}. \tag{4b}$$

The $g_{4\rho}$ and $g_{\rho 4}$ are fixed by condition 3, $r$ stands for the quantity $+\sqrt{x_1^2 + x_2^2 + x_3^2}$, and $\alpha$ is a constant determined by the mass of the sun.

One immediately sees that condition 3 is satisfied to first order. For a simple way to see that the field equations (1) are also satisfied in first approximation, one only needs to note that, if quantities of second and higher order are neglected, the left-hand side of equation (1) can be replaced by[7]

$$\sum_\alpha \frac{\partial \Gamma^\alpha_{\mu\nu}}{\partial x_\alpha}$$

$$\sum_\alpha \frac{\partial}{\partial x_\alpha}\begin{bmatrix}\mu\nu \\ \alpha\end{bmatrix},$$

where $\alpha$ only runs from 1 to 3.

As one sees from Eq. (4b), our theory implies that in the case of a mass at rest, the components $g_{11}$ through $g_{33}$ already differ from zero in terms of quantities of first order.[8] As we shall see later, no discrepancy with Newton's law (in first approximation) arises from this. The theory, however, does give a slightly different result for the effect of a gravitational field on a light ray than my earlier papers. The velocity of light is determined by the equation

$$\sum g_{\mu\nu} dx_\mu dx_\nu = 0. \tag{5}$$

Applying HUYGENS's principle, we find from equations (5) and (4b), after a simple calculation, that a light ray passing by the sun at a distance $\Delta$ experiences an angular deflection of magnitude $\frac{2\alpha}{\Delta}$, while my earlier calculation, which was not based on the hypothesis $\sum T_\mu^\mu = 0$, led to the value $\frac{\alpha}{\Delta}$. A light ray grazing the surface of the sun should experience a deflection of 1.7″ (instead of 0.85″).[9] By contrast, the result concerning the shift of spectral lines by the gravitational potential, the order of magnitude of which Mr. FREUNDLICH has confirmed for the fixed stars, remains unaffected, because this depends only on $g_{44}$.[10]

Now that we have obtained the $g_{\mu\nu}$ in first approximation, we can also calculate the components $\Gamma_{\mu\nu}^\alpha$ of the gravitational field in first approximation. From equations (2) and (4b) we have

$$\Gamma_{\rho\sigma}^\tau = -\alpha\left(\delta_{\rho\sigma}\frac{x_\tau}{r^3} - \frac{3x_\rho x_\sigma x_\tau}{2r^5}\right) \tag{6a}$$

where $\rho, \sigma, \tau$ take on the values 1, 2, 3, and

$$\Gamma_{44}^\sigma = \Gamma_{4\sigma}^4 = -\frac{\alpha x_\sigma}{2r^3}, \tag{6b}$$

where $\sigma$ takes on the values 1, 2, 3. Those components in which the index 4 appears once or three times vanish.

## SECOND APPROXIMATION

As we will see later, we only need to determine the three components $\Gamma_{44}^\sigma$ to second order to determine the orbits of the planets to the same degree of accuracy. For this purpose, the last field equation suffices along with the general conditions we have imposed on our solution. If we take into account equation (6b) and neglect quantities of third and higher order, the last field equation,

$$\sum_\sigma \frac{\partial \Gamma_{44}^\sigma}{\partial x_\sigma} + \sum_{\sigma\tau} \Gamma_{4\tau}^\sigma \Gamma_{4\sigma}^\tau = 0,$$

turns into[11]

$$\sum_\sigma \frac{\partial \Gamma_{44}^\sigma}{\partial x_\sigma} = -\frac{\alpha^2}{2r^4}.$$

From this we deduce, using equation (6b) and the symmetry properties of our solution,[12]

$$\Gamma_{44}^\sigma = -\frac{\alpha}{2}\frac{x_\sigma}{r^3}\left(1 - \frac{\alpha}{r}\right). \tag{6c}$$

## § 2. The Motion of the Planets

The equations of motion of a point mass in a gravitational field given by the general theory of relativity is

$$\frac{d^2x_\nu}{ds^2} = \sum_{\sigma\tau} \Gamma^\nu_{\sigma\tau} \frac{dx_\sigma}{ds}\frac{dx_\tau}{ds}. \tag{7}$$

We first show that these equations contain the NEWTONian equations of motion in first approximation. If the motion of the planet is with a velocity that is small compared to the velocity of light, then $dx_1$, $dx_2$ and $dx_3$ are small compared to $dx_4$. We thus obtain a first approximation by keeping only the term with $\sigma = \tau = 4$ on the right-hand side. Using equation (6b), we then find:

$$\left.\begin{aligned} \frac{d^2x_\nu}{ds^2} &= \Gamma^\nu_{44} = -\frac{\alpha}{2}\frac{x_\nu}{r^3} \quad (\nu = 1,2,3) \\ \frac{d^2x_4}{ds^2} &= 0 \end{aligned}\right\}. \tag{7a}$$

These equations show that we can set $s = x_4$ in first approximation. The first three equations are thus exactly the NEWTONian equations. If we introduce polar coordinates $r$, $\varphi$ in the orbital plane, then, as is well known, the energy law and the area law give the equations[13]

$$\left.\begin{aligned} \frac{1}{2}u^2 + \Phi &= A \\ r^2\frac{d\varphi}{ds} &= B \end{aligned}\right\}, \tag{8}$$

where $A$ and $B$ are the constants of the energy law and the area law, respectively, and where we used the abbreviations

$$\left.\begin{aligned} \Phi &= -\frac{\alpha}{2r} \\ u^2 &= \frac{dr^2 + r^2 d\varphi^2}{ds^2} \end{aligned}\right\} \tag{8a}$$

We now have to evaluate the equations to the next order. The last of the equations (7) then yields, together with equation (6b),[14]

$$\frac{d^2x_4}{ds^2} = 2\sum_\sigma \Gamma^4_{\sigma 4}\frac{dx_\sigma}{ds}\frac{dx_4}{ds} = -\frac{dg_{44}}{ds}\frac{dx_4}{ds},$$

or, correct to the first order,

$$\frac{dx_4}{ds} = 1 + \frac{a}{r}. \tag{9}$$

We now turn to the first of the three equations (7).[15] The right-hand side gives:

a) for the index combination $\sigma = \tau = 4$

$$\Gamma^{\nu}_{44}\left(\frac{dx_4}{ds}\right)^2,$$

or, with the help equations (6c) and (9), to second order:

$$-\frac{\alpha}{2}\frac{x_\nu}{r^3}\left(1+\frac{\alpha}{r}\right);$$

b) for the index combination $\sigma \neq 4$, $\tau \neq 4$ (which are the only ones we still need to consider), given that the products $\frac{dx_\sigma}{ds}\frac{dx_\tau}{ds}$ should, on the basis of equation (8), be seen as quantities of first order,[1] likewise up to quantities of second order:

$$-\frac{\alpha x_\nu}{r^3}\sum\left(\delta_{\sigma\tau}-\frac{3}{2}\frac{x_\sigma x_\tau}{r^2}\right)\frac{dx_\sigma}{ds}\frac{dx_\tau}{ds}.$$

The summation gives

$$-\frac{\alpha x_\nu}{r^3}\left(u^2-\frac{3}{2}\left(\frac{dr}{ds}\right)^2\right).$$

Using these results we obtain the equations of motion in a form correct to second order,

$$\frac{d^2x_\nu}{ds^2} = -\frac{\alpha}{2}\frac{x_\nu}{r^3}\left(1+\frac{\alpha}{r}+2u^2-3\left(\frac{dr}{ds}\right)^2\right), \tag{7b}$$

which together with equation (9) determine the motion of the mass point. As an aside, I note that equations (7b) and (9) do give deviations from KEPLER's third law in the case of circular motion.

From equation (7b) follows, first of all, the exact validity of the equation

$$r^2\frac{d\varphi}{ds} = B, \tag{10}$$

where $B$ is a constant. The area law therefore remains valid to second order if we use the planet's "proper time" to measure time. To determine the secular rotation of the orbital ellipse from equation (7b), it is most convenient to replace the quantities of first order in the expression in parentheses on the right-hand side with the help of (10) and the first of equations (8), which does not affect quantities of second order on the right-hand side. The expression

[1] Because of this we only need the field components $\Gamma^{\nu}_{\sigma\tau}$ to first order as given in equation (6a).

in parentheses then takes the form[16]

$$\left(1 - 2A + \frac{3B^2}{r^2}\right).$$

Finally, if we choose $s\sqrt{(1-2A)}$ as the time variable and call that variable $s$ again, we have, with a slightly different definition of the constant $B$;[17]

$$\left.\begin{aligned} \frac{d^2 x_\nu}{ds^2} &= -\frac{\partial \Phi}{\partial x_\nu} \\ \Phi &= -\frac{\alpha}{2}\left[1 + \frac{B^2}{r^2}\right] \end{aligned}\right\}. \tag{7c}$$

To determine the equation of the orbit, we now proceed exactly as in the NEWTONian case. From equation (7c) we obtain first[18]

$$\frac{dr^2 + r^2 d\varphi^2}{ds^2} = 2A - 2\Phi.$$

If we eliminate $ds$ from this equation with the help of (10), we find

$$\left(\frac{dx}{d\varphi}\right)^2 = \frac{2A}{B^2} + \frac{\alpha}{B^2}x - x^2 + \alpha x^3, \tag{11}$$

where $x$ is defined as $1/r$. The only difference between this equation differs and the corresponding one in NEWTONian theory is the last term on the right-hand side.

The angle traversed by the radius vector between perihelion and aphelion is therefore given by the integral for the ellipse[19]

$$\varphi = \int_{\alpha_1}^{\alpha_2} \frac{dx}{\sqrt{\frac{2A}{B^2} + \frac{\alpha}{B^2}x - x^2 + \alpha x^3}},$$

where $\alpha_1$ and $\alpha_2$ are the roots of the equation

$$\frac{2A}{B^2} + \frac{\alpha}{B^2}x - x^2 + \alpha x^3 = 0,$$

which will be very close to the roots of the equation obtained if the last term is omitted.

To the degree of accuracy required, this angle is given by[20]

$$\varphi = \left[1 + \frac{\alpha(\alpha_1 + \alpha_2)}{2}\right] \cdot \int_{\alpha_1}^{\alpha_2} \frac{dx}{\sqrt{-(x-\alpha_1)(x-\alpha_2)(1-\alpha x)}},$$

or, upon expansion of $(1-\alpha x)^{-1/2}$,

$$\varphi = \left[1+\frac{\alpha(\alpha_1+\alpha_2)}{2}\right]\cdot\int\limits_{\alpha_1}^{\alpha_2}\frac{\left(1+\frac{\alpha}{2}x\right)dx}{\sqrt{-(x-\alpha_1)(x-\alpha_2)}}.$$

The result of the integration is[21]

$$\varphi = \pi\left[1+\frac{3}{4}\alpha(\alpha_1+\alpha_2)\right],$$

or if we bear in mind that $\alpha_1$ and $\alpha_2$ are the reciprocals of the maximum and minimum distance from the sun, respectively,

$$\varphi = \pi\left(1+\frac{3}{2}\frac{\alpha}{\alpha(1-e^2)}\right). \tag{12}$$

In one complete orbit, the perihelion thus advances by

$$\varepsilon = 3\pi\frac{\alpha}{\alpha(1-e^2)} \tag{13}$$

in the direction of the orbital motion, where $a$ is the semimajor axis and $e$ is the eccentricity. If we introduce the orbital period $T$ (in seconds) and the velocity of light in cm/sec., we obtain[22]

$$\varepsilon = 24\pi^3\frac{\alpha^2}{T^2c^2(1-e^2)}. \tag{14}$$

Calculation gives a perihelion advance of 43″ per century for the planet Mercury, while astronomers find an unexplained difference between observations and NEWTONian theory of $45'' \pm 5''$ per century. Hence there is now complete agreement.

For Earth and Mars, astronomers have found advances of 11″ and 9″ per century, respectively, whereas our formula gives 4″ and 1″, respectively. However, it would seem that little weight is to be attached to these values because the eccentricity of the orbits of these planets is too small. What determines the reliability of the observation of the perihelion motion is the product of this motion and the eccentricity $\left(e\frac{d\pi}{dt}\right)$. Considering the values for these quantities given by NEWCOMB,

| | $e\frac{d\pi}{dt}$ | |
|---|---|---|
| Mercury | 8.48″ | ± 0.43 |
| Venus | −0.05 | ± 0.25 |
| Earth | 0.10 | ± 0.13 |
| Mars | 0.75 | ± 0.35, |

which I owe to Dr. FREUNDLICH,[23] one gets the impression that an advance of the perihelion has clearly been demonstrated only for Mercury. A definitive verdict on these matters, however, I am happy to leave to professional astronomers.

### *6.3.3 Commentary*

This paper is the written version of a lecture delivered to the Prussian Academy of Sciences, Berlin, November 18, 1915. It was published November 25, 1915. It is presented in facsimile as Doc. 24 in CPAE6. Our commentary is based on Earman and Janssen (1993, secs. 5–7, pp. 138–158). In the translation, we corrected a number of typos in the German original. Some more important ones are noted in the annotation (see notes 7, 11, 17 and 19). Our translation follows the one by Brian Doyle used in the companion volume to CPAE6, originally published in Lang and Gingerich (1979).

## Introduction

1. In his first November 1915 paper, Einstein (1915a) had found a condition that prohibited the use of unimodular coordinates even though these are the natural coordinates to use in the theory covariant under unimodular transformations that Einstein proposed in this paper. In the second and fourth paper, Einstein (1915b, 1915d) proposed ways around this prohibition, by setting $T = 0$ and adding a term with $T$ to the field equations, respectively. This allowed him to look upon the (amended) field equations of the first paper as generally covariant equations expressed in unimodular coordinates.
2. In this paper, Einstein's protégé Freundlich (1915) criticized a paper on the perihelion problem by a leading astronomer (Seeliger 1915, cf. Sec. 4.3, note 1). For the history of the perihelion problem, see Roseveare (1982).

## § 1

3. This iterative approximation procedure follows the one used in the Einstein-Besso manuscript (see Ch. 2, [pp. 1, 3, 4, 6 and 7]).
4. In a letter to Einstein of January 1, 1916 (CPAE8, Doc. 177a in CPAE13), Ehrenfest quoted the last two sentences of this paragraph, calling them "your incantation" (*Deine Beschwörungsformel*) and adding that "Leiden takes note of this with an icy-polite smile but does not endorse it." In the Einstein-Besso manuscript, Besso already raised the question whether the solution for the static field of the sun is unique (CPAE4, Doc. 14, [p. 16]). This question may have triggered the "hole argument" in the original form in which it can be found in the Besso memo of August 28, 1913 (Janssen 2007, p. 820). Ehrenfest's skeptical comment was actually made in the context of a discussion of the hole argument. In a letter to Ehrenfest of January 16, 1916, also in the context of the hole argument, Lorentz also questioned the uniqueness of the solution of the field equations, both in the specific case of the field of the sun and in general (Kox 2018, Doc. 179). Birkhoff (1923) would show that the Schwarzschild solution, the exact solution corresponding to the approximate one Einstein gives here (see note 6) is the unique spherically symmetric vacuum solution of the Einstein field equations (for an elementary proof of this result, known as Birkhoff's theorem, see Carroll 2004, sec. 5.2, pp. 197–204).

5. The actual calculations make it clear that Einstein is assuming that $g_{\mu\nu}$ can be written as a power series

$$g_{\mu\nu} = \overset{(0)}{g}_{\mu\nu} + \overset{(1)}{g}_{\mu\nu} + \overset{(2)}{g}_{\mu\nu} + \ldots,$$

as he did in the Einstein-Besso manuscript (see our commentary on [p. 1] in Sec. 2.3 and Earman and Janssen 1993, pp. 142–148). In the Einstein-Besso manuscript, the components of the metric, the potential for the gravitational field, are calculated to second order; in this paper, the components of gravitational field itself, defined as minus the Christoffel symbols, are.

6. Derivations of the general form of a static spherically symmetric metric in Cartesian coordinates compatible with Newtonian theory for weak fields can be found, e.g., on [p. 7] of the Einstein-Besso manuscript (see Ch. 2) and in Droste (1915, pp. 999–1000). Here we follow Earman and Janssen (1993, p. 144, though our choice of the symbols $R$, $T$ and $\Phi$ below follows the notation used in the Einstein-Besso manuscript; $T$ is not to be confused with the trace of the energy-momentum tensor for matter). To find the metric at some arbitrary point $P$ with spatial coordinates $(x,y,z)$ in some arbitrary Cartesian coordinate system in a spherically symmetric space-time, we rotate the coordinate system such that the $x$-axis of the new coordinate system goes through $P$. The spatial coordinates of $P$ in this new coordinate system are $(x',y',z') = (r,0,0)$, where $r \equiv \sqrt{x^2+y^2+z^2}$. Because of spherical symmetry, the metric in primed coordinates has the simple form $g'_{\mu\nu} = \mathrm{diag}(R,T,T,\Phi)$, where $R$, $T$ and $\Phi$ are yet to be determined functions of the coordinates. Transforming back to the coordinate system we started from (cf. Sec. 2.3, commentary on [p. 7]), we find a metric of the form:

$$g_{\mu\nu} = T\,\delta_{ij} + \frac{x^i x^j}{r^2}\,(R-T), \quad g_{i4} = g_{4i} = 0, \quad g_{44} = \Phi$$

$(i,j = 1,2,3)$. Since the space-time is Minkowskian at spatial infinity, $T$ must be equal to $-1$. To recover the Newtonian theory for weak static fields, $\Phi$ must be equal to $1 - \frac{\alpha}{r}$ with $\alpha \equiv \frac{2GM}{c^2}$ (where $G$ is Newton's gravitational constant, $M$ the mass of the sun and $c$ the velocity of light). To ensure that $g = RT^2\Phi = -1$ to first order in $\frac{\alpha}{r}$, $R$ must be equal to $-\left(1+\frac{\alpha}{r}\right)$, which means that $R - T = -\frac{\alpha}{r}$. We thus arrive at Eq. (4b):

$$g_{ij} = -\delta_{ij} - \alpha\,\frac{x^i x^j}{r^3}, \quad g_{44} = 1 - \frac{\alpha}{r}. \tag{4b}$$

The line element corresponding to this metric field is

$$ds^2 = \left(1 - \frac{\alpha}{r}\right) c^2 dt^2 - \sum_{i,j=1}^{3} \left(\delta_{ij} + \alpha\,\frac{x^i x^j}{r^3}\right) dx^i dx^j.$$

In spherical coordinates $(r,\vartheta,\varphi)$, which Einstein could not use because the transformation from Cartesian to spherical coordinates does not have determinant 1, this line element becomes

$$ds^2 = \left(1 - \frac{\alpha}{r}\right) c^2 dt^2 - \left(1 + \frac{\alpha}{r}\right) dr^2 - r^2\left(d\vartheta^2 + \sin^2\vartheta\, d\varphi^2\right),$$

which, to order $\dfrac{\alpha}{r}$, is just the line element of the Schwarzschild solution,

$$ds^2 = \left(1 - \frac{\alpha}{r}\right) c^2 dt^2 - \frac{1}{1 - \dfrac{\alpha}{r}} dr^2 - r^2 \left(d\vartheta^2 + \sin^2\vartheta\, d\varphi^2\right).$$

7. To check that the metric field in Eq. (4b) is a solution of $\Gamma^{\alpha}_{\mu\nu,\alpha} = 0$, the vacuum field equations to first order in $\alpha/r$, we first need to verify Eqs. (6a) and (6b) for the components of the gravitational field by inserting Eq. (4b) for $g_{\mu\nu}$ into the definition of $\Gamma^{\alpha}_{\mu\nu}$ in Eq. (2):

$$\Gamma^{\alpha}_{\mu\nu} = -\frac{1}{2} g^{\alpha\rho} \left(g_{\rho\mu,\nu} + g_{\rho\nu,\mu} - g_{\mu\nu,\rho}\right). \tag{2}$$

As Einstein notes in the two lines below Eqs. (6a) and (6b), $\Gamma^{\alpha}_{\mu\nu}$ is zero if either one of its indices is 4 or all three of them are. Since $g^{ij} \approx -\delta^{ij}$ and $g^{44} \approx 1$ and since $g_{i4} = 0$ and $g_{\mu\nu,4} = 0$, we have

$$\Gamma^{i}_{j4} = \Gamma^{i}_{4j} = \frac{1}{2} \delta^{ik} \left(g_{kj,4} + g_{k4,j} - g_{j4,k}\right) = 0,$$

$$\Gamma^{4}_{ij} = -\frac{1}{2} \left(g_{4i,j} + g_{4j,i} - g_{ij,4}\right) = 0, \quad \Gamma^{4}_{44} = -\frac{1}{2} g_{44,4} = 0.$$

If two of the three indices of $\Gamma^{\alpha}_{\mu\nu}$ are 4, we have $\Gamma^{i}_{44} = \dfrac{1}{2}\left(2g_{i4,4} - g_{44,i}\right) = -\dfrac{1}{2} g_{44,i}$ and $\Gamma^{4}_{4i} = -\dfrac{1}{2} g_{44,i}$. Inserting $g_{44} = 1 - \dfrac{\alpha}{r}$ on the right-hand sides of these equations, we arrive at Eq. (6b):

$$\Gamma^{i}_{44} = \Gamma^{4}_{4i} = \frac{1}{2} \frac{\partial}{\partial x^i} \left(\frac{\alpha}{r}\right) = -\frac{\alpha}{2r^2} \frac{\partial r}{\partial x^i} = -\frac{\alpha}{2} \frac{x^i}{r^3}. \tag{6b}$$

That leaves the case where none of the indices are 4. In that case,

$$\begin{aligned} \Gamma^{k}_{ij} &= \frac{1}{2} \delta^{kl} \left(g_{li,j} + g_{lj,i} - g_{ij,l}\right) \\ &= -\frac{\alpha}{2} \left( \frac{\partial}{\partial x^j} \left(\frac{x^k x^i}{r^3}\right) + \frac{\partial}{\partial x^i} \left(\frac{x^k x^j}{r^3}\right) - \frac{\partial}{\partial x^k} \left(\frac{x^i x^j}{r^3}\right) \right). \end{aligned}$$

Evaluating the derivatives, we find:

$$\begin{aligned} \Gamma^{k}_{ij} = -\frac{\alpha}{2} \Bigg\{ &\left(\delta^k_j x^i + \delta^i_j x^k + \delta^k_i x^j + \delta^j_i x^k - \delta^i_k x^j - \delta^j_k x^i\right) \frac{1}{r^3} \\ &+ x^k x^i \frac{\partial}{\partial x^j} \left(\frac{1}{r^3}\right) + x^k x^j \frac{\partial}{\partial x^i} \left(\frac{1}{r^3}\right) - x^i x^j \frac{\partial}{\partial x^k} \left(\frac{1}{r^3}\right) \Bigg\}. \end{aligned}$$

The six terms in parentheses on the first line add up to $2\,\delta_{ij} x^k$. Noting that

$$\frac{\partial}{\partial x^j} \left(\frac{1}{r^3}\right) = -\frac{3}{r^4} \frac{\partial r}{\partial x^j} = -\frac{3x^j}{r^5},$$

we see that the three terms on the second line are the same, except for the minus sign in front of the third. We thus arrive at Eq. (6a) (after correcting a typo: $r^2$ in the first term should be $r^3$):

$$\Gamma^{k}_{ij} = -\alpha \left( \delta_{ij} \frac{x^k}{r^3} - \frac{3}{2} \frac{x^i x^j x^k}{r^5} \right). \tag{6a}$$

Inserting Eqs. (6a) and (6b) into $\Gamma^{\alpha}_{\mu\nu,\alpha} = 0$, we now verify that the metric in Eq. (4b) is indeed a solution of the vacuum field equations to first order in $\alpha/r$. For the index combinations $\mu\nu = 4i$ (or, equivalently, $i4$), we get $\Gamma^{\alpha}_{\mu\nu,\alpha} = \Gamma^{4}_{4i,4} + \Gamma^{j}_{4i,j}$, which vanishes, since $\Gamma^{\alpha}_{\mu\nu}$ is time-independent and $\Gamma^{j}_{4i} = 0$. For the index combination $\mu\nu = 44$, we find

$$\Gamma^{i}_{44,i} = -\frac{\alpha}{2}\frac{\partial}{\partial x^i}\left(\frac{x^i}{r^3}\right) = -\frac{\alpha}{2}\left(\frac{3}{r^3} + x^i\frac{\partial}{\partial x^i}\left(\frac{x^i}{r^3}\right)\right) = 0,$$

where in the last step we used that $\frac{\partial}{\partial x^i}\left(\frac{x^i}{r^3}\right) = -\frac{3x^i}{r^5}$ and that $\sum_{i=1}^{3}(x^i)^2 = r^2$. That leaves the index combinations $\mu\nu = ij$:

$$\Gamma^{k}_{ij,k} = -\alpha\left(\delta_{ij}\frac{\partial}{\partial x^k}\left(\frac{x^k}{r^3}\right) - \frac{3}{2}\frac{\partial}{\partial x^k}\left(\frac{x^i x^j x^k}{r^5}\right)\right).$$

The first term on the right-hand side vanishes, as we just saw. The derivative in the second term gives:

$$\frac{\partial}{\partial x^k}\left(\frac{x^i x^j x^k}{r^5}\right) = \left(\delta^{ik}x^j x^k + \delta^{jk}x^i x^k + 3x^i x^j\right)\frac{1}{r^5} - \frac{5x^i x^j x^k}{r^6}\frac{\partial r}{\partial x^k}.$$

The first term on the right-hand side gives $\frac{5x^i x^j}{r^5}$, which cancels against the last term, since $\frac{\partial r}{\partial x^k} = \frac{x^k}{r}$ and $\sum_k (x^k)^2 = r^2$. Hence, $\Gamma^{\alpha}_{\mu\nu,\alpha} = 0$ for all values of $\mu$ and $\nu$. The metric in Eq. (4b) is a solution of the vacuum field equations to order $\alpha/r$.

8. Einstein mentioned this feature in three of the four postcards to Besso in which he reported his success in accounting for the perihelion motion of Mercury after the failure of their efforts to do so on the basis of the *Entwurf* theory recorded in the Einstein-Besso manuscript (see Ch. 2 and Pt. I, Ch. 6). As he told Besso in the fourth of these, on January 3, 1916: "That the effect is so much larger than in our calculation is because in the new theory the $g_{11}$–$g_{33}$ appear in first order and thus contribute to the perihelion motion" (CPAE8, Doc. 178). Ever since he had started working on his metric theory of gravity, Einstein had expected weak static fields to be represented by a spatially flat metric (see, e.g., Norton 1984). His calculation of the perihelion motion of Mercury showed him that this need not be the case. This realization removed obstacles to (a) the use of unimodular coordinates (see Sec. 6.2.3, note 4) and (b) the addition of a term with the trace of the energy-momentum tensor to the field equations (see Sec. 6.4.3, note 3).
9. Einstein (1911) derived the original value for the bending of light in the field of the sun directly from the equivalence principle. In the review article on *Entwurf* theory the previous year, Einstein (1914, p. 1084) indicated how this result is recovered in that theory. In the review article on the new theory the following year, Einstein (1916, pp. 821–822) went through the derivation of the result of the new theory. At this point, Einstein assumed that the use of unimodular coordinates required that he set the trace $T$ of the energy-momentum tensor for matter equal to zero. He would soon discover that he could use these coordinates without putting any restrictions on $T$. For accounts of the 1919 (and subsequent) eclipse expeditions in which Einstein's new prediction for light bending was tested, see Kennefick (2019) and Crelinsten (2006).
10. For discussion of Freundlich's controversial claim to have confirmed Einstein's prediction of a gravitational redshift, see Sec. 4.3, note 1, and Hentschel (1992, 1994).
11. The 44-component of the field equations to second order in $\frac{\alpha}{r}$ is

$$\overset{(2)}{\Gamma}{}^{i}_{44,i} + \overset{(1)}{\Gamma}{}^{\alpha}_{4\beta}\overset{(1)}{\Gamma}{}^{\beta}_{4\alpha} = 0,$$

where we have enhanced Einstein's notation to indicate to what order the different $\Gamma$'s need to evaluated. Using the expressions we found in note 7 for $\overset{(1)}{\Gamma}{}^{\alpha}_{\mu\nu}$ (note that $\overset{(0)}{\Gamma}{}^{\alpha}_{\mu\nu} = 0$) and recalling, in particular, that this quantity vanishes if one or three of its indices are 4, we have:

$$\overset{(1)}{\Gamma}{}^{\alpha}_{4\beta}\overset{(1)}{\Gamma}{}^{\beta}_{4\alpha} = \overset{(1)}{\Gamma}{}^{i}_{44}\overset{(1)}{\Gamma}{}^{4}_{4i} + \overset{(1)}{\Gamma}{}^{4}_{4i}\overset{(1)}{\Gamma}{}^{i}_{44} = 2\sum_{i=1}^{3}\left(-\frac{\alpha}{2}\frac{x^i}{r^3}\right)\left(-\frac{\alpha}{2}\frac{x^i}{r^3}\right) = \frac{\alpha^2}{2r^4},$$

where we used Eq. (6b) for $\overset{(1)}{\Gamma}{}^{4}_{4i} = \overset{(1)}{\Gamma}{}^{i}_{44}$ and $\sum_i (x^i)^2 = r^2$. We thus arrive at the equation above Eq. (6c) (after correcting a typo: there is a minus sign missing on the right-hand side):

$$\Gamma^{i}_{44,i} = -\frac{\alpha^2}{2r^4}.$$

We already saw that $\overset{(1)}{\Gamma}{}^{i}_{44,i} = 0$ (see note 7) so $\Gamma^{i}_{44}$ can be replaced by $\overset{(2)}{\Gamma}{}^{i}_{44}$ in this equation.

12. We check that

$$\Gamma^{i}_{44} = \overset{(1)}{\Gamma}{}^{i}_{44} + \overset{(2)}{\Gamma}{}^{i}_{44} = -\frac{\alpha}{2}\frac{x^i}{r^3}\left(1-\frac{\alpha}{r}\right) \tag{6c}$$

is a solution of the 44-component of the field equations to second order in $\frac{\alpha}{r}$ (see note 11). Using that $\overset{(1)}{\Gamma}{}^{i}_{44,i} = 0$ (see note 7) and inserting $\overset{(2)}{\Gamma}{}^{i}_{44} = \frac{\alpha^2}{2}\frac{x^i}{r^4}$ on the left-hand side of this component of the field equations, we find:

$$\begin{aligned}\overset{(2)}{\Gamma}{}^{i}_{44,i} &= \frac{\alpha^2}{2}\sum_{i=1}^{3}\frac{\partial}{\partial x^i}\left(\frac{x^i}{r^4}\right) = \frac{\alpha^2}{2}\left(\frac{3}{r^4} - \sum_{i=1}^{3}\frac{4x^i}{r^5}\frac{x^i}{r}\right)\\ &= \frac{\alpha^2}{2}\left(\frac{3}{r^4} - \frac{4}{r^4}\right) = -\frac{\alpha^2}{2}\frac{1}{r^4},\end{aligned}$$

which is in accordance with the right-hand side of this component of the field equations (see note 11).

## § 2

13. Einstein's analysis of the slow motion of a planet in the spherically symmetric weak static field of the sun in this section closely follows the analysis of the same problem in the context of the *Entwurf* theory in the Einstein-Besso manuscript (cf. Ch. 2, [pp. 8–11, 14 and 26]). As we noted in our commentary in Sec. 2.3 on [p. 8] of the manuscript, where Besso derived the analogue of Eq. (8) in this paper for the conservation of energy and angular momentum (the area law), it follows from the virial theorem that potential and kinetic energy, and hence the quantities $\frac{\alpha}{r}$ and $\left(\frac{dx^i}{ds}\right)^2$, are of the same order of magnitude, as Einstein notes explicitly under point (b) on p. 836.

14. The only contribution to the $\nu = 4$ component of Eq. (7) of second order in $\frac{\alpha}{r}$ comes from the index combinations $\sigma\tau = i4$ and $\sigma\tau = 4i$. Using Eq. (6b) for $\overset{(1)}{\Gamma}{}^{4}_{i4}$, we find that

$$\frac{d^2x^4}{ds^2} = 2\overset{(1)}{\Gamma}{}^4_{i4}\frac{dx^i}{ds}\frac{dx^4}{ds} = -\alpha\sum_{i=1}^{3}\frac{x^i}{r^3}\frac{dx^i}{ds}\frac{dx^4}{ds}.$$

Since $\dfrac{dx^4}{ds} = 1$ to zeroth order and

$$\sum_{i=1}^{3} x^i\frac{dx^i}{ds} = \frac{1}{2}\frac{d}{ds}\Big(\sum_{i=1}^{3}(x^i)^2\Big) = \frac{1}{2}\frac{dr^2}{ds} = r\frac{dr}{ds},$$

this reduces, to first order in $\dfrac{\alpha}{r}$, to

$$\frac{d}{ds}\left(\frac{dx^4}{ds}\right) = -\frac{\alpha}{r^2}\frac{dr}{ds} = \frac{d}{ds}\left(\frac{\alpha}{r}\right),$$

from which, again to first order, Einstein's Eq. (9) follows

$$\frac{dx^4}{ds} = 1 + \frac{\alpha}{r}. \tag{9}$$

15. To second order in $\dfrac{\alpha}{r}$ the spatial components of the equations of motion in Eq. (7) are

$$\frac{d^2x^i}{ds^2} = \left(\overset{(1)}{\Gamma}{}^i_{44} + \overset{(2)}{\Gamma}{}^i_{44}\right)\frac{dx^4}{ds}\frac{dx^4}{ds} + 2\overset{(1)}{\Gamma}{}^i_{4j}\frac{dx^4}{ds}\frac{dx^j}{ds} + \overset{(1)}{\Gamma}{}^i_{jk}\frac{dx^j}{ds}\frac{dx^k}{ds}.$$

The second term on the right-hand side vanishes because $\overset{(1)}{\Gamma}{}^i_{4j} = 0$ (see note 7). Following Einstein, we evaluate the first term on the right-hand side under (a) and the third under (b).

(a) Using Eq. (6c) for $\overset{(1)}{\Gamma}{}^i_{44} + \overset{(2)}{\Gamma}{}^i_{44}$ (see note 12) and Eq. (9) for $\dfrac{dx^4}{ds}$ (see note 14), we find:

$$\left(\overset{(1)}{\Gamma}{}^i_{44} + \overset{(2)}{\Gamma}{}^i_{44}\right)\frac{dx^4}{ds}\frac{dx^4}{ds} = -\frac{\alpha}{2}\frac{x^i}{r^3}\left(1 - \frac{\alpha}{r}\right)\left(1 + 2\frac{\alpha}{r}\right) = -\frac{\alpha}{2}\frac{x^i}{r^3}\left(1 + \frac{\alpha}{r}\right).$$

(b) Using equation (6a) for $\overset{(1)}{\Gamma}{}^i_{jk}$ (see note 7) and restoring the summation signs, we have:

$$\sum_{j,k=1}^{3}\overset{(1)}{\Gamma}{}^i_{jk}\frac{dx^j}{ds}\frac{dx^k}{ds} = -\frac{\alpha x^i}{r^3}\sum_{j,k=1}^{3}\left(\delta_{jk} - \frac{3}{2}\frac{x^jx^k}{r^2}\right)\frac{dx^j}{ds}\frac{dx^k}{ds}.$$

Substituting $u^2$, the square of the velocity of the planet in its orbit (see Eq. (8a)), for $\sum_{i,j}\delta_{ij}\dfrac{dx^i}{ds}\dfrac{dx^j}{ds}$ and $r\dfrac{dr}{ds}$ for $\sum_i\dfrac{dx^i}{ds}$ (see note 14), we can rewrite this as

$$\sum_{j,k=1}^{3}\overset{(1)}{\Gamma}{}^i_{jk}\frac{dx^j}{ds}\frac{dx^k}{ds} = -\frac{\alpha x^i}{r^3}\left(u^2 - \frac{3}{2}\left(\frac{dr}{ds}\right)^2\right).$$

Combining the results found under (a) and (b), we arrive at Eq. (7b), the correction to the Newtonian Eq. (7a) once terms of order $\left(\dfrac{\alpha}{r}\right)^2$ are taken into account:

$$\frac{d^2x^i}{ds^2} = -\frac{\alpha}{2}\frac{x^i}{r^3}\left(1 + \frac{\alpha}{r} + 2u^2 - 3\left(\frac{dr}{ds}\right)^2\right). \tag{7b}$$

16. With the help of Eq. (8a) for $u^2$ in polar coordinates, the expression in parentheses on the right-hand side of Eq. (7b) can be rewritten as

$$1+\frac{\alpha}{r}+2u^2-3\left(u^2-r^2\frac{d\varphi^2}{ds^2}\right).$$

If we substitute $\frac{B}{r^2}$ for $\frac{d\varphi}{ds}$ (see Eq. (10)), $2A-2\Phi$ for $u^2$ (see Eq. (8)), and $-2\Phi$ for $\frac{\alpha}{r}$ (see Eq. (8a)), this turns into the expression above Eq. (7c):

$$1-2\Phi-2A+2\Phi+3r^2\left(\frac{B^2}{r^4}\right)=1-2A+3\frac{B^2}{r^2}.$$

17. Using the result derived in note 16, we can rewrite Eq. (7b) as

$$\frac{d^2x^i}{ds^2}=-\frac{\alpha}{2}\frac{x^i}{r^3}\left(1-2A+3\frac{B^2}{r^2}\right).$$

If both sides are divided by $1-2A$, this turns into

$$\frac{1}{1-2A}\frac{d^2x^i}{ds^2}=-\frac{\alpha}{2}\frac{x^i}{r^3}\left(1+\frac{3B^2}{(1-2A)r^2}\right).$$

Introducing the new time variable $s'\equiv s\sqrt{1-2A}$ and the new area-law constant $B'\equiv\frac{B}{\sqrt{1-2A}}$, we can rewrite this as

$$\frac{d^2x^i}{ds'^2}=-\frac{\alpha}{2}\frac{x^i}{r^3}\left(1+3\frac{B'^2}{r^2}\right).$$

As we will check below, the right-hand side is equal to minus the gradient of the potential

$$\Phi\equiv-\frac{\alpha}{2r}\left(1+\frac{B'^2}{r^2}\right). \tag{7c}$$

The factor in parentheses is a correction to the Newtonian potential, also called $\Phi$, in Eq. (8a) (there is a typo in Eq. (7c): a factor $1/r$ is missing on the right-hand side). Taking minus the gradient of this potential, we find:

$$-\frac{\partial\Phi}{\partial x^i}=\frac{\partial}{\partial x^i}\left(\frac{\alpha}{2r}+\frac{\alpha B'^2}{2r^3}\right)=-\frac{\alpha}{2}\frac{1}{r^2}\frac{\partial r}{\partial x^i}-\frac{\alpha B'^2}{2r}\frac{3}{r^4}\frac{\partial r}{\partial x^i}.$$

Substituting $\frac{x^i}{r}$ for $\frac{\partial r}{\partial x^i}$, we recover the right-hand side of the equations of motion above:

$$-\frac{\partial\Phi}{\partial x^i}=-\frac{\alpha}{2}\frac{x^i}{r^3}\left(1+3\frac{B'^2}{r^2}\right).$$

18. Energy conservation, $\frac{1}{2}u^2+\Phi=A$, allows us to write (see the equation above Eq. (11)):

$$\frac{dr^2+r^2d\varphi^2}{ds^2}=2A-2\Phi=2A+\frac{\alpha}{r}+\frac{\alpha B^2}{r^3},$$

where, following Einstein, we wrote $u^2$ in polar coordinates (see Eq. (8a)) and dropped the prime on the new area-law constant.

Using the area law $r^2 \frac{d\varphi}{ds} = B$ (see Eq. (10)) to eliminate $ds$ from the equation above, we obtain a differential equation for $\frac{d\varphi}{dr}$. Integrating this equation between the values of $r$ for perihelion and aphelion, we expect to find a result slightly deviating from $\pi$, with the small deviation giving the motion of the perihelion per half a revolution. This is how the perihelion motion was calculated in the Einstein-Besso manuscript (see Ch. 2, [pp. 9–11, 14]) and it is how Einstein calculates it here, except that he uses $x = \frac{1}{r}$ and derives (and then integrates) a differential equation for $\frac{d\varphi}{dx}$.

If $(r^2/B)\,d\varphi$ is substituted for $ds$, the left-hand side of the equation above becomes:

$$\frac{dr^2 + r^2 d\varphi^2}{\left(r^4/B^2\right) d\varphi^2} = B^2 \left( \frac{1}{r^4} \frac{dr^2}{d\varphi^2} + \frac{1}{r^2} \right).$$

If we switch from $r$ to $\frac{1}{x}$, noting that $dr = -\frac{dx}{x^2}$, this expression reduces to

$$B^2 \left( \frac{dx^2}{d\varphi^2} + x^2 \right)$$

and the equation we started from turns into Einstein's Eq. (11)

$$\frac{dx^2}{d\varphi^2} = \frac{2A}{B^2} + \frac{\alpha}{B^2} x - x^2 + \alpha x^3. \tag{11}$$

19. On [pp. 10–11] of the Einstein-Besso manuscript (see Ch. 2), Einstein performed some contour integrations to evaluate similar integrals to find a formula for the perihelion motion of a planet in the field of the sun as given by the *Entwurf* theory. The switch from $r$ to $x = 1/r$ simplifies these calculations considerably. For a modern version, see Møller (1972, pp. 495–497, Eqs. (12.56)–(12.65)).
20. We first consider the corresponding calculation in Newtonian theory. In that case, the term $\alpha x^3$ in Eq. (11) is missing. The quadratic expression that remains can be written as $\left(x - \alpha_1\right)\left(\alpha_2 - x\right)$, where $\alpha_1$ and $\alpha_2$ are the values of $x = 1/r$ at aphelion and perihelion, respectively. So we have

$$\alpha_1 + \alpha_2 = \frac{1}{r_1} + \frac{1}{r_2} = \frac{1}{a(1+e)} + \frac{1}{a(1-e)} = \frac{2}{a(1-e^2)},$$

where $a$ is the semi-major axis of the planet's elliptical orbit and $e$ is the orbit's eccentricity. Integrating the resulting differential equation,

$$\frac{d\varphi}{dx} = \frac{1}{\sqrt{\frac{2A}{B^2} + \frac{\alpha}{B^2} x - x^2}} = \frac{1}{\sqrt{-\left(x - \alpha_1\right)\left(x - \alpha_2\right)}},$$

between aphelion (where $x$ has its minimum value $\alpha_1$) and perihelion (where $x$ has its maximum value $\alpha_2$) gives $\pi$ (see Eq. 2.17 in Sec. 2.3). In Newtonian theory the perihelion of a single planet in the field of the sun will not show any movement.

If the term $\alpha x^3$ is taken into account, the third-order polynomial on the right-hand side of Eq. (11) can be written as

$$\alpha\left(x - \alpha_1\right)\left(x - \alpha_2\right)\left(x - \alpha_3\right),$$

where two of the three roots, as Einstein notes, will only differ insignificantly from $\alpha_1$ and $\alpha_2$ in the Newtonian case. Eq. (11) tells us that the coefficient of $x^2$ should

be −1. It follows that

$$\alpha(\alpha_1+\alpha_2+\alpha_3)=1.$$

This relation allows us to replace the factor $(x-\alpha_3)$ by an expression depending only on the other two roots. Using that $\alpha\alpha_3 = 1-\alpha(\alpha_1+\alpha_2)$, we can write

$$\alpha(\alpha_3-x)=\alpha\alpha_3-\frac{\alpha x}{\alpha\alpha_3}=\Big(1-\alpha(\alpha_1+\alpha_2)\Big)\left(1-\frac{\alpha x}{1-\alpha(\alpha_1+\alpha_2)}\right).$$

To first order in $\alpha$, the polynomial can thus be rewritten as

$$-(x-\alpha_1)(x-\alpha_2)(1-\alpha(\alpha_1+\alpha_2))(1-\alpha x).$$

Eq. (11), again to first order in $\alpha$, then gives the following differential equation:

$$\frac{d\varphi}{dx}=\frac{1}{\sqrt{-(x-\alpha_1)(x-\alpha_2)(1-\alpha(\alpha_1+\alpha_2))(1-\alpha x)}}$$

$$=\frac{1+\dfrac{\alpha(\alpha_1+\alpha_2)}{2}+\dfrac{\alpha}{2}x}{\sqrt{-(x-\alpha_1)(x-\alpha_2)}},$$

which leads to the first two of the three equations for $\varphi$ above Eq. (12) (after correction of a typo in both of them: there is a factor $\frac{1}{2}$ missing in the term $\alpha(\alpha_1+\alpha_2)$ in the factor in front of the integral).

21. To find an expression for the angle $\varphi$ between aphelion and perihelion, we need the values of two integrals:

$$\int_{\alpha_1}^{\alpha_2}\frac{dx}{\sqrt{-(x-\alpha_1)(x-\alpha_2)}}=\pi$$

(see note 20 and Sec. 2.3, Eq. 2.17) and

$$\int_{\alpha_1}^{\alpha_2}\frac{x\,dx}{\sqrt{-(x-\alpha_1)(x-\alpha_2)}}=\frac{\alpha_1+\alpha_2}{2}\pi.$$

Splitting this second integral into two parts,

$$\int_{\alpha_1}^{\alpha_2}\left(\frac{\left(x-\dfrac{\alpha_1+\alpha_2}{2}\right)dx}{\sqrt{-(x-\alpha_1)(x-\alpha_2)}}+\frac{\dfrac{\alpha_1+\alpha_2}{2}\,dx}{\sqrt{-(x-\alpha_1)(x-\alpha_2)}}\right),$$

we see that the first part vanishes (the integrand is anti-symmetric around the middle of the interval $[\alpha_1,\alpha_2]$), while the second part is a constant times the first integral. To first order in $\alpha$, $\varphi$ is thus given by

$$\varphi=\left(1+\frac{\alpha(\alpha_1+\alpha_2)}{2}\right)\pi+\frac{\alpha}{2}\left(\frac{\alpha_1+\alpha_2}{2}\pi\right)=\pi\left(1+\frac{3}{4}\alpha(\alpha_1+\alpha_2)\right).$$

Substituting $\dfrac{2}{a(1-e^2)}$ for $\alpha_1+\alpha_2$ (see note 20), we arrive at Eq. (12).

22. Using Kepler's third law—which, as Einstein noted under Eq. (7b) on p. 837, continues to hold in his new theory—to substitute $\frac{8\pi^2 a^3}{T^2 c^2}$ for $\alpha$ (cf. Eq. (2.15) in our commentary on [p. 14] of the Einstein-Besso manuscript in Sec. 2.3), we get from Eq. (13) to Eq. (14):

$$\varepsilon = 3\pi \frac{\alpha}{a(1-e^2)} = 24\pi^3 \frac{a^2}{T^2 c^2 (1-e^2)}. \tag{13, 14}$$

This is more than double (i.e., $\frac{12}{5}$ times) the size of the effect found in the *Entwurf* theory (see Eq. 2.16 in Sec. 2.3). See our commentary on [p. 28] of the Einstein-Besso manuscript in Sec. 2.3 for the conversion of the perihelion advance from radians per revolution to seconds of arc per century.

23. The values listed here are part of a table in Newcomb (1895, p. 119) and are reproduced in Freundlich (1915, Col. 52). For discussion, see Roseveare (1982, pp. 51–52).